\documentclass[lettersize,journal]{IEEEtran}
\usepackage{amsmath,amsfonts}
\usepackage{algorithmic}
\usepackage{algorithm}
\usepackage{array}
\usepackage[caption=false,font=footnotesize,labelfont=rm,textfont=rm]{subfig}
\usepackage{textcomp}
\usepackage{stfloats}
\usepackage{url}
\usepackage{verbatim}
\usepackage{graphicx}
\usepackage{cite}

\usepackage{hyperref}

\usepackage{pifont}
\usepackage{xcolor}
\usepackage{makecell}

\usepackage{multirow}
\usepackage{xcolor,color,framed}

\begin{document}

\title{Secure Physical Layer Communications for Low-Altitude Economy Networking: A Survey}

\author{Lingyi Cai, Jiacheng Wang, Ruichen Zhang, Yu Zhang,   Tao Jiang,~\IEEEmembership{Fellow,~IEEE}, Dusit Niyato,~\IEEEmembership{Fellow,~IEEE},\\ Xianbin Wang,~\IEEEmembership{Fellow,~IEEE},  Abbas Jamalipour,~\IEEEmembership{Fellow,~IEEE}, and Xuemin Shen,~\IEEEmembership{Fellow,~IEEE}
        % <-this % stops a space

%\thanks{This work was supported in part by the xx, xx, xx.}

%\thanks{Lingyi Cai, xx, and Tao Jiang are with the Research Center of 6G Mobile Communications, School of Cyber Science and Engineering, Huazhong University of Science and Technology, Wuhan, 430074, China (e-mail: lingyicai@hust.edu.cn; yueyuedai@ieee.org; qiweihu@hust.edu.cn; tao.jiang@ieee.org).}% <-this % stops a space
\thanks{Lingyi Cai is with the Research Center of 6G Mobile Communications, School of Cyber Science and Engineering, Huazhong University of Science and Technology, Wuhan, 430074, China, and also with
the College of Computing and Data Science, Nanyang Technological University, Singapore (e-mail: lingyicai@hust.edu.cn).}
\thanks{Jiacheng Wang, Ruichen Zhang, and Dusit Niyato are with the College of Computing and Data Science, Nanyang Technological University, Singapore (e-mails: jiacheng.wang@ntu.edu.sg; ruichen.zhang@ntu.edu.sg; dniyato@ntu.edu.sg).}
\thanks{Yu Zhang and Tao Jiang are with the Research Center of 6G Mobile Communications, School of Cyber Science and Engineering, Huazhong University of Science and Technology, Wuhan, 430074, China (e-mail: yuzhang123@hust.edu.cn; tao.jiang@ieee.org).}
\thanks{Xianbin Wang is with the Department of Electrical and Computer Engineering, Western University, London, ON, N6A 5B9, Canada (e-mail: xianbin.wang@uwo.ca).}
\thanks{Abbas Jamalipour is with the School of Electrical and Computer Engineering,
University of Sydney, Australia (e-mail: a.jamalipour@ieee.org).}
\thanks{Xuemin Shen is with the Department of Electrical and Computer Engineering, University of Waterloo, Waterloo, ON N2L 3G1, Canada (e-mail: sshen@uwaterloo.ca).}
}

%\markboth{Journal of \LaTeX\ Class Files,~Vol.~14, No.~8, August~2021}


\maketitle

\begin{abstract}
The Low-Altitude Economy Networking (LAENet) is emerging as a transformative paradigm that enables an integrated and sophisticated communication infrastructure to support aerial vehicles in carrying out a wide range of economic activities within low-altitude airspace. However, the physical layer communications in the LAENet face growing security threats due to inherent characteristics of aerial communication environments, such as signal broadcast nature and channel openness. These challenges highlight the urgent need for safeguarding communication confidentiality, availability, and integrity. In view of the above, this survey comprehensively reviews existing secure countermeasures for physical layer communication in the LAENet. We explore core methods focusing on anti-eavesdropping and authentication for ensuring communication confidentiality. Subsequently, availability-enhancing techniques are thoroughly discussed for anti-jamming and spoofing defense. Then, we review approaches for safeguarding integrity through anomaly detection and injection protection. Furthermore, we discuss future research directions, emphasizing energy-efficient physical layer security, multi-drone collaboration for secure communication, AI-driven security defense strategy, space-air-ground integrated security architecture, and 6G-enabled secure UAV communication. This survey may provide valuable references and new insights for researchers in the field of secure physical layer communication for the LAENet.

\end{abstract}


\section{Introduction}

\begin{figure*}[!t]
\centering
\includegraphics[width=7.1in]{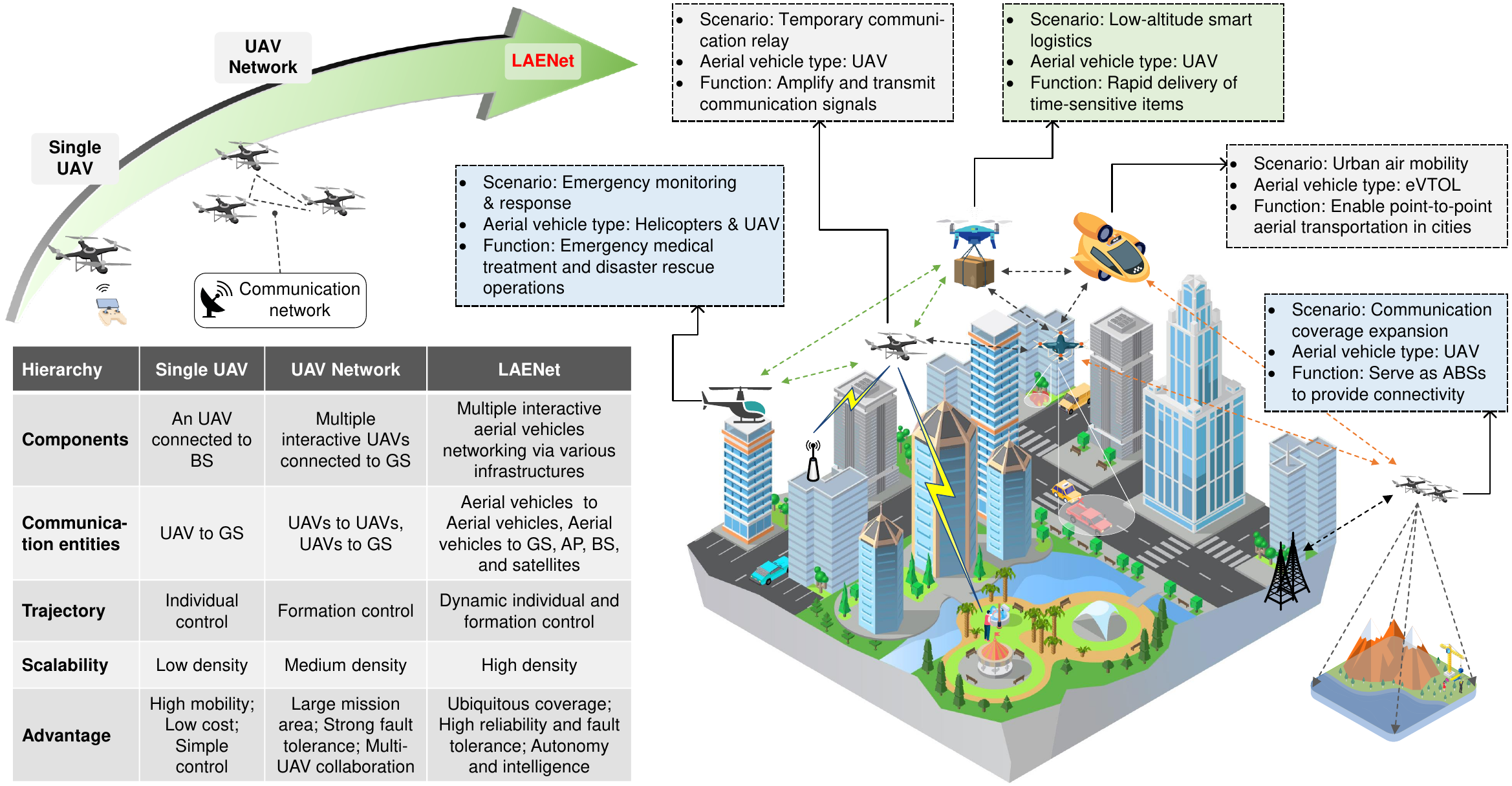}
\caption{The overall architecture of the LAENet covers the main application scenarios, including emergency monitoring and response, temporary communication relay, communication coverage expansion, low-altitude smart logistics, and urban air mobility. The table compares the similarities and differences between the LAENet, single UAV, and UAV networks, representing the evolution of the LAENet.}
\label{LAENetjiagoutu}
\end{figure*}

\IEEEPARstart{W}{ith} the rapid development of aerial vehicle technologies and communication networks, the concept of Low-Altitude Economic Networking (LAENet) has emerged to enable more comprehensive, large-scale, and intelligent connectivity to support various low-altitude activities \cite{10759668,zhao2025generative,9765519,china2023central2}, such as intelligent transportation, logistics delivery, communication enhancement, disaster monitoring, and emergency response \cite{8869712,9645207,7572034,10833672}, as shown in Fig. \ref{LAENetjiagoutu}. The LAENet is built upon earlier frameworks of single Unmanned Aerial Vehicle (UAV) operation and multi-UAV networks. A single UAV typically maintains a direct link to a ground station or base station, operating with simple control procedures and delivering cost-effective services but with limited range and scalability \cite{8742658}. The UAV network focuses on formation control and multi-UAV collaboration, enabling broader mission areas and stronger fault tolerance \cite{8742658,10916952,9745441}. Advancing from these foundations, the LAENet integrates various aerial vehicles into a high-density communication network, connecting them not only to ground stations but also to other platforms such as base stations, access points, and even satellites \cite{9200679,10879807}. Thus, the LAENet can enable ubiquitous coverage, high reliability, robust fault tolerance, greater autonomy, and intelligence.

Specifically, the LAENet refers to an integrated network system that connects various low-altitude flight operations, including general aviation, drones, electric vertical take-off and landing (eVTOL) aircraft, and other aerial platforms, within the designated low-altitude airspace (typically below 1,000 meters, and in some cases extending up to 3,000 meters) \cite{10759668,10879807}. The LAENet serves as a vital bridge between ground-based economies and airspace resources, which will drive technological innovation and unlock substantial social and economic benefits \cite{10681882,zheng2025uav}. The Civil Aviation Administration of China estimates that the country's low-altitude market will soar from 500 billion Chinese yuan (about 70 billion US dollars) in 2023 to 1.5 trillion Chinese yuan (about 200 billion US dollars) in 2025 and as much as 3.5 trillion Chinese yuan (about 480 billion US dollars) in 2035 \cite{CNLAEHS}. Currently, research institutions and enterprises across multiple regions in China are continuously advancing and expanding innovative research and commercial applications of UAVs and eVTOLs in low-altitude activities \cite{CNLAETFNE}. Meanwhile, in the United States, the Federal Aviation Administration has confirmed its commitment to actively promoting the development of electric air taxis and integrating this type of aircraft into the national airspace \cite{USFATMC}.

\begin{table}[htbp]
\centering
\caption{SUMMARY OF RELATED SURVEYS}
\label{tab:relevant_surveys}
\begin{tabular}{|c|p{0.36\textwidth}|}
\hline
\multicolumn{1}{|c|}{\textbf{References}} & \multicolumn{1}{c|}{\textbf{Focus}} \\
\hline
\cite{10.1145/3485272} & A review of cybersecurity threats, countermeasures, and research gaps in UAV networks, with a focus on emerging attack surfaces and commercial UAV applications \\
\hline
\cite{10.1145/3703625} & A survey of security threats, vulnerabilities, and countermeasures in UAV swarm networks, with a focus on classifying attack types and reviewing emerging defense technologies \\
\hline
\cite{10793113} & A review of security threats, vulnerabilities, and countermeasures in UAVs and Flying Ad Hoc Networks with attack surface analysis with simulation-based evaluation \\
\hline
\cite{HADI2023103607} & A survey of vulnerabilities across software, hardware, and communication layers in UAV systems, and an exploration of emerging defense technologies \\
\hline
\cite{9488323} & A survey of security challenges in drone communication and a review of emerging technologies used to enhance the speed, reliability, and security of UAV networks \\
\hline
\cite{10.1145/3673225} & A review of UAV security challenges, existing controls, and future research directions, with an emphasis on the transformative role of AI in enabling secure UAV systems \\
\hline
\cite{10599134} & A review of security threats classified from a cyberspace security perspective and countermeasures in UAV systems \\
\hline
\cite{9946859} & A survey of security threats, requirements, and countermeasures in UAV-aided Internet of Things (IoT) applications \\
\hline
\cite{KUMAR2024110695} & A survey of cybersecurity vulnerabilities and countermeasures in UAV systems, integrating threat classification, communication protocols, and emerging techniques \\
\hline
\cite{9900257} & A survey of PLS in UAV communications, focusing on key challenges, methodologies, and recent advancements for both static and mobile UAV deployment scenarios \\
\hline
\cite{8675384} & A review of security challenges, practical deployment aspects, and standardization progress associated with integrating UAVs into cellular networks \\
\hline
\cite{10236463} & A survey of layer-wise cybersecurity threats and AI-enabled countermeasures in UAV-assisted IoT applications \\
\hline
%[3] & Opportunities, challenges, open problems, and mathematical tools for UAV base stations and cellular-connected drone-UEs. \\
%\hline
\end{tabular}
\end{table}

In the LAENet, physical layer communication serves as a critical foundation for wireless communication between aerial vehicles and between aerial vehicles and communication infrastructure \cite{10003076,9900257,10916952}. The physical layer converts digital data from higher protocol layers into signals suitable for transmission over aerial communication channels \cite{10496520,8233654,10599123}. This process encompasses encoding data into bit sequences, modulating them onto carrier waves, and ensuring reliable signal propagation through the wireless medium \cite{10496520,10110330,9927483}. At the receiver side, the physical layer performs inverse operations, including demodulating the incoming signals, decoding the bit sequences, and passing the data to upper layers for further processing \cite{LIU2021589,10461999,10163877}. Therefore, the physical layer supports the core communication mechanisms in the LAENet and plays a crucial role in its aerial deployment. For example, aerial vehicles deployed as aerial base stations (ABSs) or aerial relays can overcome interference, signal distortion, and environmental variations inherent in communication links by using physical layer functionalities such as channel access, multiplexing, and channel equalization \cite{8233654,9453787,10812989}.

However, physical layer communication in the LAENet is exposed to a variety of security threats due to the inherent characteristics of aerial communication environments \cite{10.1145/3715319}. The broadcast nature of wireless signals and the prevalence of line-of-sight (LoS) propagation make aerial links particularly vulnerable to eavesdropping, jamming, and spoofing attacks \cite{7875081,10759668}. These attacks can compromise communication confidentiality, disrupt communication, or deceive aerial vehicles by impersonating legitimate transmitters \cite{10678860,10352334}. Furthermore, the openness of wireless channels and weak authentication mechanisms increase the risk of unauthorized access and injection attacks, allowing adversaries to infiltrate the network or inject malicious signals \cite{8883128,10474115}. Additionally, the open medium and dynamic spectrum access may cause anomalous behaviors to disrupt normal communication operations in the LAENet \cite{8883124,8883127}.

Confronted with these substantial security challenges, this paper conducts a comprehensive analysis on physical layer communications of the LAENet and provides a thorough survey of technologies and solutions to address communication confidentiality, availability, and integrity. Table \ref{challengeandsolutionsbiaoge} gives a clear structure for showing existing efforts on secure physical layer communications for the LAENet.

\subsection{Related Surveys}

Recently, a number of excellent survey and tutorial papers have overviewed security issues in UAV networks and communications and have summarized corresponding countermeasures and solutions, as shown in Table \ref{tab:relevant_surveys}. Some works consider security issues at the system level including intrusion, privacy, and trust issues. The work in \cite{10.1145/3485272} provides a comprehensive review of security threats facing UAVs and UAV networks, including communication vulnerabilities, sensor spoofing, jamming, and malware attacks. It examines various countermeasures such as encryption, global positioning system (GPS) spoofing mitigation, and firmware signing. A gap analysis is performed to identify remaining security vulnerabilities and provide recommendations for future UAV development. The study in \cite{10.1145/3703625} conducts a comprehensive review of security issues in UAV swarm networks, examining various potential attacks such as communication attacks, identity-based attacks, resource attacks, routing attacks, data attacks, and machine learning (ML) attacks. It categorizes these threats and presents corresponding security technologies and countermeasures, including cryptography, physical layer security techniques, blockchain, machine learning, and intrusion detection systems. The authors in \cite{10793113} provide a detailed examination of security challenges in UAVs and FANETs, covering various attack vectors including communication, identity-based, resource, routing, data, and machine learning attacks. The study in \cite{HADI2023103607} examines security and privacy vulnerabilities in UAV systems across hardware, software, and communication layers. It discusses various threats such as eavesdropping and jamming attacks, and presents defense mechanisms including blockchain, machine learning-based intrusion detection, and secure communication protocols.

Some studies emphasize cyber security challenges within UAV networks. The study in \cite{9488323} comprehensively reviews security issues in drone communication, including Denial of Service (DoS), GPS spoofing, and man-in-the-middle attacks. It examines vulnerabilities across different drone applications and presents countermeasures using blockchain, software-defined networks, machine learning, and fog computing.  The authors of \cite{10.1145/3673225}  provide a comprehensive survey of security challenges in UAV systems, including various types of attacks, privacy concerns, and trust issues. It identifies current research trends and gaps while establishing a future roadmap with a focus on artificial intelligence (AI)'s potential to enhance UAV security. The authors in \cite{10599134} provide a comprehensive review of security issues in UAV networks, examining various potential attacks such as spoofing, replay, jamming, and eavesdropping attacks. It categorizes these threats and presents corresponding security technologies and countermeasures. The study in \cite{9946859} provides a comprehensive review of security issues in UAV-aided IoT applications and presents corresponding security technologies and countermeasures The work in \cite{KUMAR2024110695} reviews cybersecurity threats affecting UAV systems and evaluates existing countermeasures in enhancing UAV security.

In addition, some surveys analyze the challenges faced by UAV systems from a layered perspective (e.g., physical layer, link layer, network layer, application layer). The work in \cite{9900257} deeply reviews the current state of physical layer security (PLS) in UAV communications, examining unique air-to-ground channel characteristics, static and mobile UAV deployment scenarios, and various security enhancement techniques. The work in \cite{8675384} presents a comprehensive overview of UAV cellular communications, covering the classification of consumer drones,the concept and potential of UAV-mounted flying base stations. It explores the integration of UAVs into cellular networks as novel user equipment and addresses key challenges related to interference, regulatory compliance, and security. The authors of \cite{10236463} review the cybersecurity landscape of UAV-assisted IoT applications, examining layer-wise security threats from physical to application layers. It explores how AI, ML, deep learning (DL), and reinforcement learning (RL) techniques have been employed to address authentication, data privacy, and attack prevention challenges.

\begin{table}[t]
\centering
\caption{Challenges and Solutions \\ Red circles describe the security issues; green circles represent the overall countermeasures for the security issues; green check markers indicate different types of solutions under each countermeasure}
\label{challengeandsolutionsbiaoge}
\begin{tabular}{|c|p{0.38\textwidth}|}
\hline
\multicolumn{2}{|l|}{Section III, Challenge 1: Communication confidentiality} \\
\hline
\multirow{2}{*}{\begin{tabular}[c]{@{}c@{}}Issues\end{tabular}}  & 
\begin{tabular}[t]{@{}l@{}}
\textcolor{red}{\ding{108}} Eavesdropping attack \cite{8758975,8883128} \\
\textcolor{red}{\ding{108}} Unauthorized access \cite{10003076,9200889,9279294} 
\end{tabular} \\
\hline
\multirow{7}{*}{\begin{tabular}[c]{@{}c@{}}Solutions\end{tabular}} & 
\begin{tabular}[t]{@{}l@{}}
\textcolor{green}{\ding{108}} Anti-eavesdropping strategies \\
\textcolor{green!30}{\ding{51}} Convex optimization-based strategies \cite{9203867,10636964,10572013,9453748,9417318,9450021,10310294} \\
\textcolor{green!30}{\ding{51}} Reinforcement learning-based strategies \cite{10153699,10325641,9801656,9161257,10287142,10734220} \\
\textcolor{green!30}{\ding{51}} Deep learning-based strategies \cite{9713997,10114676,10194980,9584882,10700928,10759093} \\
\textcolor{green}{\ding{108}} Communication authentication \\
\textcolor{green!30}{\ding{51}} PUFs-based authentication \cite{9237145,10146461,10436373} \\
\textcolor{green!30}{\ding{51}} Channel based-authentication \cite{9672766,10233023,10834505} 
\end{tabular} \\
\hline
\multicolumn{2}{|l|}{Section IV, Challenge 2: Communication availability} \\
\hline
\multirow{2}{*}{\begin{tabular}[c]{@{}c@{}}Issues\end{tabular}}  & 
\begin{tabular}[t]{@{}l@{}}
\textcolor{red}{\ding{108}} Jamming attack \cite{5751298,8883124,10464352} \\
\textcolor{red}{\ding{108}} Spoofing attack \cite{5751298,8883127,8758975,9279294}
\end{tabular} \\
\hline
\multirow{7}{*}{\begin{tabular}[c]{@{}c@{}}Solutions\end{tabular}} & 
\begin{tabular}[t]{@{}l@{}}
\textcolor{green}{\ding{108}} Anti-jamming strategies \\
\textcolor{green!30}{\ding{51}} Convex optimization \cite{9271902,9200570,9454372} \\
\textcolor{green!30}{\ding{51}} Single-agent RL \cite{7925694,8023829,8314744,9264659} \\
\textcolor{green!30}{\ding{51}} Multi-agent RL \cite{9816050,10750022,10614297} \\
\textcolor{green}{\ding{108}} Spoofing defense \\
\textcolor{green!30}{\ding{51}} PLA \cite{9739860,10851372,10834505} \\
\textcolor{green!30}{\ding{51}} GNSS spoofing detection \cite{8946587,10685447,9845684}
\end{tabular} \\
\hline
\multicolumn{2}{|l|}{Section V, Challenge 3: Communication Integrity} \\
\hline
\multirow{2}{*}{\begin{tabular}[c]{@{}c@{}}Issues\end{tabular}}  & 
\begin{tabular}[t]{@{}l@{}}
\textcolor{red}{\ding{108}} Anomalous behaviors \cite{8999433,8113526,10325641} \\
\textcolor{red}{\ding{108}} Injection attacks \cite{8883128,9900257,10368002}
\end{tabular} \\
\hline
\multirow{7}{*}{\begin{tabular}[c]{@{}c@{}}Solutions\end{tabular}} & 
\begin{tabular}[t]{@{}l@{}}
\textcolor{green}{\ding{108}} Anomaly detection \\
\textcolor{green!30}{\ding{51}} Jamming anomaly detection \cite{9741304,9858012,9829873,9634169} \\
\textcolor{green!30}{\ding{51}} Abnormal power detection \cite{8688501} \\
\textcolor{green!30}{\ding{51}} Eavesdropping anomaly detection \cite{8854240} \\
\textcolor{green}{\ding{108}} Injection defense \\
\textcolor{green!30}{\ding{51}} Jamming signal injection defense \cite{9741304,9634169,10530539} \\
\textcolor{green!30}{\ding{51}} Spoofing signal injection defense \cite{10.1145/3395351.3401703,sathaye2022semperfi,8045998}
\end{tabular} \\
\hline
\end{tabular}
\end{table}

\begin{table*}[htbp]
\centering
\caption{LIST OF ABBREVIATIONS}
\label{LISTOFABBREVIATIONS}
\begin{tabular}{|c|c||c|c|}
\hline
\textbf{Abbreviation} & \textbf{Description} & \textbf{Abbreviation} & \textbf{Description} \\
\hline\hline
A2G & Air-to-ground  & ABS & Aerial Base Station  \\
\hline
AN & Artificial Noise  & AI & Artificial Intelligence  \\
\hline
BCD & Block Coordinate Descent  & BS & Base Station  \\
\hline
CNN & Convolutional Neural Network  & CSI & Channel State Information  \\
\hline
DDPG & Deep Deterministic Policy Gradient  & DDQN & Double-deep Q-Learning   \\
\hline
DL & Deep Learning  & DNN & Deep Neural Network  \\
\hline
DQN & Deep Q-Network  & eVTOL & Electric Vertical Take-off and Landing  \\
\hline
DRL & Deep Reinforcement Learning  & FAR & False Alarm Rate  \\
\hline
G2A & Ground-to-air  & G2U & Ground-to-UAV   \\
\hline
GAN & Generative Adversarial Network & GNSS & Global Navigation Satellite System  \\
\hline
GPS & Global Positioning System  & GS & Ground Station  \\
\hline
IoT & Internet of Things  & LAENet & Low-Altitude Economy Networking  \\
\hline
LSTM & Long Short-Term Memory  & LoS & Line-of-sight  \\
\hline
MARL & Multi-agent  Reinforcement Learning  & MDP & Markov Decision Process  \\
\hline
MDR & Miss Detection Rate  & MEC & Mobile Edge Computing  \\
\hline
ML & Machine Learning   & MSE & Mean Square Error  \\
\hline
NOMA & Non-orthogonal Multiple Access  & PLA & Physical-layer Authentication  \\
\hline
PLS & Physical Layer Security   & PUF & Physical Unclonable Function   \\
\hline
QoE & Quality of Experience  & RF & Radio Frequency  \\
\hline
RIS & Reconfigurable Intelligent Surfaces   & RL &  Reinforcement Learning  \\
\hline
RNN & Recurrent Neural Network  & RSS & Received Signal Strength  \\
\hline
SCA & Successive Convex Approximation  & SDNR & Signal-to-disturbance-plus-noise Ratio  \\
\hline
SNR & Signal-to-noise Ratio  & SOC & Second-Order Cone  \\
\hline
TDMA & Time-division Multiple Access  & THz & Terahertz  \\
\hline
U2G & UAV-to-ground Communication  & UAV & Unmanned Aerial Vehicle   \\
\hline
\end{tabular}
\end{table*}

%\begin{figure}[!t]
%\centering
%\includegraphics[width=3in]{LAEfig1-v3.pdf}
%\caption{The overall structure of this paper.}
%\label{lunwenjiegoutu}
%\end{figure} 

\subsection{Contributions of Our Survey}

The related surveys and tutorials primarily focus on the classification of overall security threats and corresponding countermeasures in UAV networks or UAV-assisted applications, with relatively little attention given to security issues of communication in the physical layer. Different from existing studies, our survey uniquely concentrates on the security challenges specific to physical layer communications in the LAENet, as summarized in Table \ref{challengeandsolutionsbiaoge}. It fills a critical gap in the literature by conducting an in-depth analysis of threats in physical layer communications that were previously underexplored or only briefly mentioned in prior studies. By offering a comprehensive and systematic analysis of these underexplored issues, our work brings new insights to seek effective solutions to enhance physical layer security in communications of the LAENet. The key contributions of this paper are summarized as follows:

The key contributions of this paper are summarized as follows:

\begin{itemize}

\item A thorough discussion of the six main security issues in the physical layer communication of the LAENet is presented, namely, eavesdropping attack, unauthorized access, jamming attack, spoofing attack, anomalous behaviors, and injection attack. We analyze these attacks in the context of their potential occurrence throughout the entire operation of LAENet, providing essential references for ensuring the security of physical layer communication in the future LAENet deployments.

\item We review countermeasures against various attacks in detail and offer a comprehensive tutorial on achieving communication confidentiality, communication availability, and communication integrity in LAENet. In addition, the lessons learned for each security issue are presented to emphasize the limitations of existing works and provide high-level insights for improvements.

\item Several potential future research directions for secure physical layer communication in LAENet are proposed, including energy-efficient physical layer security, multi-drone collaboration for secure communication, AI-driven security defense strategy, space-air-ground integrated security architecture, and 6G-enabled secure UAV communication. These diverse perspectives offer new guidance for future research on secure physical layer communication in LAENet.

\end{itemize}

%The structure of this survey is outlined in Fig. \ref{lunwenjiegoutu}. 
The remainder of this paper is given as follows. Section II introduces the background of the LAENet and security issues in physical layer communication of the LAENet. In Section III, a comprehensive exploration of achieving communication confidentiality for the LAENet is presented. Section IV reviews the solutions for communication availability in the LAENet. In Section V, countermeasures on communication integrity for the LAENet are discussed. Section VI provides future research directions, and Section VII concludes this paper. Additionally, Table \ref{LISTOFABBREVIATIONS} lists the abbreviations commonly employed throughout this survey.

%Background and Motivation:
%Definition and significance of the Low-Altitude Economy Networking (LAENet), highlighting its rapid development and extensive attention from global industries and academia. 
%Overview of LAENet's applications across various fields, including transportation, logistics, agriculture, and smart cities. \cite{9765519,9200679,8869712,9645207,MITTAL2020104046,drones7090582,s19214779}

%Role of secure physical layer communications in LAENet: 
%LAENet faces significant security challenges in physical layer communications. Open and shared wireless channels are vulnerable to eavesdropping, jamming, spoofing, and intrusion threats. Ensuring secure physical layer communications is essential to addressing these challenges, as it can safeguard the confidentiality, availability, and integrity of communication in LAENet.

\section{Background Knowledge}

In this section, we introduce the background of the LAENet, including its definition and application scenarios. Subsequently, the concept of physical layer communication in the LAENet and its security threats are elaborated in detail.

\begin{figure*}[!t]
\centering
\includegraphics[width=7.1in]{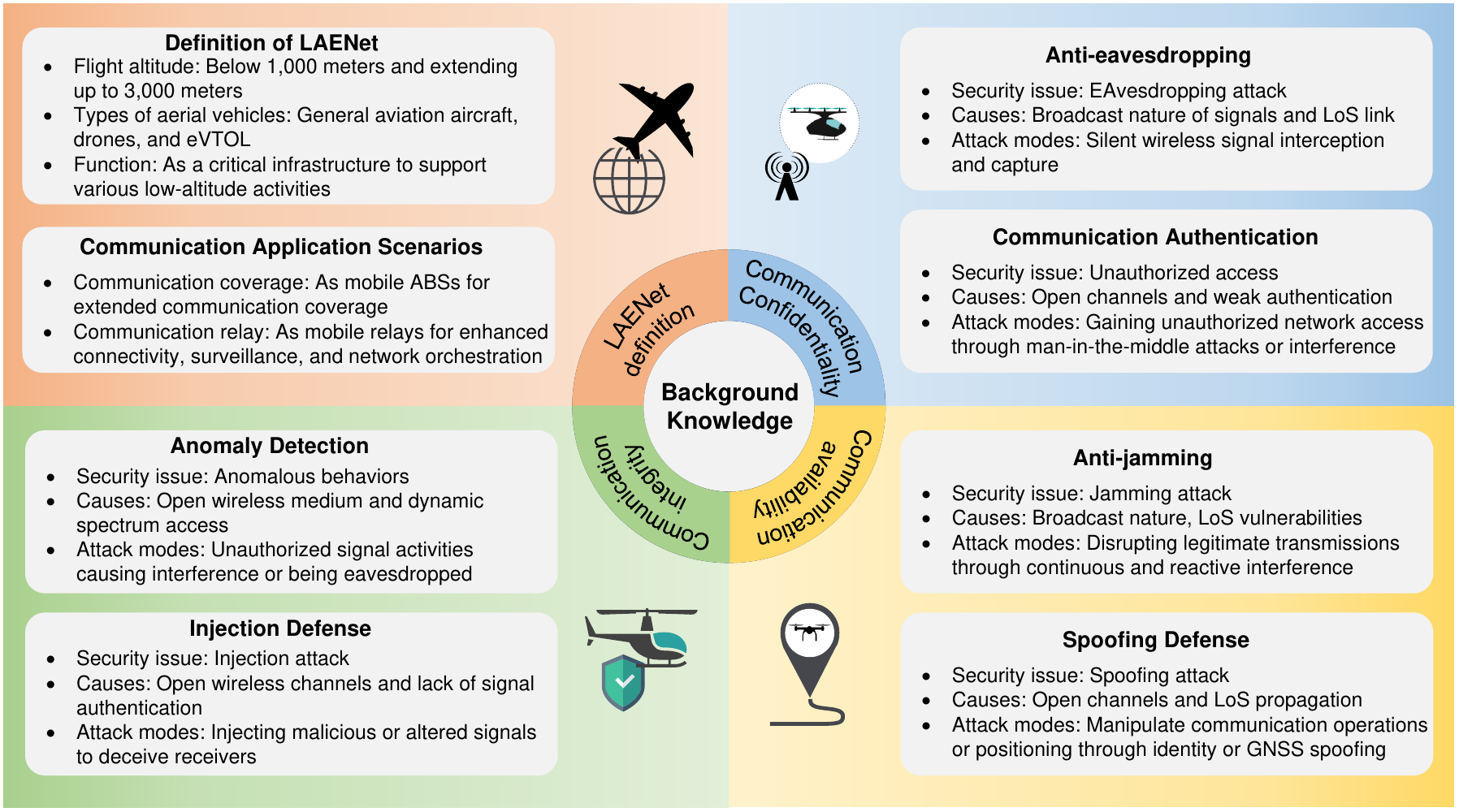}
\caption{Background knowledge of the LAENet and security issues in its physical layer communication. Describe the definition of the LAENet and its communication application scenarios. Elaborate on three key metrics for secure physical layer communication: communication confidentiality, which combats eavesdropping attacks and unauthorized access; anti-jamming strategies and spoofing defense for ensuring communication availability; and anomaly detection and injection defense to prevent adversaries from compromising communication integrity.}
\label{BackgroundofLAENet}
\end{figure*}

\subsection{Background of LAENet}

The LAENet is a sophisticated and dynamic system that integrates various aerial and terrestrial technologies to enable seamless communication, coordination, and management of diverse aerial operations within low-altitude airspace \cite{10759668,ye2024integrated}. The LAENet includes numerous different types of constituents, such as flight equipment, base stations, and other communication platforms. Specifically, the LAENet connects various aerial vehicles, including general aviation aircraft for passenger transport and emergency rescue, drones for surveillance and logistics, and eVTOL designed for urban air mobility and last-mile cargo delivery \cite{HUANG2024100694,yang2024embodied}. These aerial vehicles can incorporate ground and aerial base stations, further high-altitude platforms, such as weather balloons and satellites, to receive environmental information and precise navigation \cite{10879807}.

Different from traditional aviation networks that rely on centralized air traffic control, the LAENet can independently construct communication and networking by seamlessly interconnecting a variety of aerial and ground-based systems, which enables continuous information exchange, flight path optimization, and autonomous operations \cite{10833672,li2025aerial}. Therefore, the LAENet has opened opportunities for various application scenarios and plays key roles from the perspective of communication coverage and relay \cite{10749978,10089851,10720878}. Specifically, the LAENet can extend the communication coverage by deploying aircraft as ABSs in areas lacking communication infrastructure \cite{Jiang20236GNN,8869876,10884737}. For instance, these ABSs deployed at optimal altitudes can provide connectivity and network services in remote or disaster-stricken areas \cite{9453811,9454158}. Moreover, if the direct communication links between ground base stations and user equipment are unreliable, such as in mountainous regions and densely populated areas, the aircraft can act as mobile relays to improve connectivity by capturing, amplifying, and transmitting communication signals \cite{10097717,9635590,AHMAD2020107122}. It also can be regarded as a surveillance unit to monitor airspace dynamics while simultaneously functioning as a low-altitude network orchestrator to optimize communication and computing resources \cite{9562043,9453811,9094586}.

To integrate and evolve these capabilities, the LAENet needs to establish effective communication infrastructure to ensure reliable connectivity and efficient interaction across various environments \cite{10003076,8690950}. Physical layer communication, as the bottom layer in the network architecture, may directly influence the communication performance of the LAENet across aerial and terrestrial networks \cite{8883128,7875081}. For example, it governs how signals are generated, transmitted, and received between aircraft and base stations\cite{10003076}. Building on this, it manages the channel and spectrum resources to enhance signal transmission quality and maintain stable connectivity \cite{7875081}. Therefore, ensuring the security of physical layer communication in the LAENet is crucial for supporting a wide range of applications in low-altitude domains.

\subsection{Security Issues in Physical Layer Communication of LAENet}

Based on previous studies \cite{samonas2014cia,10623395}, we discuss the security issues in the physical layer communication of the LAENet from three aspects: confidentiality, availability, and integrity of communications. The details of each measurement are described as follows.

\begin{itemize}

\item The confidentiality of physical layer communications in the LAENet can be compromised by security threats such as eavesdropping and unauthorized access \cite{8509094}. Eavesdropping arises primarily from the broadcast nature of wireless signals and LoS link, making transmissions highly susceptible to interception \cite{8883128}. An eavesdropper silently capturing or intercepting signals can lead to the exposure of confidential information. Meanwhile, unauthorized access threats exploit the open and broadcast nature of UAV communications \cite{10003076}. Attackers may gain illegal access to the LAENet by disguising themselves as legitimate UAVs or ground stations, thereby deceiving or interfering with the normal operation of UAVs \cite{9200889}.

\item Similarly, the open nature of wireless channels and LoS propagation bring jamming and spoofing security issues for communication availability \cite{5751298}. Specifically, jammers can continuously transmit interference signals to disrupt communication, where a jammer can be a drone or a base station \cite{8883124}. The spoofing attack can not only achieve identity spoofing by forging legitimate transmission identities but also launch signal deception attacks to disrupt UAV communications and positioning \cite{8883127}. Therefore, jamming and spoofing lead to unauthorized access and signal disruptions or errors, making communication unavailable in the LAENet.

\item Integrity as a microscopic metric measures the deviations of signals, channels, and spectrum in communication under adversaries' influence \cite{shakiba2021physical}. The communication integrity of the LAENet can be affected by anomalous behaviors and injection attacks. Anomalous behaviors often use dynamic spectrum access and the open wireless medium, including abnormal jamming, abnormal transmission power, and covert eavesdropping 
\cite{8999433}. These anomalous behaviors can introduce harmful interference, violate spectrum policies, and expose sensitive information to eavesdroppers \cite{8113526,10325641}. Moreover, the injection attack exploits the open nature of wireless channels to alter signals or inject illegal signals, such as spoofing signals or malicious GNSS signals, to deceive receivers and interfere with communication, thereby leading to degraded signal quality, false navigation, and network congestion \cite{8883128,9900257,10368002}.

\end{itemize}

Overall, as illustrated in Fig. \ref{BackgroundofLAENet}, this survey reviews existing research on achieving communication confidentiality, availability, and integrity for the LAENet. Specifically, the investigation of anti-jamming strategies and communication authentication schemes aims to enhance communication confidentiality. Studies on anti-jamming techniques and spoofing defense mechanisms have been explored to ensure communication availability. Furthermore, research on communication integrity has focused on anomaly detection and injection attack mitigation approaches.

%\section{Research Challenges of Joint Security and Trajectory Optimization in LAENet}

%\subsection{Non-convex and Coupled Problems}

%Trajectory control, resource allocation, and security constraints for low-altitude vehicles often lead to non-convex and coupled optimization problems.

%\subsection{Dynamic and Uncertain Decision Making}

%Low-altitude vehicles must make adaptive real-time decisions in rapidly changing environments, such as avoiding collisions or modifying trajectories to prevent eavesdropping.

%\subsection{High-dimensional and Nonlinear Problems}

%Coordination among multiple low-altitude vehicles involves numerous variables, with potential nonlinear relationships between them.

%\section{Detailed Methods}

\section{Communication Confidentiality for LAENet}

\subsection{Anti-eavesdropping Strategy}

%Low-altitude aircraft trajectory optimization is a key technology in anti-eavesdropping communication. By dynamically adjusting its flight path, a low-altitude aircraft can actively shape the communication environment, enabling legitimate receivers to obtain better channel conditions while simultaneously degrading the eavesdropper's reception ability. 

%The primary approach is to combine trajectory optimization with variables such as transmission power, interference power, and time allocation to maximize the secrecy rate and counter eavesdroppers.

The LAENet faces significant eavesdropping threats due to the inherent vulnerabilities of UAV-enabled wireless communications. The openness of wireless channels, especially the LoS links in air-to-ground (A2G) and ground-to-air (G2A) communications, increases susceptibility to interception by eavesdroppers that disrupt legitimate communications compared to traditional terrestrial channels \cite{8758975}. Traditional cryptographic methods, while effective in many scenarios, are less suitable for UAV communications due to their computational complexity and the dynamic mobility of UAVs \cite{9201322}. This highlights the critical need for robust security measures to ensure the confidentiality and reliability of the LAENet communications. To address these limitations, leveraging PLS techniques to counter eavesdropping threats effectively has emerged as a promising solution \cite{8543573,9070153,9463880,7944621}.

In the LAENet, anti-eavesdropping solutions can leverage the controllable mobility of low-altitude aircraft to enhance physical layer security. By dynamically optimizing their trajectories, low-altitude aircraft can actively adapt their flight paths to shape the communication environment \cite{10013691}. This approach allows them to fly closer to legitimate ground nodes, strengthening communication links and improving channel conditions for intended receivers, while simultaneously distancing themselves from potential eavesdroppers. In this subsection, we present a critical role of UAV trajectory in forming the communication environment, and how PLS can be enhanced through trajectory optimization and resource allocation to mitigate eavesdropping risks. Our analysis focuses on three prominent methodologies in this domain: convex optimization, deep learning, and reinforcement learning.

\begin{table*}[]

\caption{Summary of Convex Optimization for Anti-Eavesdropping Strategy \\ CIRCLES DESCRIBE THE METHODS; CORRECT MARKERS AND CROSS MARKERS REPRESENT PROS AND CONS RESPECTIVELY.}
\label{tab:my-table}
\begin{tabular}{|c|c|m{2cm}|m{2.4cm}|l|}
\hline
Ref & \begin{tabular}[c]{@{}c@{}}Optimization\\ Objectives\end{tabular}           & \multicolumn{1}{c|}{\begin{tabular}[c]{@{}c@{}}Eavesdropper and\\ Jammer Type\end{tabular}} & \multicolumn{1}{c|}{\begin{tabular}[c]{@{}c@{}}Optimization\\ Constraints\end{tabular}}                         & \multicolumn{1}{c|}{Pros \& Cons}                         \\ \hline
{\cite{9203867}}   & \begin{tabular}[c]{@{}c@{}}Secure calculation\\ capacity$^{1}$\end{tabular} & UAV jammer and fixed ground eavesdropper                                                    & Transmit power, time allocation, and computation capacity                                                       & \begin{tabular}[c]{@{}l@{}} \textcolor{blue!20}{\ding{108}} BCD and P-BCD for secure calculation capacity maximization\\ \textcolor{green}{\ding{51}} Secure capacity of NOMA and TDMA has been significantly improved\\ \textcolor{red}{\ding{55}} High complexity for NOMA due to dual-loop iterations\end{tabular} \\ \hline
{\cite{10636964}}   & \begin{tabular}[c]{@{}c@{}}Secure calculation\\ capacity\end{tabular}       & Base station jammer and fixed ground eavesdropper                                           & Transmission power, time allocation, and CPU processing frequency                                               & \begin{tabular}[c]{@{}l@{}} \textcolor{blue!20}{\ding{108}} JDPB algorithm with SCA and BCD for secure task offloading\\ \textcolor{green}{\ding{51}} Reduce complexity via region division\\ \textcolor{red}{\ding{55}} Fixed UAV altitude limits 3D trajectory optimization\end{tabular}      \\ \hline
{\cite{10572013}}   & \begin{tabular}[c]{@{}c@{}}Average secrecy\\ rate$^{2}$\end{tabular}        & Antenna jammer and fixed aerial eavesdropper                                                & Transmit power and jamming power                                                                                & \begin{tabular}[c]{@{}l@{}} \textcolor{blue!20}{\ding{108}} BCD and SCA optimization with hybrid FSO/RF links\\ \textcolor{green}{\ding{51}} Enhance communication security via hybrid FSO/RF links and AN\\ \textcolor{red}{\ding{55}} Rely on simplified channel models (e.g., free-space path loss)\end{tabular}      \\ \hline
{\cite{9453748}}   & \begin{tabular}[c]{@{}c@{}}Worst-case\\ secrecy rate\end{tabular}           & UAV jammer and fixed ground eavesdropper                                                    & UAV speed, collision avoidance, positioning error, and energy harvesting                            & \begin{tabular}[c]{@{}l@{}}\textcolor{blue!20}{\ding{108}} Robust 3D trajectory and time switching optimization \\ \textcolor{green}{\ding{51}} Full mobility of UAVs in 3D for improving secrecy rate\\ \textcolor{red}{\ding{55}} The performance may degrade with flying eavesdroppers\end{tabular}      \\ \hline
{\cite{9417318}}   & \begin{tabular}[c]{@{}c@{}}Average secrecy\\ rate\end{tabular}              & None and flying eavesdropper                                                                & Transmit power control and user scheduling                                                                      & \begin{tabular}[c]{@{}l@{}}\textcolor{blue!20}{\ding{108}} Joint trajectory and communication design against mobile eavesdroppers\\ \textcolor{green}{\ding{51}} Initial trajectory design for keeping away from eavesdroppers \\ \textcolor{red}{\ding{55}} Security performance relies on the initial trajectory design\end{tabular}      \\ \hline
{\cite{9450021}}   & \begin{tabular}[c]{@{}c@{}}Secure calculation\\ capacity\end{tabular}       & Ground jammer and flying eavesdropper                                                       & Transmit power, time slot, computation capacity, UAV speed, and collision avoidance & \begin{tabular}[c]{@{}l@{}}\textcolor{blue!20}{\ding{108}} Integrate a dual-UAV system with a ground jammer in MEC\\ \textcolor{green}{\ding{51}} Incorporate the  UAV server and UAV eavesdropper with a ground jammer\\ \textcolor{green}{\ding{51}} Allow a UAV server to hover near ground users for secure offloading \\ \textcolor{red}{\ding{55}} Numerous flight constraints may require extensive tuning\end{tabular}      \\ \hline
{\cite{10310294}}   & Secrecy rate                                                                & Coastal jammer and flying eavesdropper                                                      & Transmit power, time slot, computation capacity, UAV speed, and collision avoidance       & \begin{tabular}[c]{@{}l@{}}\textcolor{blue!20}{\ding{108}} A secure communication for UAV-relay-assisted maritime MEC\\ \textcolor{green}{\ding{51}} Simultaneously optimize multiple parameters for improved secrecy rate\\ \textcolor{red}{\ding{55}} Iterative decomposition increases the computational burden \\ \textcolor{red}{\ding{55}} Assume prior knowledge of Channel State Information (CSI) of devices \end{tabular}      \\ \hline
\end{tabular}

\raggedright
\footnotesize $^{1}$Secure calculation capacity is defined as the average number of secure calculation bits in UAV flying time \cite{10636964}.\\
\footnotesize $^{2}$Secrecy rate is defined as the difference between the achievable rate of legitimate UAV’s channel and the rate of eavesdropper channel \cite{IRRAM2022103431}.

\end{table*}

\begin{figure}[!t]
\centering
\includegraphics[width=3.5in]{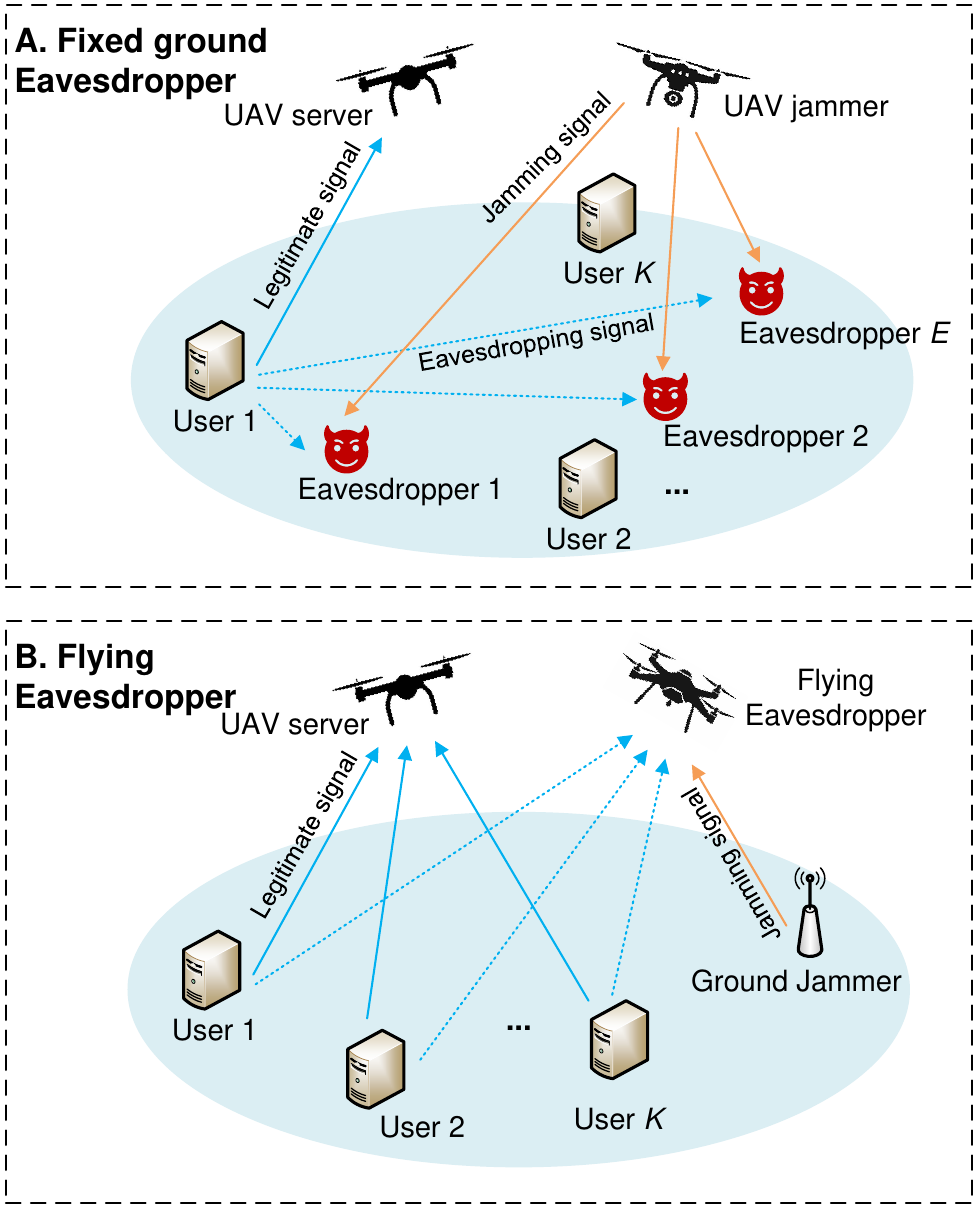}
\caption{The overall architecture of the anti-eavesdropping strategy. Part A illustrates the system model against fixed ground eavesdroppers. In this setup, one UAV operates as a mobile server, while another UAV serves as a jammer to emit jamming signals to disrupt the eavesdroppers' interception capabilities. Part B presents the system model for flying eavesdroppers, where one UAV acts as the server, and another UAV functions as a mobile eavesdropper. To mitigate eavesdropping risks, a ground-based jammer actively emits interference signals to secure communications.}
\label{Anti-eavesdroppingsystemmodel}
\end{figure} 

%This dual-part architecture highlights distinct strategies for addressing different types of eavesdropping threats.

Convex optimization plays a crucial role in addressing anti-eavesdropping challenges in UAV-enabled communication networks, particularly for solving the joint optimization of trajectory and resource allocation \cite{Lu2021}. Due to the inherent non-convex nature of these problems, advanced convex optimization techniques such as Successive Convex Approximation (SCA) and Block Coordinate Descent (BCD) are widely utilized \cite{10013691}. These methods enable UAVs to enhance physical layer security by optimizing flight paths and resource utilization, minimizing the risk of eavesdropping while ensuring secure and efficient communication. Additionally, the decision variables may be discrete, which requires the application of various relaxation methods to transform the complex optimization problem into a more tractable form to obtain efficient solutions \cite{10.1145/3560261}.

The study in \cite{9203867} explores physical-layer security in UAV-assisted Mobile Edge Computing (MEC) systems in the presence of multiple ground-based eavesdroppers. The proposed system utilizes dual UAVs for task execution and anti-eavesdropping measures. One UAV operates as a mobile MEC server, while the other emits jamming signals to disrupt eavesdroppers, as shown in Fig. \ref{Anti-eavesdroppingsystemmodel}. The time-division multiple access (TDMA) scheme and non-orthogonal multiple access (NOMA) scheme are proposed to maximize the minimum secure computing capacity by jointly optimizing communication resources, computation resources, and UAV trajectories. To address the non-convexity of the optimization problem, the problem is transformed into tractable forms via auxiliary variables and decomposition. Specifically, for the TDMA scheme, the problem is decoupled into two sub-problems using BCD. The communication and computation resources are optimized via second-order cone (SOC) constraints and SCA, while UAV trajectories are iteratively updated via first-order Taylor approximations to handle non-convex terms. For the NOMA scheme, a penalized BCD (P-BCD) algorithm is proposed to tackle binary constraints. The problem is split into three blocks that are penalty parameter adjustment, resource allocation via SOC and SCA, and trajectory optimization with convex relaxations. The experimental results demonstrate that the proposed algorithms significantly enhance secure computing capacity, with the NOMA scheme achieving up to about 4.3 Mbps and the TDMA scheme reaching about 4.2 Mbps under optimal conditions. Compared to baselines including the straight flight design and no power control, the proposed strategies improve secure computing capacity by about 20\% to 30\%, particularly in scenarios with lower power budgets (e.g., 0.2 W) and higher required computing bits (e.g., 1 Mbps). The convergence of the algorithms is achieved within 20 iterations, which indicates the efficiency in optimizing UAV trajectories and resource allocation for anti-eavesdropping.

The study in \cite{9203867} mainly foucuses on a dual-UAV-assisted secure MEC system. In some cases, multi-UAV systems hold great promise for collaboratively executing complex tasks while enhancing the secure communications \cite{8883127,10636964}. In the work \cite{10636964}, the joint optimization of task offloading, trajectory planning, and resource allocation for secure communications in multi-UAV MEC systems is studied. Firstly, a base station emits jamming signals to protect against fixed-location ground eavesdroppers. Then, it investigated the joint optimization of task offloading, trajectory planning, and resource allocation for secure communications in multi-UAV MEC systems. The problem is decomposed into two sub-problems: (1) resource allocation and trajectory planning, addressed via SCA and BCD algorithms; (2) offloading decisions, solved through Joint Dynamic Programming and Bidding (JDPB) method. For the first sub-problem, non-convex constraints related to transmission power and UAV trajectory are transformed into convex forms using first-order Taylor expansion and relaxation techniques. Specifically, the transmission power optimization sub-problem is approximated via SCA, while the trajectory planning sub-problem is iteratively solved by introducing auxiliary variables and convex approximations. For the second sub-problem, a bidding mechanism is integrated with dynamic programming to reduce computational complexity by grouping dynamic users into sub-regions. The experimental results demonstrate that the proposed JDPB algorithm achieves a sum average secure calculation capacity of 10.1 Mbps in the first time slot. Additionally, under different settings of time slot sizes, transmission power, and flying speed, the sum average secure calculation capacity achieved by JDPB consistently outperforms baseline schemes such as the Greedy Strategy and the Random Strategy.

Unlike the above studies that deal with ground eavesdroppers, the work in \cite{10572013} targets threats from aerial eavesdroppers and explores secure communication in a hybrid Free Space Optical (FSO) and Radio Frequency (RF) system. The UAV acts as both a relay and a jammer, emitting artificial noise (AN) during RF transmission to confuse a fixed-position aerial eavesdropper. The work introduces a novel perspective on protecting space-air-ground networks from eavesdropping by leveraging FSO for its inherent resistance to interception and jointly optimizing trajectory design and power allocation to maximize the secrecy rate with two transmission schemes. The first scheme is the slot-based scheme for delay-sensitive data. % The problem is divided into UAV trajectory optimization and power allocation sub-problems. 
The trajectory sub-problem is convexified using first-order Taylor expansion to approximate elevation angle and channel gain constraints, while the power allocation sub-problem is transformed into a convex form by introducing a lower bound on transmit power to ensure convexity. The second scheme is the period-based scheme for delay-insensitive data, in which the relaxed constraints on sum secrecy rates over the entire flight period are adopted. A similar SCA method \cite{10636964} is applied to convexly approximate the non-convex terms in the constraints. Compared to benchmark schemes without jamming power optimization, both methods achieve approximately 0.4 Mbps higher secrecy rates by integrating AN transmission and hybrid FSO/RF links. %The slot-based scheme ensures per-time-slot secrecy rate balancing, while the period-based scheme prioritizes total secrecy rate maximization across the flight period, demonstrating superior performance under relaxed constraints.

\begin{figure*}[!t]
\centering
\includegraphics[width=7.1in]{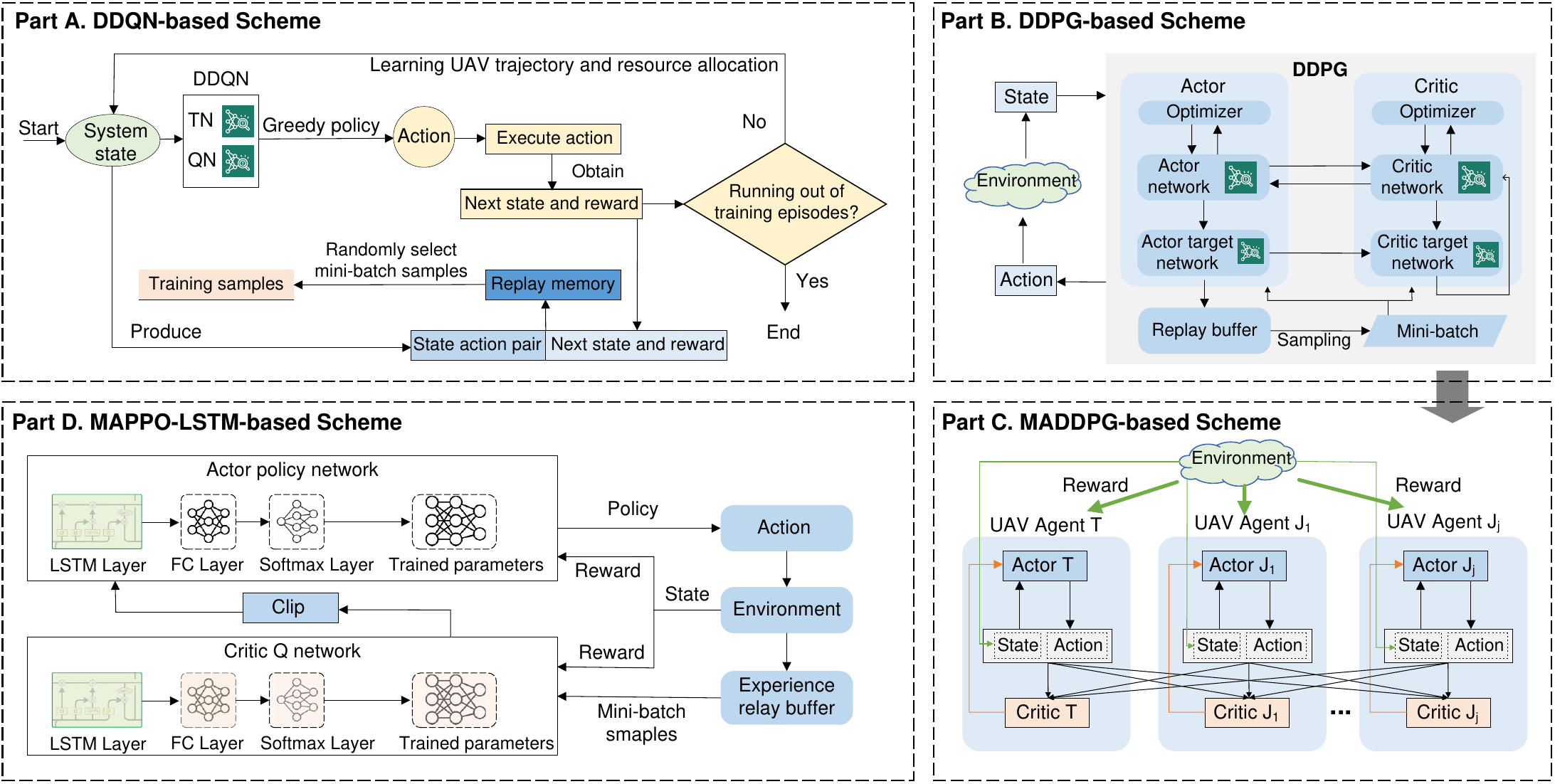}
\caption{ The overall architecture of the RL for anti-eavesdropping. Part A describes the DDQN-based scheme, where the system state is used to generate actions through the DDQN network, followed by action execution and obtaining the next state and reward. An experience replay mechanism is employed to store and randomly sample training data. Part B presents the DDPG-based scheme, where actions are generated through Actor and Critic networks, interacting with the environment to obtain rewards. An experience replay buffer is used to store and sample mini-batches. Part C describes the MADDPG-based scheme, involving multiple UAV agents, each with its own Actor and Critic networks, interacting with the environment and sharing rewards. Part D showcases the MAPPO-LSTM-based scheme, where Actor and Critic networks with LSTM layers process time-series data and train through an experience replay buffer.}
\label{Anti-eavesdroppingDRL}
\end{figure*}

It is worth noting that most existing studies consider optimizing UAV trajectories on a 2D plane. However, optimizing UAV 3D trajectories may be more pratical\cite{9061039}. The study in \cite{9453748} considers the UAV's 3D flight trajectory and imperfect knowledge of eavesdroppers' locations, while formulating an optimization approach to maximize the worst-case secrecy rate under various practical constraints, including maximum UAV speed, UAV collision avoidance, UAV positioning error, and UAV energy harvesting. To address the non-convexity of the optimization problem, the original problem is decomposed into multiple sub-problems using BCD and SCA techniques similar to studies in \cite{10636964} and \cite{10572013}. By incorporating the additional degree of freedom in the vertical dimension, the proposed approach improves the ability to avoid fixed eavesdropping zones, outperforming 2D trajectory models in maintaining secure communication links under dynamic conditions. Simulation results show that the average secrecy rate of the proposed 3D optimization scheme outperforms that of the fixed-height 2D benchmarks (set at 100 m) by over 20\%.

Unlike the above studies that focus on fixed ground eavesdroppers, mobile eavesdroppers, such as hostile UAVs, introduce more complex threats due to their ability to maneuver, track, and position for intercept communications \cite{HADI2023103607,9417318}. For example, the authors in \cite{9417318} address the challenges caused by a flying eavesdropper that exploits UAV LOS communication. This work focuses on jointly optimizing the UAV's trajectory, transmit power control, and user scheduling to maximize the minimum average secrecy rate, which enables dynamic adjustments to ensure secure communication even against an mobile eavesdropper.

Compared to the anti-eavesdropping strategies in \cite{9417318} that rely heavily on accurate trajectory optimization and resource allocation, the studies in \cite{9450021,10310294} propose using a jammer to actively emit jamming signals, effectively reducing the interception capability of flying eavesdroppers during the computational offloading process of relay UAVs,  as shown in Fig. \ref{Anti-eavesdroppingsystemmodel}. Meanwhile, with the support of SCA and BCD methods similar to \cite{9453748}, the joint optimization problem of UAV trajectories, resource allocation (including transmit power, time slot allocation, and computation capacity), and jamming strategies can be solved while ensuring practical constraints such as flight speed and anti-collision requirements. Importantly, compared to systems targeting fixed ground eavesdroppers, the works in \cite{9450021,10310294} enhance secure calculation capacity or secrecy rate by modeling the trajectories of both the relay UAV and the mobile eavesdropper as dynamic variables optimized over discrete time slots. Specifically, simulation results in \cite{9450021} demonstrate that the secure calculation capacity of the proposed scheme converges to approximately 2.78 Mbps within 4 iterations, which is significantly higher than the baseline strategy (where only the location of the relay UAV, transmit power, and jamming power are optimized) by approximately 1.6 Mbps. %In addition, the secrecy rate achieved in \cite{10310294} is approximately 1.42 Mbps, representing a 33.96\% improvement over the benchmark scheme that optimizes only the relay UAV location, transmit power, and local computation.}

{\bf Lesson Learned.} Convex optimization has emerged as a fundamental tool for developing anti-eavesdropping strategies in UAV-enabled communication systems, particularly for addressing the inherent non-convexity of joint trajectory and resource allocation problems.  For fixed eavesdroppers, simpler optimization models with fewer dynamic variables (e.g., 2D trajectory optimization) can achieve secure communication effectively. However, mobile eavesdroppers require more sophisticated formulations, including 3D trajectory optimization and robust constraints to account for uncertainties in eavesdropper positions. Another important insight is the adaptability of convex optimization when combined with complementary methods like artificial noise jamming and resource allocation strategies. By leveraging convex optimization, systems can balance secrecy performance with energy efficiency, ensuring practical applicability in real-world UAV operations. Techniques such as SCA and BCD have proven highly effective in decoupling complex optimization problems into solvable subproblems, allowing iterative refinement toward locally optimal solutions. Overall, convex optimization offers a flexible and mathematically rigorous approach to securing UAV-enabled communication systems for anti-eavesdropping.

%\subsubsection{Reinforcement Learning} 
As the number of ground devices increases, along with UAV flight time and the number of optimization variables, the computational complexity of conventional algorithms grows exponentially, leading to infeasibility or suboptimal solutions \cite{8643815,9171468}. Moreover, these methods struggle to adapt to real-time scenarios where UAVs must communicate with mobile users and operate in environments with uncertain or partial information \cite{8589002,8643815}. RL enables UAVs to interact with the environment and autonomously learn optimal policies based on real-time observations \cite{10283826}, as shown in Fig. \ref{Anti-eavesdroppingDRL}. By leveraging Deep RL (DRL), UAVs can efficiently adapt to changing eavesdropping conditions, optimize secure trajectories, and dynamically allocate resources \cite{10413638,10283889}. This learning-driven approach significantly enhances PLS by ensuring adaptive, scalable, and intelligent anti-eavesdropping strategies in UAV communication networks.

\begin{table*}[]

\caption{Summary of RL for anti-eavesdropping strategy \\ CIRCLES DESCRIBE THE METHODS; CORRECT MARKERS AND CROSS MARKERS REPRESENT PROS AND CONS RESPECTIVELY.}
\label{tab:my-table}
\begin{tabular}{|c|c|c|l|}
\hline
\textbf{Techniques} & \textbf{Reference} & \textbf{Algorithm} & \multicolumn{1}{c|}{\textbf{Pros \& Cons}} \\ \hline
\multirow{6}{*}{\begin{tabular}[c]{@{}c@{}}Value-based\\ RL\end{tabular}} 
& \cite{10153699} & DQN & \begin{tabular}[c]{@{}l@{}}\textcolor{blue!20}{\ding{108}} DQN algorithm for UAV trajectory optimization to maximize the secrecy capacity \\ \textcolor{green}{\ding{51}} Low computational complexity, making it easy to train \\ \textcolor{red}{\ding{55}} Q-value overestimation, leading to suboptimal action selection \end{tabular} \\ \cline{2-4} 
& \cite{10325641} & DDQN & \begin{tabular}[c]{@{}l@{}}\textcolor{blue!20}{\ding{108}} DDQN-based joint trajectory, time allocation, and offloading optimization \\ \textcolor{green}{\ding{51}} Accelerated convergence via action space pruning \\ \textcolor{green}{\ding{51}} Real-time optimization of trajectory and resources \\ \textcolor{red}{\ding{55}} DDQN is restricted to discrete action spaces \\ \textcolor{red}{\ding{55}} DDQN is not suitable for continuous action control \end{tabular} \\ \hline
\multirow{11}{*}{\begin{tabular}[c]{@{}c@{}}Policy Gradient-\\ based RL\end{tabular}} 
& \cite{9801656} & CAA-MADDPG & \begin{tabular}[c]{@{}l@{}}\textcolor{blue!20}{\ding{108}} Multi-Agent DRL with attention mechanisms (CAA-MADDPG) for secrecy rate maximization\\ \textcolor{green}{\ding{51}} Handle complex multi-agent with the attention mechanism \\ \textcolor{red}{\ding{55}} Assume prior knowledge of eavesdropper locations \\ \textcolor{red}{\ding{55}} Assume ground devices are static\end{tabular} \\ \cline{2-4} 
& \cite{9161257} & DDPG & \begin{tabular}[c]{@{}l@{}}\textcolor{blue!20}{\ding{108}} DDPG-based RL for enhancing bi-directional UAV communication security \\ \textcolor{green}{\ding{51}} Support mobile devices and ensure bi-directional securit \\ \textcolor{red}{\ding{55}} Computational overhead increases with device density \\ \textcolor{red}{\ding{55}} performance may be sensitive to hyperparameter selection \end{tabular} \\ \cline{2-4} 
& \cite{10287142} & PPO+DCCN & \begin{tabular}[c]{@{}l@{}} \textcolor{blue!20}{\ding{108}} Hybrid DCCN and PPO  for secrecy rate maximization\\ \textcolor{green}{\ding{51}} The PPO optimizes the UAV trajectory based on the results from DCCN\\ \textcolor{red}{\ding{55}} The performance is sensitive to the choice of clipping factor in PPO\end{tabular} \\ \cline{2-4}
& \cite{10734220} & MAPPO+LSTM & \begin{tabular}[c]{@{}l@{}}\textcolor{blue!20}{\ding{108}} MAPPO for multi-agent cooperative anti-eavesdropping and LSTM-enhanced sequential learning \\ \textcolor{green}{\ding{51}}  The MAPPO+LSTM improves the learning capability of sequential sample data\\ \textcolor{red}{\ding{55}}  Assume perfect knowledge of CSI may be challenging in real-world scenarios \end{tabular} \\ \hline
\end{tabular}

\end{table*}

The study in \cite{10153699} proposes a Deep Q-Network (DQN)-based approach to address the challenge of securing UAV-assisted multi-user wireless communications against passive eavesdropping attacks. The UAV trajectory optimization is formulated as a Markov Decision Process (MDP), where the state space includes the UAV's 3D coordinates and the positions of users. The action space consists of discrete movements in the $x$, $y$, and $z$ directions, with each action representing a step change in position. The reward function is designed to maximize the legitimate users' rates, defined as the sum of the channel capacities of users served by the UAV. Unlike many prior works that assume perfect knowledge of eavesdropper CSI \cite{9203867,10310294}, this study focuses on optimizing legitimate user rates and using the DQN-based approach without requiring full knowledge of the eavesdropping channels. The DQN iteratively optimizes the UAV's trajectory, beamforming matrix, and transmit power allocation, ensuring the UAV dynamically adjusts its position to maximize secrecy capacity. Numerical results show that the secrecy capacity improves with the number of users. The proposed method converges an order of magnitude faster than the Q-learning method and achieves around 35\% higher secrecy capacity than Q-learning after 20,000 episodes.

However, the DQN method may face the issue of Q-value overestimation, leading to suboptimal action selection \cite{9448360}. The authors in \cite{10325641} propose a double-deep Q-learning (DDQN)-based scheme to jointly optimize the UAV trajectory, time allocation, and offloading decision strategy, aiming to maximize the average secure computing capacity for anti-eavesdropping in UAV-aided MEC. The system model involves one legitimate UAV server, one illegitimate UAV eavesdropper, one ground jammer, and ground users. The proposed DDQN-based scheme models the optimization problem as an MDP with states, actions, and rewards. The states include the coordinates of the UAVs, while the actions involve offloading decisions, time allocation, and trajectory adjustments. The reward function maximizes secure computing capacity. The DDQN model includes a deep Q-network (QN) and a deep target network (TN) to generate actions and evaluate their values. The parameters of the QN are updated by minimizing the loss function, and the parameters of the TN are periodically updated. The proposed scheme reduces the action space size by deleting illegal actions, such as those that violate time allocation constraints or result in resource waste. Unlike prior works \cite{9417318,10153699} that rely on conventional optimization or DQN with limited consideration of task queues, this approach integrates real-time resource allocation and trajectory optimization while ensuring dynamic constraints. The proposed DDQN scheme converges in 2000 episodes, half the episodes required by DQN (4000 episodes), and achieves a 0.02 Mbits higher average secure computing capacity.

The value-based RL method (e.g., DQN) mainly focuses on dealing with discrete action spaces that may lead to the loss of optimal solutions \cite{9403369}. The policy gradient-based RL method (e,g., Deep Deterministic Policy Gradient (DDPG)) can handle continuous action spaces \cite{9606690}, which are more suitable for UAV trajectory and transmit power optimization problems.

The authors in \cite{9801656} propose a multi-agent DRL framework to address the challenge of secure UAV communications in the presence of eavesdroppers. The system model is similar to Part A of Fig. \ref{Anti-eavesdroppingsystemmodel}, where the UAV server sends confidential information to ground users, and UAV jammers send AN signals to ground eavesdroppers using 3D beamforming. This study designs the Multi-Agent Deep Deterministic Policy Gradient with a continuous action attention mechanism (CAA-MADDPG) to maximize the system’s secrecy rate. %The state space includes UAV 3D positions, channel path loss, and ground user/eavesdropper locations. Actions involve continuous 3D velocity adjustments and power allocation. The reward function incentivizes maximizing the secure capacity and penalizing eavesdropping success. 
The attention mechanism dynamically prioritizes relevant agents' observations (e.g., jammers focusing on eavesdroppers) to reduce the exploration space and accelerate convergence, thereby enhancing the system's ability to counteract eavesdropping attempts. The simulation results show that CAA-MADDPG achieves a secure rate of 4.5 bps/Hz and converges in 1000 episodes with three UAV jammers, outperforming MADDPG (around 4 bps/Hz and 1400 episodes) and DDPG schemes.

However, the study in \cite{9801656} just considers UAV-to-ground communication (U2G) and assumed the ground devices are static. The work in \cite{9161257} addresses the challenge of securing bi-directional ground-UAV communications in a dynamic environment with mobile ground devices and eavesdroppers. Different from prior works that assume static ground eavesdroppers \cite{10636964,9453748}, this study considers mobile ground eavesdroppers for more practical real-world scenarios. The communication in U2G and ground-to-UAV (G2U) cases is modeled, considering factors such as channel gains and distances. The problem of maximizing the worst-case average secrecy rate is formulated as a constrained MDP (CMDP) under the constraints of UAV flight space, flight speed, energy capacity, anti-collision, and peak transmit power. To solve the CMDP, the authors design a DDPG-based RL algorithm. The algorithm includes three key components: the primary network (actor and critic networks), the target network, and the replay buffer. They also adopt state normalization and exploration noise to speed up the training convergence of the DDPG. The proposed joint optimization scheme achieves a secrecy rate over 40\% higher compared to baselines that optimize only trajectory or only power. In addition, DDPG outperforms DQN by approximately 15\% in secrecy rate due to its ability to handle continuous actions.

The DDPGG methods form a fixed mapping from states to actions, which is not suitable for large state spaces that require more exploration and uncertainty \cite{10545344}. The PPO alleviates this limitation by introducing proximal policy optimization, which allows for more exploration in the large action space.

%The study in \cite{10287142} proposes a hybrid framework to maximize the secrecy capacity in the network: a Double Cascade Correlation Network (DCCN) for RIS reflection coefficient optimization and a PPO-based DRL algorithm for UAV trajectory design. DCCN bypasses the need for labeled training data by cascading two neural networks, one mapping channel coefficients to RIS configurations and another predicting secrecy rates, to maximize the secrecy channel rate. The PPO algorithm optimizes UAV trajectory by formulating it as an MDP, where the state incorporates real-time channel coefficients (e.g., outdated CSI, interference, and RIS configurations), and actions define continuous UAV movement parameters (azimuth angle and distance). The PPO dynamically adjusts the UAV’s position to maximize secrecy rates by using clipped surrogate objectives to stabilize policy updates and advantage estimation to prioritize high-reward actions. Simulation results show that the proposed scheme (DCCN+PPO) achieves an average secrecy rate of 0.73 bps/Hz, outperforming the benchmarks DCCN+DDPG (0.67 bps/Hz) and Random RIS+PPO (0.13 bps/Hz).

The study in \cite{10287142} proposes a hybrid framework (Double Cascade Correlation Network (DCCN)+PPO) to maximize the secrecy capacity. DCCN bypasses the need for labeled training data by cascading two neural networks to maximize the secrecy channel rate. The PPO dynamically adjusts the UAV’s position by using clipped surrogate objectives to stabilize policy updates and advantage estimation to prioritize high-reward actions. Simulation results show that the proposed scheme (DCCN+PPO) achieves an average secrecy rate of 0.73 bps/Hz, outperforming the benchmarks DCCN+DDPG (0.67 bps/Hz) and Random RIS+PPO (0.13 bps/Hz). However, the average secrecy continues to decline when the transmit power is higher than 2 W, since the jamming signals transmitted by the secondary source against the eavesdropper will also affect the primary users.

The study in \cite{10287142} considered only the use of one UAV to assist in secure communication. However, in low-altitude economic networks, it may be more important for multiple UAVs to collaborate to implement anti-eavesdropping strategies. The study in \cite{10734220} considers a system model treats all secondary transmitters and multiple UAV jammers as multi-agents. A Multi-Agent PPO algorithm combined with Long Short-Term Memory (LSTM) networks, named MAPPO-LSTM, is proposed to maximize the secure communication rate by jointly optimizing the UAV trajectory, transmission power, and energy harvesting coefficients. The problem is formulated as a non-convex MDP consisting of an action space, state space, observation, and reward (which consists of the sum of the secure communication rate, SINR information, and battery capacity). The MAPPO algorithm introduces counterfactual baselines to address the credit assignment problem in centralized learning and combines with the LSTM network to enhance the learning capability of sequential sample data. Compared to benchmark schemes MAPPO and MADDPG, the proposed MAPPO-LSTM method achieved around 17\%-20\% higher average secrecy rate in large-scale scenarios, with convergence speeds 1.37 times and 1.93 times faster, respectively. In addition, the reward is sensitive to the discount factor, where setting factor to 0.99 enables faster and more stable convergence. Deviations from this value result in more pronounced fluctuations in the reward and secrecy rate.

%The study in \cite{10734220} considers a system model including a primary transmitter, multiple secondary transmitters, multiple UAV jammers, multiple receivers, and an eavesdropper, where secondary transmitters and multiple UAV jammers are all treated as multi-agents. A Multi-Agent PPO algorithm combined with Long Short-Term Memory (LSTM) networks, named MAPPO-LSTM, is proposed to maximize the secure communication rate by jointly optimizing the UAV trajectory, transmission power, and energy harvesting coefficients. The problem is formulated as a non-convex MDP consisting of an action space, state space, observation, and reward. The MAPPO algorithm introduces counterfactual baselines to address the credit assignment problem in centralized learning and combines with the LSTM network to enhance the learning capability of sequential sample data. The LSTM network manages information flow through forget gates, input gates, and output gates, effectively capturing the long-term dependencies in the time-series data of jamming signals. Simulation results show that the proposed MAPPO-LSTM method achieved a 17\% higher average secure communication rate compared to standard MAPPO in large (2000×2000 m²) areas. In addition, the convergence speed of MAPPO-LSTM was 1.37 times and 1.93 times faster than the benchmark schemes MAPPO and MADDPG, respectively.

{\bf Lesson Learned.} The RL has emerged as a powerful yet challenging tool for anti-eavesdropping in UAV-assisted secure communications. A key lesson is that multi-agent cooperation significantly enhances security compared to single-agent approaches, enabling adaptive trajectory control, power allocation, and jamming coordination in dynamic environments. However, scalability and convergence efficiency remain critical bottlenecks, especially in high-dimensional, time-varying settings, as many studies unrealistically assume perfect channel information, and deep RL's convergence time leaves the system vulnerable before optimization completes. Furthermore, key limitations demand further attention, such as RL's computational complexity, which restricts its use in resource-limited settings requiring real-time security, and its sensitivity to hyperparameter tuning that requires meticulous configuration to ensure optimal performance. Future advancements should focus on developing generalizable and robust learning architectures that can dynamically adapt to evolving threats while maintaining computational feasibility, addressing practical deployment challenges, exploring hybrid approaches, prioritizing security in system design, and balancing security performance with energy consumption.

\begin{table*}[]
\centering

\caption{Summary of Deep Learning for anti-eavesdropping strategy \\ CIRCLES DESCRIBE THE METHODS; CORRECT MARKERS AND CROSS MARKERS REPRESENT PROS AND CONS RESPECTIVELY.}
\label{tab:my-table}
\centering
\begin{tabular}{|c|c|c|l|}
\hline
\textbf{Techniques} & \textbf{Reference} & \textbf{Algorithm} & \multicolumn{1}{c|}{\textbf{Pros \& Cons}} \\ \hline
\multirow{9}{*}{\begin{tabular}[c]{@{}c@{}}Neural network\\ model\end{tabular}} 
& \cite{9713997} & DNN & \begin{tabular}[c]{@{}l@{}}\textcolor{blue!20}{\ding{108}} Use DNN to optimize UAV deployment and jamming strategies for secure communication \\ \textcolor{green}{\ding{51}} The DNN model reduces the complexity of exhaustive searches \\ \textcolor{red}{\ding{55}} Rely on complete statistical channel knowledge \\ \textcolor{red}{\ding{55}} require intensive resources to generate a large amount of training data \end{tabular} \\ \cline{2-4} 
& \cite{10114676} & CNN-LSTM & \begin{tabular}[c]{@{}l@{}}\textcolor{blue!20}{\ding{108}} CNN-LSTM-based SEM prediction for dynamic secure UAV trajectory optimization \\ \textcolor{green}{\ding{51}} Efficient spatiotemporal feature extraction via CNN-LSTM \\ \textcolor{red}{\ding{55}} Assume fixed UAV height and passive eavesdropper \\ \textcolor{red}{\ding{55}} Training CNN-LSTM network requires a substantial amount of data
 \end{tabular} \\ \cline{2-4} 
& \cite{10194980} & FL-DNN & \begin{tabular}[c]{@{}l@{}}\textcolor{blue!20}{\ding{108}} FL-DNN-RL integration (FairLearn) for maximizing fairness in secrecy rates \\ \textcolor{green}{\ding{51}} Collaborative learning via FL improves generalization in anti-eavesdropping strategies \\ \textcolor{red}{\ding{55}} Involving multiple learning mechanisms requires significant computational resources \\ \textcolor{red}{\ding{55}} Assuming perfect CSI and eavesdropper localization may be impractical \end{tabular} \\ \hline
\multirow{7}{*}{\begin{tabular}[c]{@{}c@{}}Generative AI\\ model\end{tabular}} 
& \cite{9584882} & MD-GAN  & \begin{tabular}[c]{@{}l@{}}\textcolor{blue!20}{\ding{108}} MD-GAN with unknown CSI as model weights\\ \textcolor{green}{\ding{51}} Adapt to dynamic environments via gradient-based training \\ \textcolor{green}{\ding{51}} Do not require knowledge of the eavesdropper’s detection threshold \\ \textcolor{red}{\ding{55}} Training a GAN can be computationally intensive \end{tabular} \\ \cline{2-4} 
& \cite{10700928} & DD-GAN & \begin{tabular}[c]{@{}l@{}}\textcolor{blue!20}{\ding{108}}DD-GAN uses genetic algorithm-generated datasets for GAN training
 \\ \textcolor{green}{\ding{51}} Achieve an effective trade-off between covert rate and detection probability \\ \textcolor{red}{\ding{55}} Training relies on the quality and quantity of the genetic algorithm-generated data \end{tabular} \\ \cline{2-4} 
& \cite{10759093} & GDMTD3 & \begin{tabular}[c]{@{}l@{}}\textcolor{blue!20}{\ding{108}} GDMTD3 integrates generative diffusion models into TD3
 \\ \textcolor{green}{\ding{51}}  Handle high-dimensional action spaces to adapt mobile eavesdroppers \\ \textcolor{red}{\ding{55}}  Computational complexity from diffusion denoising steps \end{tabular} \\ \hline
\end{tabular}

\end{table*}

%\subsubsection{Deep Learning} 
Deep learning, with its strong learning capabilities, parallel processing, and comprehensive reasoning \cite{8382166,10348530,hou2025split}, has huge potential to enhance anti-eavesdropping strategies in UAV communications, especially in environments with rapidly changing conditions and complex interactions \cite{10477870}. Given the intricate problem of UAV trajectory variation and its nonlinear characteristics in time and space \cite{10463689,9416564}, deep learning networks, such as neural networks and generative models, are emerging as potential solutions to improve the security and performance of UAV communication systems.

\begin{figure*}[!t]
\centering
\includegraphics[width=7.1in]{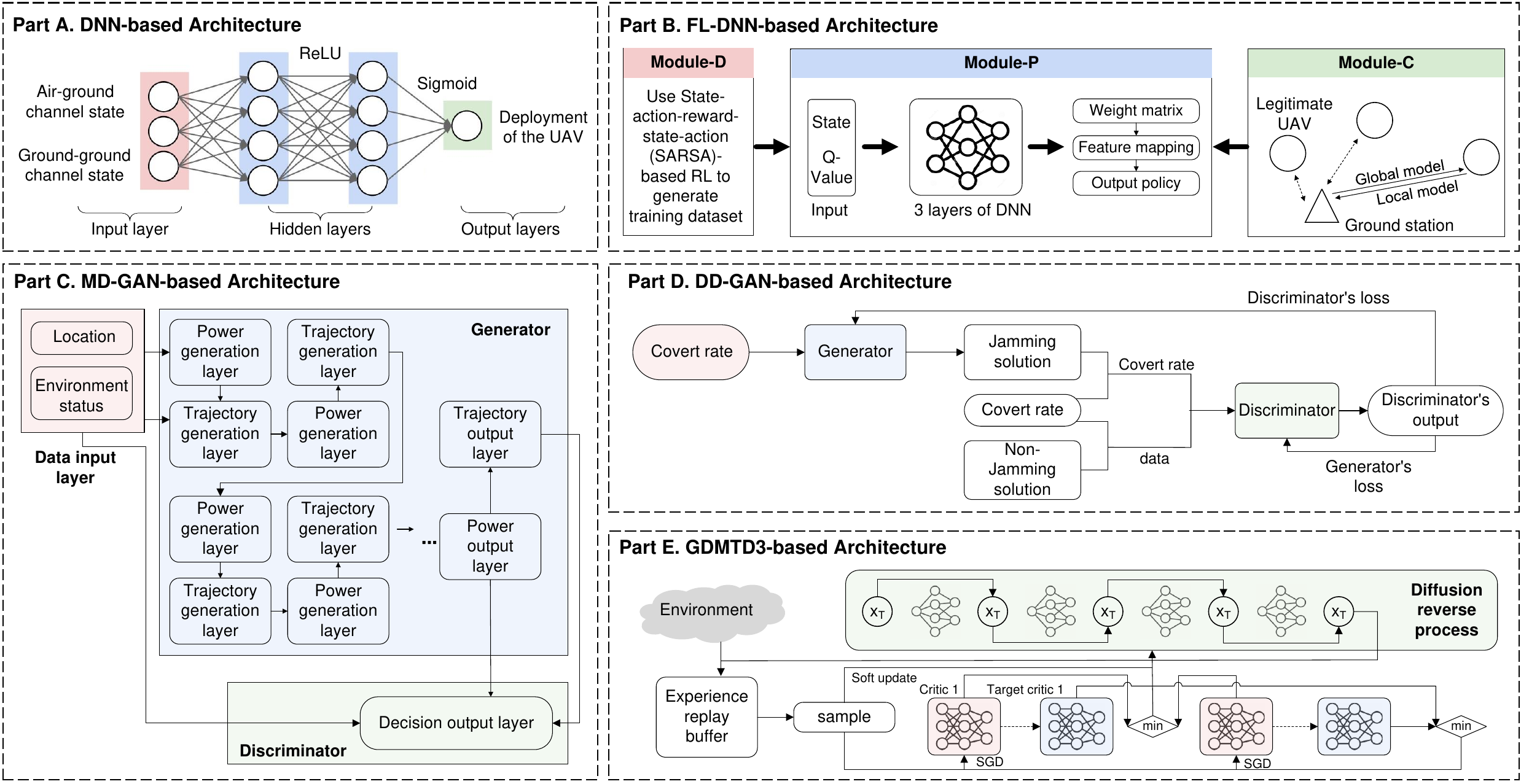}
\caption{ The overall architecture illustrates various deep learning-based architectures designed to enhance anti-eavesdropping capabilities in UAV deployment scenarios. Part A presents a DNN-based architecture that processes air-ground and ground-ground channel states to determine UAV deployment. Part B shows an FL-DNN-based architecture, incorporating modules for reinforcement learning, DNN-based feature mapping, and FL. Part C depicts an MD-GAN-based architecture, where a generator produces trajectories and power outputs based on location and environment status, while a discriminator evaluates the decisions. Part D introduces a DD-GAN-based architecture, focusing on generating jamming solutions to maximize covert rates, with a discriminator distinguishing between jamming and non-jamming solutions. Part E illustrates a GDMTD3-based architecture, utilizing an experience replay buffer and diffusion reverse process to optimize UAV deployment strategies.}
\label{Anti-eavesdroppingdeeplearning}
\end{figure*}

The study in \cite{9713997} explores the use of deep learning to optimize UAV deployment and jamming strategies against eavesdroppers to maximize the secrecy rate in the complete CSI scenario. The optimization problem is decomposed into two layers: the inner layer optimizes jamming power for a fixed UAV location, and the outer layer optimizes UAV deployment. The inner problem is solved using a bi-section search algorithm, while the outer problem is addressed using a deep neural network (DNN) to approximate the optimal UAV deployment. The DNN is designed as a fully connected structure, which includes an input layer, two hidden layers, and an output layer, as shown in part A of Fig. \ref{Anti-eavesdroppingdeeplearning}. The DNN is trained using a dataset generated by simulating different UAV deployments and corresponding secrecy rates. The final optimal deployment of the UAV can be approximated when the mean square error of weights between neurons is minimized. The DNN model achieves an average distance error of 2.2 meters compared to the optimal deployment found by the exhaustive search baseline.

The fully connected neural network used in \cite{9713997} is suited for problems where inputs and outputs are fixed-dimensional vectors without inherent spatial or sequential relationships \cite{8382166}. Moreover, convolutional neural networks (CNNs) and recurrent neural networks (RNNs) can also contribute to anti-eavesdropping. In contrast to fully connected networks, CNNs are particularly effective for exploring spatial features from images or spatial maps \cite{8755300}. RNNs, on the other hand, focus on handling sequential data by maintaining a memory of previous inputs through recurrent connections \cite{8382166}. The authors in \cite{10114676} propose a CNN-LSTM-based secure efficiency map (SEM) framework, which is constructed by calculating each subarea's security-efficiency index using a weighted exponential coefficient to combine normalized secure spectrum efficiency (secrecy rate per unit bandwidth) and secure energy efficiency (secrecy rate per unit power). Historical SEMs are fed into a CNN-LSTM network to predict future SEMs by leveraging spatial-temporal feature extraction and time-series correlation. Based on predicted SEMs, a trajectory planning algorithm dynamically guides the UAV to subareas with the highest security-efficiency indices. The proposed SEM-enabled trajectory planning achieves an average security-efficiency index of 0.81, outperforming baseline schemes (e.g., static trajectory \cite{8589002} or non-predictive methods \cite{9801656,9816078}) by over 30\%.

Previous deep learning-based architectures \cite{9713997,10114676} are centralized, lacking collaboration and knowledge sharing among UAVs, while also facing challenges in privacy preservation and scalability. To address these limitations and optimize secrecy rate maximization under constraints such as UAV mobility, power budgets, and scheduling fairness, the authors in \cite{10194980} propose a federated learning (FL)-based framework (FairLearn). As shown in part B of Fig. \ref{Anti-eavesdroppingdeeplearning}, the FairLearn employs three learning modules: (1) Module-D uses RL to dynamically generate training datasets by exploring UAV trajectories, power allocation, and scheduling policies; (2) Module-P employs a DNN trained on these datasets to predict optimal 3D trajectory, transmit power, and user scheduling, maximizing proportional fairness in secrecy rates (defined as the difference between legitimate UAV-user rates and eavesdropper rates); (3) Module-C applies FL to aggregate DNN models across UAVs, enabling collaborative learning while preserving data privacy. Simulation results show that FairLearn’s secrecy rate is 26.6\% higher than BCD at 1.4W transmit power. After 100s of execution, FairLearn achieves 14.34\%, 24.56\%, and 108\% higher secrecy rates than BCD, MAQ, and QCQP baselines, respectively.

It is worth noting that UAVs can only obtain limited prior environmental information without knowing perfect channel information and the eavesdropper's detection threshold or exact location. Some previous methods \cite{10310294,10734220,10194980} may find it difficult to solve the optimization problem in such scenarios. In contrast, the generative adversarial network (GAN) has emerged as a new model for solving optimization problems with limited prior information \cite{8907392,9069247}. GAN can effectively model and approximate unknown distributions (such as channel coefficients, detection thresholds, and environmental parameters) through adversarial learning, where the generator continuously improves its strategy by learning from the feedback from the discriminator \cite{8907392}.

The work in \cite{9584882} addresses challenges related to partial channel distribution information and unknown eavesdropper detection thresholds by proposing a model-driven GAN (MD-GAN) framework. The unknown channel coefficients and detection thresholds are treated as trainable weights in the MD-GAN. The MD-GAN transforms the joint trajectory and power optimization problem into a dynamic game between a generator (UAV) and a discriminator (eavesdropper), where the UAV acts as a jammer to protect secondary users from eavesdroppers. The generator optimizes the UAV’s 3D trajectory and jamming power, while the discriminator evaluates detection errors. Then, a GAN-based joint trajectory and power optimization (GAN-JTP) algorithm is developed to achieve Nash equilibrium (i.e., maximizing the covert rate and the probability of detection errors). As shown in part C of Fig. \ref{Anti-eavesdroppingdeeplearning}, the GAN-JTP algorithm consists of two stages: network learning and network training. In the network learning stage, the generator optimizes the UAV’s trajectory and transmit power based on the current state and environment. In the network training stage, the generator and discriminator are alternately trained using gradient backpropagation to update their weights. Simulation results show that increasing the training of the discriminator accelerates the convergence of the generator (e.g., when the training step is 10, convergence is achieved within 30 iterations, compared to 89 iterations when the training step is 1). For a flight period of 100 seconds, the GAN-JTP algorithm achieves a 0.47\% increase in covert rate with a 0.15\% reduction in detection error probability compared to the BCD-based scheme \cite{8456560}.

\begin{table*}[]
\centering

\caption{Summary of authentication for Communication Confidentiality \\ CIRCLES DESCRIBE THE METHODS; CORRECT MARKERS AND CROSS MARKERS REPRESENT PROS AND CONS RESPECTIVELY.}
\label{tab:my-table}
\centering
\begin{tabular}{|c|c|c|l|}
\hline
\textbf{Techniques} & \textbf{Reference} & \textbf{Algorithm} & \multicolumn{1}{c|}{\textbf{Pros \& Cons}} \\ \hline
\multirow{9}{*}{\begin{tabular}[c]{@{}c@{}}PUFs-based\\ authentication\end{tabular}} 
& \cite{9237145} & PUFs$^{1}$ & \begin{tabular}[c]{@{}l@{}}\textcolor{blue!20}{\ding{108}} PUF-based dynamic session key generation and mutual authentication protocol \\ \textcolor{green}{\ding{51}}  Lightweight design with no stored secrets \\ \textcolor{red}{\ding{55}} Potential overhead during temporary identity updates  \end{tabular} \\ \cline{2-4} 
& \cite{10146461} & PUF-fuzzy extractor & \begin{tabular}[c]{@{}l@{}}\textcolor{blue!20}{\ding{108}} PUF-fuzzy extractor mutual authentication with TS-based dynamic session adaptation \\ \textcolor{green}{\ding{51}} Dynamic session time adaptation minimizes idle periods and enhancing security \\ \textcolor{red}{\ding{55}} Higher computational cost due to fuzzy extractors
 \end{tabular} \\ \cline{2-4} 
& \cite{10436373} & {\begin{tabular}[c]{@{}c@{}}PUFs-fuzzy\\ extractor-AEGIS\end{tabular}}  & \begin{tabular}[c]{@{}l@{}}\textcolor{blue!20}{\ding{108}} Integration of PUFs, fuzzy extractors, and AEGIS for mutual authentication \\ \textcolor{green}{\ding{51}} The proposed password/biometric update mechanism reduces server dependency \\ \textcolor{red}{\ding{55}} Multiple cryptographic operations and protocols may be cause delay in the implementation  \end{tabular} \\ \hline
\multirow{7}{*}{\begin{tabular}[c]{@{}c@{}}Channel based-\\ authentication\end{tabular}} 
& \cite{9672766} & Rician channel  & \begin{tabular}[c]{@{}l@{}}\textcolor{blue!20}{\ding{108}} Physical-layer fingerprinting authentication based on Rician channel characteristics\\ \textcolor{green}{\ding{51}} Optimal power allocation balances data, AN, and authentication tag transmission \\ \textcolor{red}{\ding{55}} Reliance on encrypted tags requires high demand on UAV processing capabilities \end{tabular} \\ \cline{2-4} 
& \cite{10233023} & Rayleigh channel & \begin{tabular}[c]{@{}l@{}}\textcolor{blue!20}{\ding{108}}SNR difference-based PLA scheme \\ \textcolor{green}{\ding{51}} The SNR-based design can be implemented without additional hardware infrastructure \\ \textcolor{red}{\ding{55}} The simplified Rayleigh channel model may limit to real-world propagation environments \end{tabular} \\ \cline{2-4} 
& \cite{10834505} & {\begin{tabular}[c]{@{}c@{}}Rayleigh/Rician\\ channels\end{tabular}}  & \begin{tabular}[c]{@{}l@{}}\textcolor{blue!20}{\ding{108}} AD metric$^{2}$ for authentication under Rayleigh/Rician channels \\ \textcolor{green}{\ding{51}}  AD metric-based method improves the detection accuracy of authentication \\ \textcolor{green}{\ding{51}}  Detailed analysis of authentication performance under different propagation conditions \\ \textcolor{red}{\ding{55}}  Computational complexity in Rician channels due to hypergeometric functions \end{tabular} \\ \hline
\end{tabular}

\raggedright
\footnotesize $^{1}$Physical Unclonable Functions (PUFs) are hardware-based security primitives that exploit inherent and unique physical variations in devices to generate unclonable and unpredictable responses for communication authentication.\\

\footnotesize $^{2}$Authentication Distance (AD) is a metric proposed in \cite{10834505} to distinguish legitimate and illegitimate signals for communication authentication.

\end{table*}

The MD-GAN \cite{9584882} relies on model-driven methods where the unknown channel information and detection threshold are treated as trained weights. Differently, the authors in \cite{10700928} propose a data-driven GAN (DD-GAN) framework that focuses on generating data consisting of environmental parameters and optimal solutions to train the GAN. Specifically, the DD-GAN transforms the optimization process into an interactive game between the UAV and eavesdropper, where the UAV aims to maximize the covert rate, and the eavesdropper aims to detect the presence of covert communication. To address the non-convexity and lack of eavesdropper detection threshold information in the optimization process, the DD-GAN trains a generator (UAV) and discriminator (eavesdropper) adversarially, using genetic algorithm-generated samples as prior data, as shown in part D of Fig. \ref{Anti-eavesdroppingdeeplearning}. The generator produces power and trajectory solutions, while the discriminator evaluates the detectability. The loss function of the discriminator is designed to maximize the probability of correctly identifying real data and minimize the probability of being fooled by generated data. The generator's loss function aims to maximize the probability that the generated data is mistaken for real data by the discriminator. %The training process continues until the loss functions have converged.
%Experimental results show that DD-GAN reduces detection error probability by 8\% and UAV transmit power by 55\%, while increasing the covert rate by 10.4\% compared to the baseline (UAV knows the exact location of eavesdropper) at 100s. %Additionally, DD-GAN converges within 200 training steps, showing stable adversarial training dynamics.

Besides GANs \cite{9584882,10700928}, another generative model, the diffusion model, has advanced the effective representation of multi-dimensional data distributions \cite{10419041}. The diffusion model can better capture the complex dynamics and the trade-off in the multi-objective optimization problem concerning secure communication \cite{10749978}. For example, The diffusion model captures complex state-action distributions, enabling adaptive beamforming and UAV repositioning under eavesdropper mobility. To tackle dynamic environments and high-dimensional action spaces in secure communication and energy efficiency multi-objective optimization problem, the authors in \cite{10759093} propose GDMTD3, a Twin Delayed Deep Deterministic Policy Gradient (TD3) algorithm enhanced with generative diffusion models. Key innovations include integrating diffusion-based reverse processes into the actor network for robust policy generation and addressing continuous action spaces, as shown in part E in Fig. \ref{Anti-eavesdroppingdeeplearning}. The training process of GDMTD3 involves initializing the online critic and actor networks, interacting with the environment, and updating the network parameters based on the collected experiences. The actor network uses a generative diffusion model to sample actions, while the critic networks evaluate the actions using twin critic networks to reduce overestimation bias. Simulation results show that GDMTD3 outperforms DRL-based benchmarks (including PPO, TD3, and DDPG), achieving about 50\% higher cumulative rewards and around 21\% higher average secrecy rate than TD3. In addition, when the number of UAVs increases from 4 to 8, the average secrecy rate increases accordingly. However, increasing the number of UAVs from 8 to 16 raises energy consumption but only marginally improves secrecy rates, highlighting a performance-energy trade-off.

{\bf Lesson Learned} A key lesson learned is that deep learning, particularly through advanced architectures such as GANs \cite{9584882,10700928
} and diffusion models \cite{10759093}, can address complex, dynamic environments with partial channel state information and unknown eavesdropper locations, while demonstrating superior performance over traditional methods \cite{10310294,10734220,9713997}. These approaches demonstrate that deep learning not only strengthens the resilience of secure communications but also enables autonomous, real-time decision-making to counteract evolving eavesdropping threats in UAV networks.

\subsection{Communication Authentication}

In the LAENet, as UAVs operate in open environments and rely on wireless communication, they are highly vulnerable to security threats such as node capture and man-in-the-middle attacks \cite{8883128}. Ensuring secure and reliable authentication between UAVs and ground stations/users or among UAVs is critical to preventing unauthorized access \cite{9279294,8382275}. Traditional cryptographic authentication schemes often impose significant computational and memory overheads and incur considerable lantency, making them unsuitable for resource-constrained UAVs \cite{9551779}. Recently, advancements such as PUFs and Physical-layer Authentication (PLA) mechanisms % based on channel characteristics
have opened new possibilities for lightweight and effective authentication in the LAENet.

PUFs are a class of hardware security primitives that leverage the inherent manufacturing variations (such as variations in circuit delay or RF properties) in semiconductor devices to generate unique and unpredictable responses \cite{8390918}. When a specific input is applied to a PUF, the device generates a corresponding response, forming a challenge-response pair that is unique to this device \cite{8390918}. Such uniqueness and unpredictability make PUFs highly resistant to cloning and tampering, making them as a secure means for device authentication and key generation \cite{9018072}. %Different from traditional cryptographic keys stored in memory, PUFs do not require external storage since their responses are derived from intrinsic physical properties. 
In addition, employing a PUF in a UAV allows for secure authentication without the need for complex cryptographic operations, making it an efficient solution for resource-constrained scenarios \cite{9745033}.

\begin{figure*}[!t]
\centering
\includegraphics[width=7.1in]{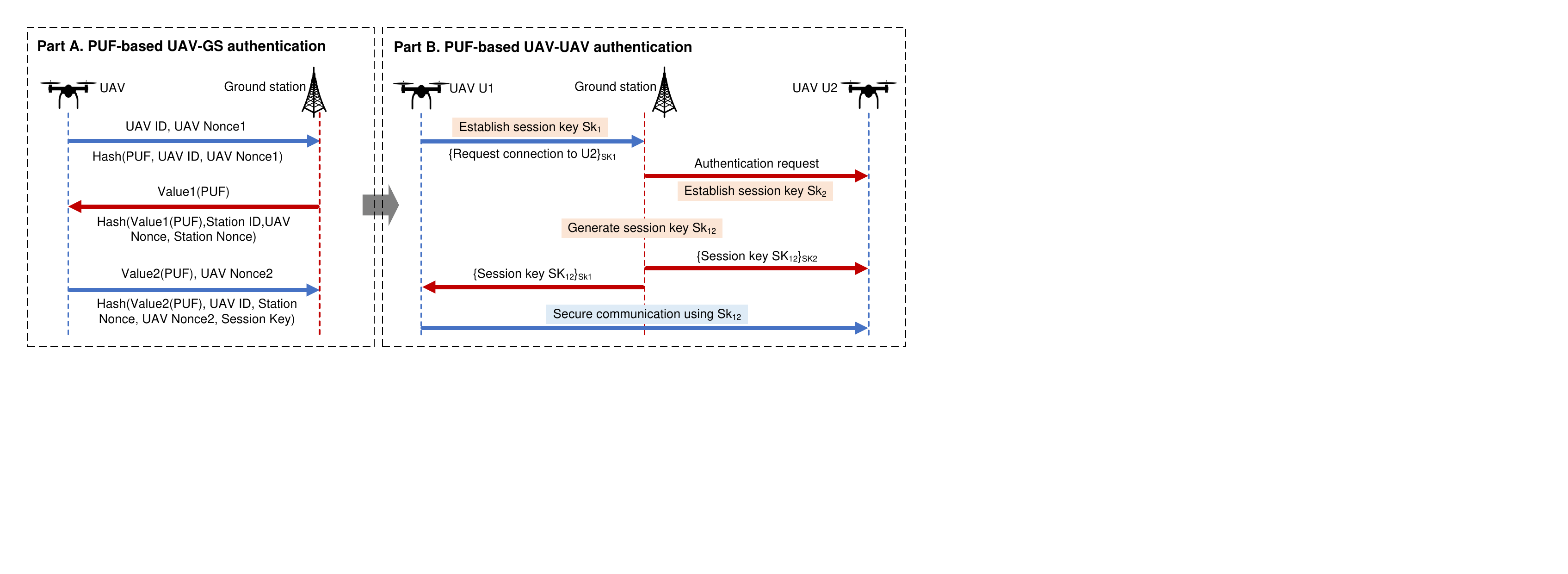}
\caption{ The overall architecture of the PUF-based authentication schemes for UAV-GS and UAV-UAV communication in \cite{9237145}. Part A illustrates the PUF-based authentication process between a UAV and a ground station (GS). The UAV sends its ID and a nonce to the GS, which responds with a hash value based on the PUF, UAV ID, and nonce. The UAV then sends a value derived from the PUF and another nonce, and the GS verifies the authentication by comparing hash values. Part B shows the PUF-based authentication between two UAVs (U1 and U2) through the GS. After establishing a session key with the GS, U1 requests a connection to U2. The GS facilitates the authentication by generating a new session key, which is securely shared between U1 and U2.}
\label{PUF authenticationfig}
\end{figure*}

The work in \cite{9237145} proposes a lightweight mutual authentication protocol, named SecAuthUAV, for securing UAV-ground station and UAV-UAV communications. SecAuthUAV employs PUFs in each UAV to generate a unique, unclonable session key that functions as a non-reproducible fingerprint. The protocol consists of three phases, as shown in Fig. \ref{PUF authenticationfig}. During UAV registration, a challenge-response pair from the UAV's PUF is stored, and a temporary identity is generated. In the UAV-ground station authentication phase, the UAV and ground station authenticate each other using challenge-response pairs and nonces, establish a session key, and update their identities. Lastly, in the UAV-UAV authentication phase, the GS facilitates secure communication by authenticating a second UAV and generates a session key for both UAVs. %Simulation results show that SecAuthUAV achieves a computation cost of 361 $\mu$s, outperforming baseline protocols \cite{9153239} (667 $\mu$s) by about 45\%. Communication overhead is reduced to 1600 bits, 22\% lower than \cite{9153239} (1952 bits), and storage cost is 352 bits, 9\% higher than \cite{9153239} but significantly lower than others (e.g., 640 bits for \cite{8693567}).

However, the work in \cite{9237145} ignores the fact that the noise in PUFs can result in significant deviation in the output for the same input at different time points. In addition, \cite{9237145} does not adjust the session time after establishing an authenticated session between two parties, which may lead to the active session remaining idle for a long time and thus give an opportunity for an adversary to interfere with the communication link. In light of this, the authors in \cite{10146461} propose an UAV Authentication with Adaptive Session (UAAS) framework to address these challenges. Firstly, they combine PUFs and fuzzy extractors to address PUF noise. The fuzzy extractors consist of two phases: the $Gen(.)$ phase creates a key and non-sensitive helper data, and the $Rep(.)$ phase reconstructs the key from a noisy PUF response using the helper data while tolerating minor deviations. Then, the Thompson Sampling (TS)-based scheme is proposed to dynamically adapt the session time. TS is a probabilistic approach that balances exploration and exploitation, determining the session time based on the fraction of busy time to minimize idle periods and reduce the risk of adversarial interference. Although the security analysis demonstrates that UAAS improves the security level in the mutual authentication mechanism, its throughput is 20.38\% lower and computational cost is 126 ms higher than the baseline \cite{9237145} due to security overhead.

\begin{figure}[!t]
\centering
\includegraphics[width=3in]{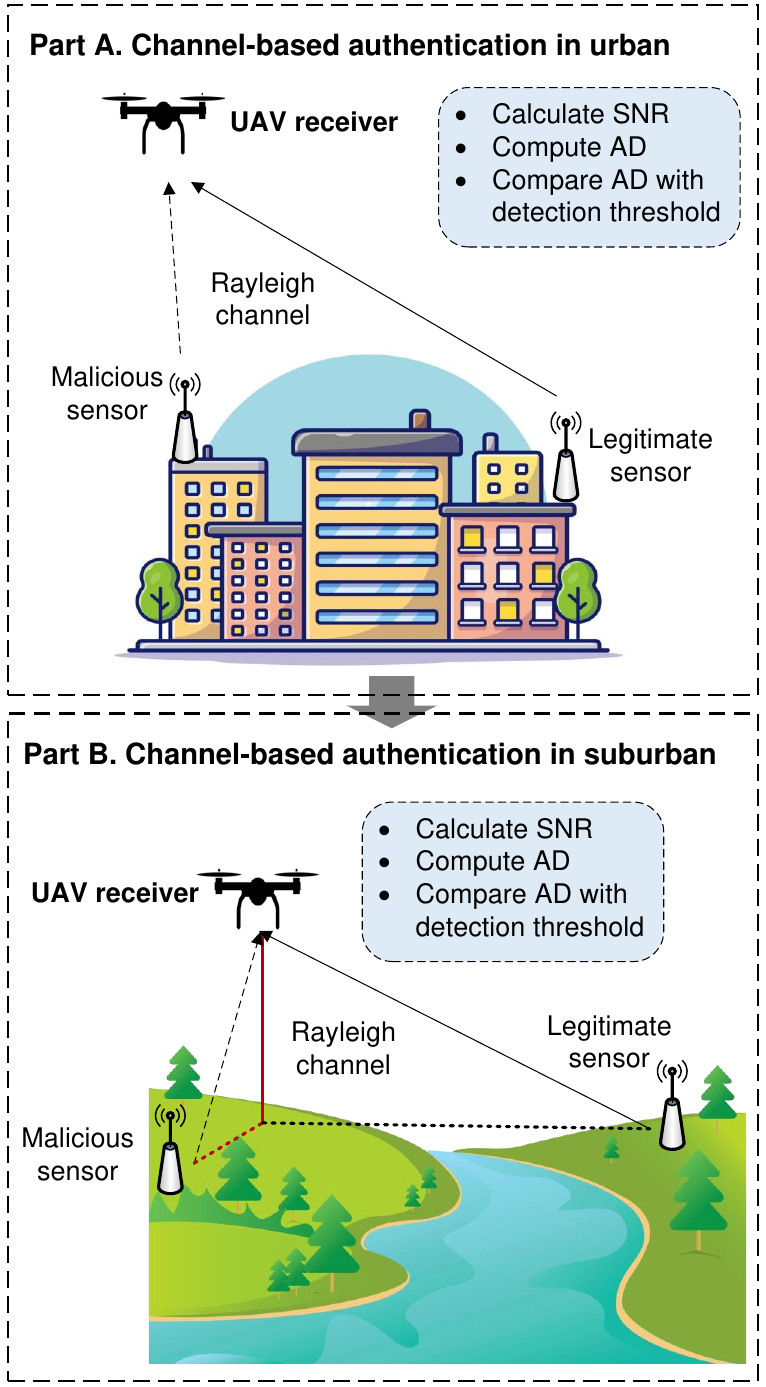}
\caption{ The overall architecture of the channel-based authentication in urban and suburban environments in \cite{10834505}. Part A depicts the authentication process in an urban environment under Rayleigh channel conditions. The UAV receiver calculates the SNR, computes the AD, and compares it with a detection threshold to distinguish between legitimate and malicious sensors. Part B illustrates the authentication process in a suburban environment, where the UAV receiver performs similar steps to authenticate legitimate sensors and detect malicious ones under Rayleigh channel conditions.}
\label{Channel authenticationfig}
\end{figure}

In the LAENet, while establishing mutual authentication between UAVs and ground stations is critical, it is also important to incorporate role-specific access controls for users to ensure communication confidentiality and preventing unauthorized access \cite{ZHANG2024110118}. The work in \cite{10436373} proposes an authentication framework PUF-enabled authentication framework for Internet of Drones (PAF-IoD) to establish mutual authentication among users, UAVs, and ground stations. %UAVs collect sensitive information from the environment and transmit it to ground stations. 
Users need to authenticate with stations to access the stored data or communicate directly with UAVs, where the users' authentication mechanism includes three factors (identity, password, and biometric data). % and employs XOR operations and SHA-256 for enhanced security. 
Similar to \cite{10146461}, PAF-IoD uses PUFs and a fuzzy extractor in the authentication process to generate a unique and tamper-proof session key while tolerating the noise in PUFs. Furthermore, the designed authenticated encryption with associative data (AEAD)-based encryption algorithm is utilized for encrypting and decrypting messages exchanged between the user, ground station server, and UAVs. %PAF-IoD achieves a communication cost of 2400 bits during authentication, outperforming baseline frameworks (\cite{9272616,alzahrani2021resource,SURESHKUMAR2019938}) by 9.64\%–41.69\%. Its authentication phase completes in 13.725 ms, offering a 52.32\%–73.83\% improvement in computational efficiency.

%The aforementioned works \cite{9237145,10146461,10436373} leverage the intrinsic physical properties of hardware for authentication design through PUFs, ensuring device uniqueness and resistance to cloning. However, authentication mechanisms are required to adapt to the dynamic conditions of complex communication environments in the LAENet scenario. 

In addition to leveraging intrinsic physical properties of hardware for authentication design through PUFs \cite{9237145,10146461,10436373}, the characteristics of communication channels can be used for authentication. The PLA mechanism authenticates devices by exploiting the unique physical characteristics of wireless communication channels, such as CSI, received signal strength (RSS), and signal-to-noise ratio (SNR) \cite{7420748}. The main reason is that the wireless channel between two communicating entities exhibits location-specific and time-varying properties due to multipath propagation, fading, and environmental factors \cite{9592624}. These diverse physical channel conditions, which provide a robust set of features for authentication, have been investigated in terrestrial communication networks \cite{7420748,9592624,9335644}. Furthermore, the source of received signals can be accurately and promptly detected \cite{9335644}, making PLA particularly advantageous in the dynamic and complex communication environments of the LAENet.

The authors in \cite{9672766} leverage the unique properties of the physical layer channel, Rician channel, to develop a PLA for UAV-ground station communication. Given that UAVs receive signals subject to the Rician fading model, the ground station integrates authentication directly into the transmission process. It employs a one-way collision-resistant function (e.g. cryptographic hash function) that combines data symbols with a shared secret key to generate a low-power authentication tag for UAV and seamlessly embeds it into the transmitted signal. The authentication tag is validated by the correlation shaped by the Rician statistical characteristics of the fading channel, i.e., the correlation between the estimated tag (derived from the received signal) and the expected tag (generated using the secret key and decoded data symbols).

%However, the work in \cite{9672766} still partially relies on cryptographic tag generation for authentication, which may not be suitable for UAVs with limited processing capabilities. The study in \cite{10233023} leverages channel characteristics and geographical locations for PLA design, where the signal-to-noise ratio (SNR) differences between consecutive transmissions are utilized as the authentication metric. For single UAV scenarios, closed-form expressions for the probability density function of SNR differences, false alarm rate (FAR), and miss detection rate (MDR) are derived under Rayleigh fading channels. A non-convex optimization problem of minimizing MDR under FAR constraints is formulated and solved by using SCA algorithms. Furthermore, this authentication mechanism is extended to double UAVs, where averaged SNR differences are analyzed, and closed-form FAR/MDR expressions are derived. Simulation results show that with FAR constraint equal to 0.2, the proposed authentication scheme achieves an MDR less than 0.005, outperforming the RSS-based baseline [20] by about 40\% in MDR reduction. %In addition, theoretical FAR/MDR predictions match Monte Carlo points well, which confirms the correctness and accuracy of derived FAR/MDR expressions.

However, the work in \cite{9672766} still partially relies on cryptographic tag generation for authentication, which may not be suitable for UAVs with limited processing capabilities. The study in \cite{10233023} leverages channel characteristics and geographical locations for PLA design, where the SNR differences between consecutive transmissions are utilized as the authentication metric. Specifically, a legitimate transmitter and a jammer have distinct channel variations due to differences in their geographical locations. The UAV authenticates the legitimate transmitter or jammer by formulating a binary hypothesis test based on the SNR difference between two successive transmissions. If the difference falls within a predefined threshold, the transmission is authenticated as from the legitimate transmitter; otherwise, it is classified as a jammer. The closed-form expressions for the probability density function of SNR differences, false alarm rate (FAR), and miss detection rate (MDR) are derived under Rayleigh fading channels in single-UAV and dual-UAV scenarios. The non-convex optimization problem of minimizing MDR under FAR constraints is solved using an SCA algorithm, which outperforms the RSS-based baseline \cite{9739860} by about 40\%.

It is worth noting that study \cite{9672766} may lack a comprehensive analysis of the UAV-PLA performance under different propagation conditions. Additionally, the detection performance may be further improved with other indicators. As shown in Fig. \ref{Channel authenticationfig}, the work in \cite{10834505} proposes a novel PLA framework under different propagation conditions, including dense urban and suburban environments modeled by Rayleigh and Rician channels, respectively. A new metric, Authentication Distance (AD), is proposed as the normalized difference in received SNR between adjacent transmissions. For Rayleigh channels, closed-form expressions for FAR and MDR are derived using convolution and integration-by-parts, while Rician channels employ doubly non-central $F$ distributions to model AD statistics. Similar to study \cite{10233023}, this authentication framework minimizes MDR under FAR constraints. In dense urban settings, MDR depends on path loss and transmitter-UAV geometry. For suburban environments, it incorporates elevation angle-dependent Rician factors and path loss exponents to improve discriminability between legitimate and illegitimate signals. The proposed AD-based method outperforms the SNR-difference baseline \cite{10293903}, achieving 40–60\% lower MDR. %Simulation results validate that theoretical FAR/MDR expressions align with Monte Carlo results. 

{\bf Lesson Learned.} Leveraging physical-layer characteristics, such as PUFs and channel properties, in conjunction with communication models and optimization algorithms, has proven effective in enhancing authentication accuracy and reducing detection errors. However, some methods also reveal limitations. For instance, the assumptions of ideal channel conditions and the neglect of practical implementation constraints may limit the applicability of the proposed solutions \cite{10233023,10834505}. Future research should focus on addressing these limitations by exploring more practical channel models and considering the trade-offs between security and system complexity.

\section{Communication Availability for LAENet}

\subsection{Anti-Jamming Strategy}

\begin{table*}[]
\centering

\caption{Summary of Anti-Jamming Strategy for Communication Availability \\ CIRCLES DESCRIBE THE METHODS; CORRECT MARKERS AND CROSS MARKERS REPRESENT PROS AND CONS RESPECTIVELY.}
\label{tab:my-table}
\centering
\begin{tabular}{|c|c|c|l|}
\hline
\textbf{Techniques} & \textbf{Reference} & \textbf{Algorithm} & \multicolumn{1}{c|}{\textbf{Pros \& Cons}} \\ \hline
\multirow{7}{*}{\begin{tabular}[c]{@{}c@{}}Convex \\ optimization\end{tabular}} 
& \cite{9271902} & BCD, SCA & \begin{tabular}[c]{@{}l@{}}\textcolor{blue!20}{\ding{108}} BCD and SCA for UAV 3D trajectory optimization for anti-jamming \\ \textcolor{green}{\ding{51}} Probabilistic LoS performs well in real-world scenarios such as urban environments \\ \textcolor{red}{\ding{55}} High computational complexity may be challenging in resource-constrained environments
 \end{tabular} \\ \cline{2-4} 
& \cite{9200570} & {\begin{tabular}[c]{@{}c@{}}SCA,\\ Dinkelbach\end{tabular}} & \begin{tabular}[c]{@{}l@{}}\textcolor{blue!20}{\ding{108}} SCA and Dinkelbach algorithm for energy-efficient trajectory optimization under malicious jammers \\ \textcolor{green}{\ding{51}} Balance between throughput and energy consumption in anti-jamming \\ \textcolor{red}{\ding{55}} Assume static and LoS-dominated channels 
 \end{tabular} \\ \cline{2-4} 
& \cite{9454372} & BCD, SCA & \begin{tabular}[c]{@{}l@{}}\textcolor{blue!20}{\ding{108}} BCD and SCA for joint UAV trajectory and transmit power optimization under jamming
 \\ \textcolor{green}{\ding{51}} Improve throughput by considering transmit power optimization against jammers
 \\ \textcolor{red}{\ding{55}} Assume a fixed UAV altitude and a static channel environment \end{tabular} \\ \hline
\multirow{7}{*}{\begin{tabular}[c]{@{}c@{}}Multi-agent \\ RL\end{tabular}} 
& \cite{9816050} & MALQL  & \begin{tabular}[c]{@{}l@{}}\textcolor{blue!20}{\ding{108}} Collaborative MALQL algorithm for anti-jamming with channel and power allocation\\ \textcolor{green}{\ding{51}} Accelerate convergence compared to single-agent Q-learning \\ \textcolor{red}{\ding{55}} Assume predefined UAV trajectories limits to adaptability \end{tabular} \\ \cline{2-4} 
& \cite{10750022} & MARL & \begin{tabular}[c]{@{}l@{}}\textcolor{blue!20}{\ding{108}}MARL with adversarial pre-training for dynamic and generalized jamming \\ \textcolor{green}{\ding{51}} Generalize to various jamming patterns via adversarial populations for pre-training
 \\ \textcolor{red}{\ding{55}} Pre-training for generalized jamming may require significant offline resources \end{tabular} \\ \cline{2-4} 
& \cite{10614297} & MATD3 & \begin{tabular}[c]{@{}l@{}}\textcolor{blue!20}{\ding{108}} MATD3 algorithm with PER for dynamic resource management under jamming attacks \\ \textcolor{green}{\ding{51}}  Handle high-dimensional continuous action spaces \\ \textcolor{red}{\ding{55}}  The integration of PER and spectrum sensing may increase the computational complexity \end{tabular} \\ \hline
\end{tabular}

\end{table*}

Jamming attacks pose significant challenges to communication availability in the LAENet by disrupting legitimate communication links and degrading the performance of aircraft communication networks \cite{10464352,10591496}. As shown in Fig. \ref{Anti-jammingfig}, these attacks can exploit the openness and broadcasting nature of UAV networks, making them particularly vulnerable to interference \cite{10464352}. Malicious jammers can transmit strong signals that weaken signal strength, degrade signal quality, and increase communication delays, leading to unreliable coverage and potential paralysis of the entire network \cite{9493713,10591496}. This vulnerability underscores the urgent need for effective anti-jamming technologies to ensure reliable communication in the LAENet.

Various anti-jamming strategies have been explored to safeguard the LAENet against malicious jamming, mainly focusing on trajectory adjustment, as well as channel and power control. Overall, by adjusting the trajectory in the spatial domain, an UAV can evade jamming signals while maintaining reliable communication with legitimate devices \cite{9493713,9271902}. Besides the spatial-domain anti-jamming strategy, the UAV can implement a frequency-domain-based anti-jamming scheme. The UAV can select legitimate channels while avoiding jamming signals and control transmit power to minimize energy consumption and latency under jamming attacks \cite{7925694,8023829}.

Convex optimization methods can be used to adjust the UAV’s trajectory to achieve anti-jamming by strategically guiding its movement to reduce interference and enhance communication reliability \cite{9493713,9271902}. It provides a systematic and efficient approach to handle the complex, non-convex problems that arise when optimizing UAV trajectories and various constraints under malicious jamming conditions \cite{8543573}. The work in \cite{9271902} investigates anti-jamming 3D trajectory design for UAV-enabled wireless sensor networks under a probabilistic LoS channel model. The probabilistic LoS model accounts for elevation angle-dependent shadowing effects in urban environments compared with simplified LoS models. The BCD and SCA algorithms are employed to optimize the UAV's horizontal and vertical trajectory, allowing the UAV to move closer to the ground station for improved transmission rates while dynamically adjusting its elevation angle relative to the jammer to mitigate interference.

%The BCD and SCA algorithms are used to maximize the minimum average expected rate by jointly optimizing GS transmission scheduling, UAV horizontal trajectory, and vertical trajectory. 

However, the anti-jamming trajectory optimization in \cite{9271902} under the probabilistic LoS model does not consider the energy consumption issue. The study in \cite{9200570} utilizes SCA and Dinkelbach’s algorithm to adjust the UAV's trajectory to avoid areas with jammers while maximizing energy efficiency, which is defined as the ratio of total throughput to propulsion energy consumption during flight. Compared to hovering-centric benchmarks, the optimized trajectory reduced energy consumption by 82\% while maintaining 73.16\% of the sum throughput. It is worth noting that the transmit power of the UAV and station is fixed in \cite{9200570}, whereas power optimization is also an important factor for energy efficiency. The authors in \cite{9454372} use the SCA and BCD algorithms to maximize throughput by iteratively optimizing power allocation (via convex reformulation of throughput bounds) and UAV trajectory (via slack variables for distance constraints and jamming mitigation) to avoid jamming signals. The proposed scheme achieves 40\% higher throughput compared to the "Line trajectory with fixed power" baseline.

%It is worth noting that the transmit power of the UAV and the source node were both fixed in \cite{9200570}, indicating that joint UAV trajectory and transmit power design under malicious jamming has not been solved. The authors in \cite{9454372} aim to maximize end-to-end throughput by jointly optimizing the UAV trajectory and transmit power of both the UAV and the ground source node. They use the SCA and BCD algorithms to iteratively optimize power allocation (via convex reformulation of throughput bounds) and trajectory (via slack variables for distance constraints and jamming mitigation), ensuring convergence to a suboptimal solution. The proposed scheme achieves 40\% higher throughput compared to the "Line trajectory" baseline (fixed path and power).

While convex optimization methods \cite{9271902,9200570,9454372} work well for fixed jamming patterns, they may struggle to handle dynamic, intelligent jamming \cite{9367220} in real-time due to their reliance on global information and the challenges inherent in solving non-convex problems with increased optimized variables \cite{8589002}. In contrast, RL and DRL offer significant advantages by enabling autonomous, adaptive decision-making \cite{10283826,9403369}. These approaches can continuously adjust to environmental changes, learn from past interactions, and optimize performance in real-time \cite{10413638,9989422}. The RL-based anti-jamming methods have emerged as a promising solution due to their ability to operate without excessive prior information (such as unknown environment, CSI, and jamming mode) \cite{9403369}. Single-agent RL algorithms have been used in previous works to develop anti-jamming strategies in communication networks by regarding jammers and other legitimate users as part of the environment, including independent anti-jamming channel selection methods \cite{7925694,8023829,8314744,9264659}. However, these single-agent approaches may fail to converge when dealing with a large number of agents or a high-dimensional action-state space \cite{9816050}, making them impractical for complex, multi-agent scenarios in the LAENet. To address these limitations, multi-agent RL (MARL) methods have been proposed to allow each agent to make decisions based on local information and exchange data with others (such as observations or model parameters).%, which not only enhances scalability but also ensures that the system can adapt to the dynamic nature of jamming threats in real-time. %Thus, the multi-agent RL-based approaches have become a promising direction for anti-jamming.

The study in \cite{9816050} proposes a collaborative multiagent layered Q-learning (MALQL) algorithm for anti-jamming communication in UAV networks by jointly optimizing channel and power allocation to maximize system Quality of Experience (QoE). The problem is modeled as a local interaction Markov game based on the constructed interference graph. The MALQL divides the problem into two subgames of channel selection (Layer 1) and power allocation (Layer 2), as shown in part B of Fig. \ref{Anti-jammingfig}. The channel layer uses a graph-based interference model to capture mutual interference among UAVs. Each UAV is represented as a node, and edges are formed between UAVs that are within a predefined interference distance. This model allows UAVs to identify and avoid channels that are being used by neighboring UAVs or jammed by external attackers, thereby reducing the jamming likelihood. The power layer optimizes transmit power to meet rate thresholds. Theoretical analysis confirms that MALQL can converge to a pure strategy Nash equilibrium.

\begin{figure*}[!t]
\centering
\includegraphics[width=7.1in]{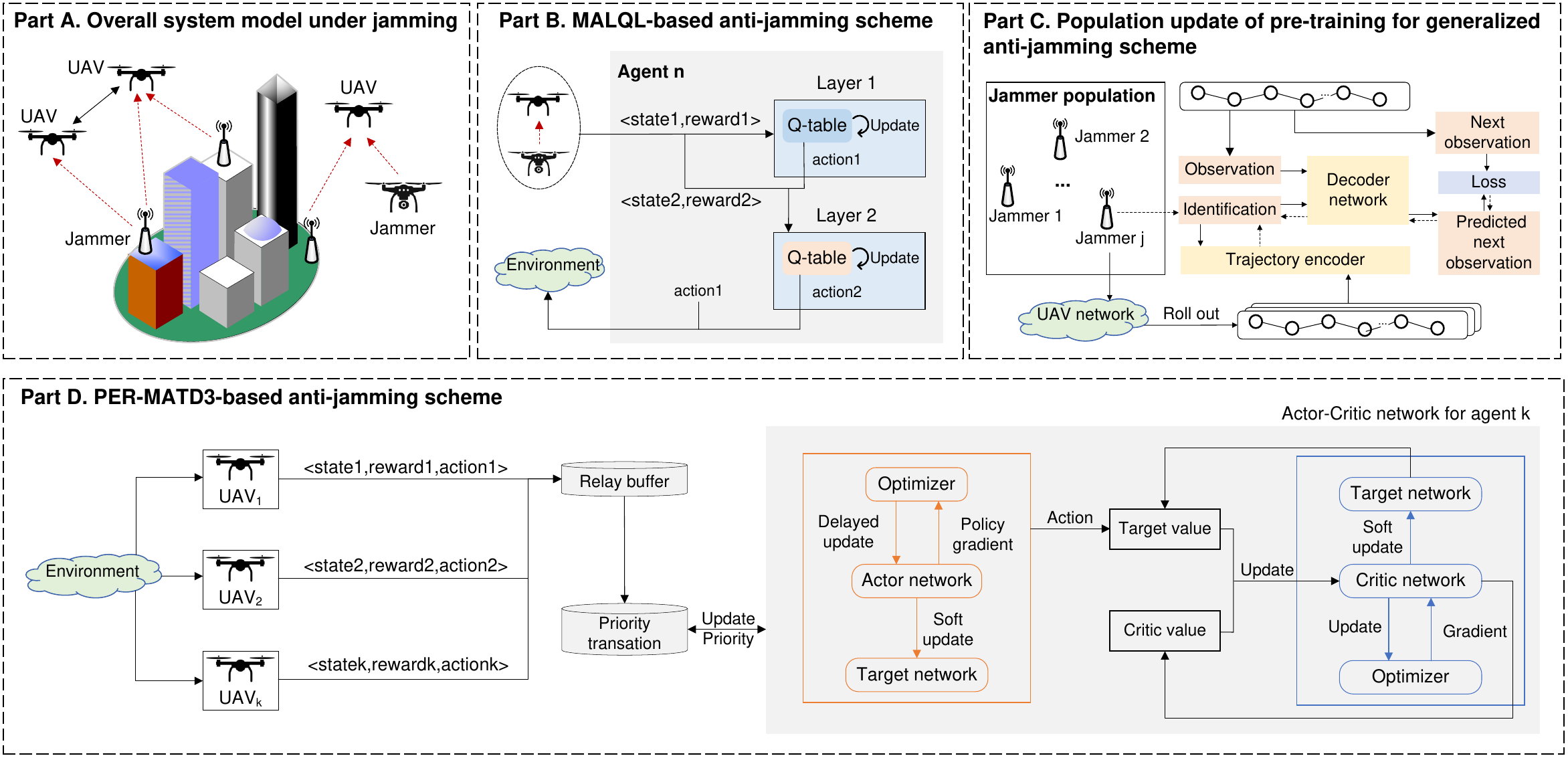}
\caption{The overall architecture illustrates various reinforcement learning-based anti-jamming schemes designed to enhance communication reliability in UAV-assisted MEC systems under jamming attacks. Part A presents the overall system model, depicting UAVs and jammers interacting within a dynamic environment. Part B shows the MALQL-based anti-jamming scheme, where agents use layered Q-learning to determine actions based on local observations and rewards. Part C depicts the population update mechanism for pre-training a generalized anti-jamming scheme, involving a jammer population, trajectory encoder, and decoder network to optimize jamming policies. Part D introduces the PER-MATD3-based anti-jamming scheme, incorporating a priority experience replay buffer and actor-critic networks to dynamically allocate resources and optimize UAV deployment strategies.}
\label{Anti-jammingfig}
\end{figure*}

Nevertheless, there are still some issues with the anti-jamming mechanism in \cite{9816050}. Considering that the rapid mobility of UAVs may expose them to various and unknown jamming patterns due to frequent transitions to new scenarios, the anti-jamming methods need to be generalized \cite{9777258}, especially in the LAENet. The work \cite{9816050} randomly initializes strategies and learns from scratch for a particular deployment environment with no pretraining, which may lead to a reduction in the generalization ability of the anti-jamming strategy. In light of this, the authors in \cite{10750022} introduce an adversarial pre-training stage in the proposed two-stage MARL with a decentralized partially observable Markov decision process. Specifically, the adversarial pre-training stage uses a quality-diverse jammer population (e.g., fixed, random, sweeping, statistic, and RL-based jamming) to bootstrap generalized anti-jamming strategies instead of directly initializing the agents with random anti-jamming policies, as shwon in part C of Fig. \ref{Anti-jammingfig}. This pre-training ensures that UAVs are not overfitted to specific jamming patterns and can generalize to new jamming attacks in real-world deployments. %Then, in the fine-tuning stage, the pre-trained policies are deployed and further optimized using online learning% (designed graph convolutional MARL with parallel Q-networks and an information temporal smoothing mechanism) to adapt to dynamic environments. 
The pre-trained policies are deployed in the fine-tuning stage, where a graph convolutional-based MARL algorithm is proposed to jointly optimize channel selection and power allocation for anti-jamming similar to \cite{9816050}. Simulation results show that the proposed solution achieves 20–30\% higher cumulative rewards than collaborative multi-agent Q-learning \cite{8664589} and independent Q-learning \cite{7925694} under fixed and sweeping jamming.

\begin{figure}[!t]
\centering
\includegraphics[width=3.5in]{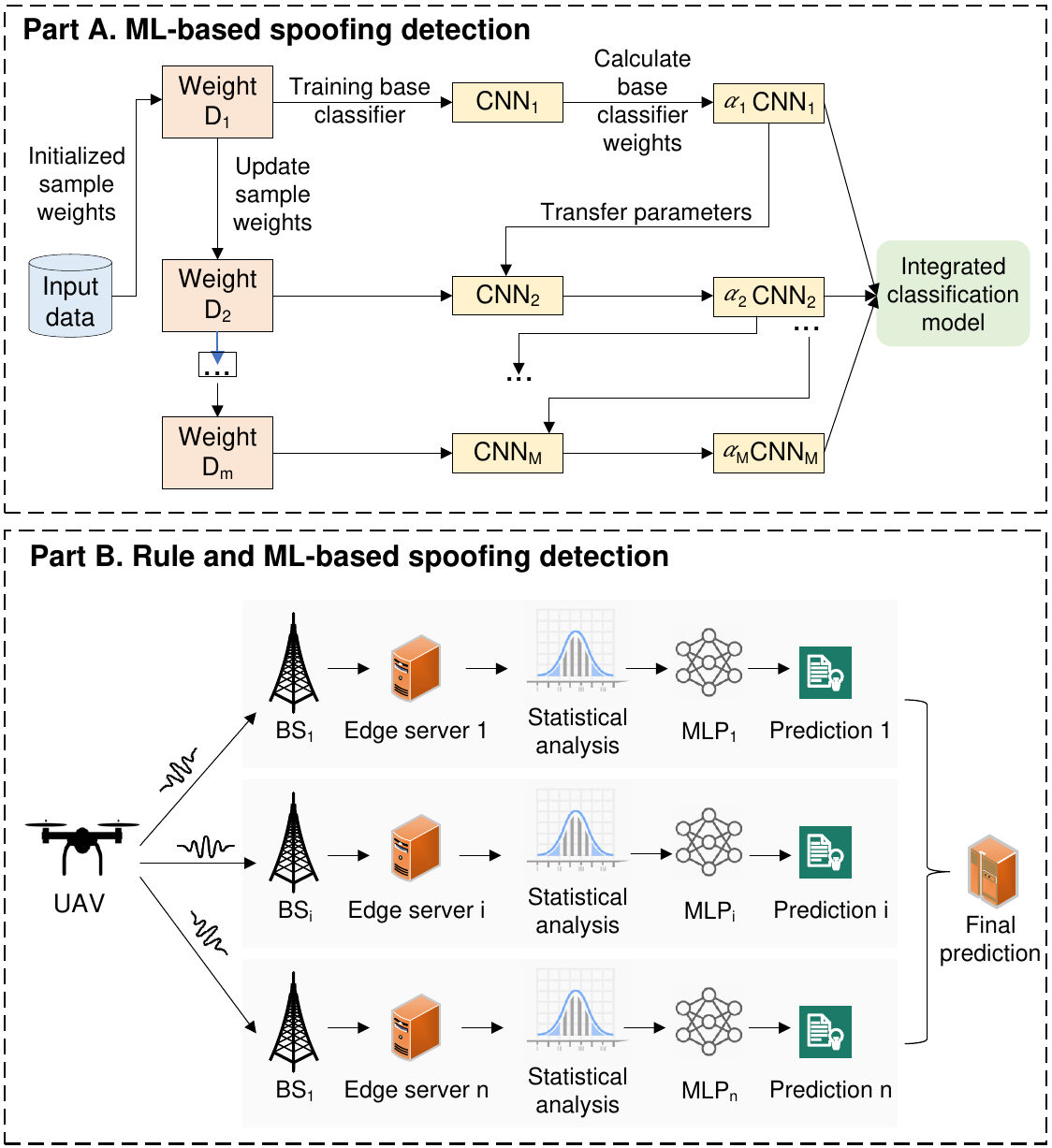}
\caption{The overall framework of ML and rule-based spoofing detection for GPS spoofing detection in the LAENet. Part A depicts an ML-based spoofing detection mechanism in \cite{10685447}, where multiple CNN classifiers are trained with updated sample weights to form an integrated classification model. Each CNN transfers its optimized parameters to subsequent classifiers, enhancing the model's robustness. Part B presents a hybrid rule and ML-based approach in \cite{9845684}, where statistical analysis of path losses between UAVs and multiple base stations (BSs) is performed by edge servers. The analyzed data is processed through MLPs to generate individual predictions, which are aggregated to produce a final spoofing detection decision.}
\label{spoofingfiglabel}
\end{figure}

Note that previous RL-based anti-jamming strategies \cite{9816050,10750022} mainly rely on the Q-learning method, which is suitable for discrete action spaces but may be limited in dealing with high-dimensional continuous spaces \cite{9403369,9606690}. The authors in \cite{10614297} propose a PER-MATD3 algorithm against jamming by integrating spectrum-aware channel selection and prioritized experience replay (PER) into an MADRL framework, as shown in part D of Fig. \ref{Anti-jammingfig}. The proposed spectrum-aware intelligent channel selection uses energy detection-based spectrum sensing, enabling UAVs to identify and avoid jammed channels. The TD3 is specifically designed to handle continuous-valued states and actions, where two critic networks, target policy smoothing, and delayed policy updates are used to further stabilize DRL training. By leveraging PER, the agents can learn from high-error experiences, thereby accelerating adaptation to time-varying CSI, imperfect jamming detection, and co-channel interference. By jointly optimizing CPU frequency, bandwidth allocation, and channel selection to minimize the impact of jamming, PER-MATD3 reduces system cost (a linear combination of latency and energy consumption) by approximately 16.7\%, 9.1\%, and 1.2\% compared to the baselines of Q-learning, MATD3-JSC (without PER), and PER-MATD3 (without channel selection), respectively.

{\bf {Lesson Learned.}} Recent advancements in anti-jamming strategies show that intelligent decision-making for trajectory control, channel selection, and power control is essential for effective jamming mitigation. A key takeaway is the successful integration of MARL to develop dynamic and adaptive anti-jamming solutions \cite{9672766}. By employing intelligent algorithms such as adversarial pre-training and decentralized decision-making, UAV networks can generalize anti-jamming strategies across diverse environments \cite{10233023,10834505}. However, challenges persist in the generalization of these strategies across various jamming types and environmental conditions, as well as balancing the trade-offs between energy consumption, latency, and throughput. Future research could delve into the integration of more adaptive learning frameworks (such as deep learning) into the LAENet for anti-jamming, enabling it to better manage partial or imperfect environmental observations for low-latency, real-time decision-making in multi-UAV systems.

\subsection{Spoofing Defense}

\begin{table*}[]
\centering

\caption{Summary of Spoofing Defense for Communication Availability \\ CIRCLES DESCRIBE THE METHODS; CORRECT MARKERS AND CROSS MARKERS REPRESENT PROS AND CONS RESPECTIVELY.}
\label{tab:my-table}
\centering
\begin{tabular}{|c|c|c|l|}
\hline
\textbf{Techniques} & \textbf{Reference} & \textbf{Algorithm} & \multicolumn{1}{c|}{\textbf{Pros \& Cons}} \\ \hline
\multirow{8}{*}{\begin{tabular}[c]{@{}c@{}}PLA \\ \end{tabular}} 
& \cite{9739860} & RSS & \begin{tabular}[c]{@{}l@{}}\textcolor{blue!20}{\ding{108}} Spatial correlations of RSS distances in PLA against spoofing attacks \\ \textcolor{green}{\ding{51}} Use RSS-based channel characteristics to reduce PLA computational complexity \\ \textcolor{red}{\ding{55}} Assume an ideal transmission scenario without external interference \end{tabular} \\ \cline{2-4} 
& \cite{10851372} & {\begin{tabular}[c]{@{}c@{}}Rayleigh\\ channel\end{tabular}} & \begin{tabular}[c]{@{}l@{}}\textcolor{blue!20}{\ding{108}} Defend against spoofing attacks by considering channel randomness and jamming \\ \textcolor{green}{\ding{51}} Simultaneously address spoofing and jamming attacks via PLA \\ \textcolor{red}{\ding{55}} Assume static UAVs and a known jamming distribution \end{tabular} \\ \cline{2-4} 
& \cite{10834505} & {\begin{tabular}[c]{@{}c@{}}Rayleigh and\\ Rician channels \end{tabular}}  & \begin{tabular}[c]{@{}l@{}}\textcolor{blue!20}{\ding{108}} AD-based PLA for spoofing defense under Rayleigh and Rician channels \\ \textcolor{green}{\ding{51}} Provide a thorough analysis of spoofer identification in urban and suburban environments \\ \textcolor{red}{\ding{55}} Assume perfect CSI in channel modeling \end{tabular} \\ \hline
\multirow{8}{*}{\begin{tabular}[c]{@{}c@{}}GNSS spoofing \\ detection\end{tabular}} 
& \cite{8946587} & {\begin{tabular}[c]{@{}c@{}}Rule-based\\ detection \end{tabular}}  & \begin{tabular}[c]{@{}l@{}}\textcolor{blue!20}{\ding{108}} Combine cooperative localization mechanism with Stackelberg game against spoofing attacks \\ \textcolor{green}{\ding{51}} Spoofing detection is based on neighboring UAV signal sources without predefined thresholds \\ \textcolor{red}{\ding{55}} Extending to larger UAV groups may require complex adjustments \end{tabular} \\ \cline{2-4} 
& \cite{10685447} & {\begin{tabular}[c]{@{}c@{}}ML-based\\ detection \end{tabular}} & \begin{tabular}[c]{@{}l@{}}\textcolor{blue!20}{\ding{108}}Improved AdaBoost-CNN for multi-modal spoofing attack identification \\ \textcolor{green}{\ding{51}} High accuracy in identifying spoofing attacks with limited data samples \\ \textcolor{red}{\ding{55}} Dependence on predefined signal features may lead to model overfitting \end{tabular} \\ \cline{2-4} 
& \cite{9845684} & {\begin{tabular}[c]{@{}c@{}}Rule \& ML-\\ based detection \end{tabular}} & \begin{tabular}[c]{@{}l@{}}\textcolor{blue!20}{\ding{108}} MLP and statistical feature extraction on path-loss data for detecting GPS spoofing \\ \textcolor{green}{\ding{51}}  No additional hardware/energy burden on UAVs \\ \textcolor{green}{\ding{51}}  Robust performance under sparse base station coverage  \\ \textcolor{red}{\ding{55}} Spoofing detection performance degrades in areas with unstable cellular signals \end{tabular} \\ \hline
\end{tabular}

\end{table*}

In the LAENet, the openness of A2G transmission channels and the dynamic nature of low-altitude aircraft networks make them particularly susceptible to identity-based spoofing attacks \cite{8758975}. In such attacks, a malicious entity impersonates a legitimate transmitter using falsified identity information, such as a spoofed media access control address, to gain unauthorized access to the network \cite{9279294}. Once authenticated, the spoofer can disrupt communications among aircraft by launching more severe attacks, such as rogue access point infiltration and denial-of-service attacks, ultimately leading to network congestion and service outages \cite{9672766}. Given the limitations of conventional authentication methods that rely on complex cryptographic protocols \cite{9279294}, PLA offers a promising alternative by leveraging the inherent and unique physical characteristics of wireless transmissions for the LAENet, which is introduced in Section III-B. Overall, this type of PLA can defend against spoofing attacks in the LAENet by exploiting the unique characteristics of the wireless channel (such as RSS, Rayleigh channel, and Rician channel) to identify and separate between legitimate devices and spoofers.

The work in \cite{9739860} proposes a PLA framework to detect spoofing attacks by exploiting spatial correlations of RSS in A2G channels. The key idea is that the RSS from a legitimate transmitter will remain relatively consistent due to its fixed location, while the RSS from a spoofer will vary significantly because of its different position and channel conditions. Thus, the UAV receiver can perform a hypothesis test to authenticate incoming signals. if the RSS distance between the current signal and a previously authenticated signal is below a predefined threshold, the signal is accepted as legitimate. Otherwise, it is flagged as a spoofing attempt. However, the work \cite{9739860} is considered under an ideal transmission scenario, where the propagation environment is perfectly exempted from external interference. To address this limitation, the authors in \cite{10851372} develop a PLA framework that accounts for channel randomness and interference uncertainty. First, they model the G2A link as a Rayleigh fading channel. Then, they introduce jamming signals as external interference. By modeling the jamming power statistically and incorporating it into the analysis of detected power differences, if the difference in power exceeds a predefined threshold, it is identified as a spoofing attempt. Thus, even in real-world scenarios with interference, the proposed framework can better differentiate between natural channel fading and anomalies caused by spoofing attacks.

In addition to using the statistical properties of the Rayleigh channel to design PLA against spoofing in environments with multipath fading (such as urban areas), the channel characteristics in suburban environments should also be considered. To address this, the work \cite{10834505
} proposes a PLA framework to counter spoofing attacks in both urban (Rayleigh channel) and suburban (Rician channel) environments. As mentioned earlier (in Section III-B), a new metric AD is devised to distinguish between legitimate signals and spoofing signals based on differences in channel randomness and geographical factors, such as elevation angles and distances. Adopting the unique fading characteristics of Rayleigh and Rician channels makes it statistically difficult for a spoofer to accurately mimic a legitimate signal. By considering elevation angles and distances in channel modeling, it ensures that a spoofer cannot easily replicate a legitimate signal even if the spoofer knows the legitimate transmitter's location. Simulation results show that the probability of a successful spoofing attack is significantly reduced compared to the baseline \cite{10293903}, where the spoofing miss detection probability drops to 0.014 in suburban environments and 0.371 in dense urban areas.

In the LAENet, in addition to being vulnerable to identity-based spoofing attacks, aircrafts are also susceptible to signal spoofing attacks from the Global Navigation Satellite System (GNSS), particularly GPS spoofing, which poses a significant security threat by generating and transmitting counterfeit satellite signals resulting in severe positioning deviations \cite{10599134}. By interfering with or suppressing legitimate GNSS signals, attackers can manipulate UAV locations in an imperceptible manner to mislead UAVs, causing deviations from intended flight paths, violations of no-fly zone regulations, or increased collision risks \cite{8883128}. Given a critical role of GNSS in UAV operations, effective detection and mitigation strategies for spoofing attacks are essential to ensure flight safety and prevent security breaches in the LAENet. Currently, studies on signal spoofing attack recognition in the LAENet mostly focuses on recognizing GNSS spoofing attack detection, which primarily falls into two categories with respect on rule-based and ML-based methods \cite{10599134,10.1145/3485272}. Rule-based detection methods typically assess the relative distance and positional deviations of UAVs to determine if they are under GNSS spoofing attack. On the other hand, the ML-based methods pay attention to recognize different spoofing types by learning the characteristics of received signals.

Generally, the simplified rule-based methods determine whether a UAV has encountered spoofing attacks based on whether its trajectory follows a predictable path \cite{9079458,9760395}, since a UAV may exhibit deviations from this path due to the false signals imposed by the spoofer. If the measured deviations exceed predefined thresholds, it indicates a potential spoofing attack. However, relying on predefined thresholds for detecting deviations may not dynamically adapt to the spoofing attacks. The study in \cite{8946587} proposes a defense mechanism based on cooperative localization, where each UAV uses the relative distances and real locations of neighboring UAVs to detect spoofing attacks. Specifically, each UAV measures its relative distances based on alternative signal sources of neighboring UAVs and compares these results with its own GPS-derived location. If inconsistencies are detected (e.g., the GPS-derived location does not match the majority of the calculated locations), the UAV identifies itself or a neighboring UAV as being under attack. To optimize defense deployment, an equilibrium of a dynamic Stackelberg game is derived between the drone operator (leader) and the spoofer (follower). %, where the spoofer's goal is to minimize the distance between the UAV's location and the attacker's desired destination, and the operator aims to minimize the distance between the UAV's location and the real destination. 
Simulation results show that the proposed scheme can effectively prevent spoofer's capture, while random/deterministic baselines suffer from attackers capturing one to two UAVs.

Recent ML-based methods for spoofing defense primarily focus on binary classification concerning normal GPS signals and spoofing signals \cite{9844986,9760100}. However, they fail to recognize specific types of spoofing attack necessary for countermeasures in complex environments. Hence, there is an urgent need to recognize diverse GPS spoofing attack patterns for effective countermeasures for the LAENet. The authors in \cite{10685447} propose an improved AdaBoost-CNN algorithm to address the challenge of recognizing diverse GPS spoofing attack patterns for UAVs, as shown in part A of Fig. \ref{spoofingfiglabel}. Three categorized spoofing attack patterns are considered including static and dynamic spoofing based on the UAV's motion state, power-matched and overpowered spoofing based on signal power, and position and time spoofing based on the spoofing targets. The authors select key GPS spoofing signal features such as signal quality monitoring, carrier-to-noise ratio, Doppler shift, and clock error to train the classification model. The improved AdaBoost-CNN algorithm integrates multiple weak CNN classifiers into a strong classification model. Each CNN base classifier uses the updated network parameters from the previous CNN as initial values, enabling iterative refinement of network weights to enhance feature extraction and generalization. With 800 simulated samples, the improved AdaBoost-CNN achieves 100\% accuracy, outperforming original AdaBoost-CNN (94.38\%), CNN (74.38\%), DNN (60.94\%), SVM (40.63\%), and KNN (53.13\%).

\begin{table*}[]
\centering

\caption{Summary of Anomaly Detection for Communication Integrity \\ CIRCLES DESCRIBE THE METHODS; CORRECT MARKERS AND CROSS MARKERS REPRESENT PROS AND CONS RESPECTIVELY.}
\label{tab:my-table}
\centering
\begin{tabular}{|c|c|c|l|}
\hline
\textbf{Anomaly type} & \textbf{Reference} & \textbf{Algorithm} & \multicolumn{1}{c|}{\textbf{Pros \& Cons}} \\ \hline
\multirow{8}{*}{\begin{tabular}[c]{@{}c@{}}Jamming \\ \end{tabular}} 
& \cite{9741304} & HDBN & \begin{tabular}[c]{@{}l@{}}\textcolor{blue!20}{\ding{108}} SA module based on HDBN for detecting jamming anomalies \\ \textcolor{green}{\ding{51}} UAccurately characterize and detect jamming anomalies via KLD/DB metrics \\ \textcolor{red}{\ding{55}} Unstable initialization in unsupervised learning affects the performance of the HDBN \end{tabular} \\ \cline{2-4} 
& \cite{9858012} & {\begin{tabular}[c]{@{}c@{}}GDBN\end{tabular}} & \begin{tabular}[c]{@{}l@{}}\textcolor{blue!20}{\ding{108}} GDBN to model the radio environment and detect and classify jamming anomalies \\ \textcolor{green}{\ding{51}} Unsupervised learning eliminates dependency on labeled data in classification of anomalies \\ \textcolor{red}{\ding{55}} Computational complexity increases with the number of jamming categories \end{tabular} \\ \cline{2-4} 
& \cite{9829873} & {\begin{tabular}[c]{@{}c@{}}Active-GDBN \end{tabular}}  & \begin{tabular}[c]{@{}l@{}}\textcolor{blue!20}{\ding{108}} Active-GDBN used to model UAV-jammer interactions for anomaly detection \\ \textcolor{green}{\ding{51}}Actively incorporate UAV’s actions for faster adaptation and jamming detection \\ \textcolor{red}{\ding{55}} M-MJPF requires significant computational resources \end{tabular} \\ \cline{2-4} 
& \cite{9634169} & {\begin{tabular}[c]{@{}c@{}}Blind channel\\estimation \\ \& ACS\end{tabular}}  & \begin{tabular}[c]{@{}l@{}}\textcolor{blue!20}{\ding{108}} Blind channel estimation based on ACS properties to detect jammer signals \\ \textcolor{green}{\ding{51}}Does not rely on prior knowledge of the jammer's behavior \\ \textcolor{red}{\ding{55}} Assumes a specific structure of the multicarrier modulation format \end{tabular} \\ \hline
\multirow{1}{*}{\begin{tabular}[c]{@{}c@{}} Abnormal  \\ Power\end{tabular}} 
& \cite{8688501} & {\begin{tabular}[c]{@{}c@{}}Spectrum \\ surveillance \end{tabular}}  & \begin{tabular}[c]{@{}l@{}}\textcolor{blue!20}{\ding{108}} Local and cooperative detection of abnormal power emission \\ \textcolor{green}{\ding{51}} Handle both aggressive and passive power misuse \\ \textcolor{green}{\ding{51}}  Cloud-based framework enables real-time closed-loop surveillance \\ \textcolor{red}{\ding{55}} Computational complexity increases with the number of SN \end{tabular} \\ \hline
\multirow{1}{*}{\begin{tabular}[c]{@{}c@{}} Eavesdropping \end{tabular}} 
& \cite{8854240} & {\begin{tabular}[c]{@{}c@{}}SVM \&  \\ K-means \end{tabular}}  & \begin{tabular}[c]{@{}l@{}}\textcolor{blue!20}{\ding{108}} One-class SVM and K-means clustering for detecting eavesdropping anomalies \\ \textcolor{green}{\ding{51}} One-class SVM and K-means are stable under varying eavesdropper power \\ \textcolor{red}{\ding{55}} Detection performance mainly depends on the quality and quantity of the ATD \end{tabular} \\ \hline
\end{tabular}

\end{table*}

Furthermore, integrating rule-based approaches with machine learning-based methods provides an effective and robust defense against spoofing attacks. The work in \cite{9845684} leverages statistical features of path losses between UAVs and terrestrial base stations to detect a UAV's trajectory deviation due to GPS spoofing, as shown in part B of Fig. \ref{spoofingfiglabel}. The spoofing detection is formulated as a nonlinear optimization problem that aims to minimize hypothesis test errors by adjusting thresholds, statistical feature weights, and the number of base stations. To further accurately analyze path loss's statistical features for final decisions on predicting GPS spoofing probabilities, multilayer perceptron (MLP) neural networks are deployed on edge cloud servers, where individual MLP models at each BS are used to analyze statistical features of path losses. Simulation results show that the proposed method achieves 97\% accuracy with two base stations and 83\% accuracy with a single base station, outperforming baseline approaches such as adaptive trustable residence area (ATRA), which necessitates three base stations for triangulation \cite{9348030}.

{\bf Lesson Learned}. For identity spoofing in the LAENet, leveraging signal features such as received signal strength and channel randomness in PLA design is an effective approach \cite{9739860,10851372,10834505}. On the other hand, employing rule-based or ML-based techniques can detect and mitigate GNSS signal spoofing \cite{8946587,10685447,9845684}. While ML-based methods show promising performance, they are limited by factors such as computational complexity and dependency on large datasets. Rule-based methods are simpler but may struggle in dynamic or uncertain environments. Future research could explore the application of RL to develop adaptive and robust spoofing defense mechanisms in the LAENet, which has not yet been extensively studied. Different from the abovementioned approaches, RL dynamically learns from interactions with the environment, and its sequential decision-making ability enables UAVs and ground stations to optimize spoofing defense strategies based on continuous feedback \cite{9403369}, make it a promising direction for enhancing spoofing defense in the LAENet

\section{Communication Integrity for LAENet}

\subsection{Anomaly Detection}

Due to the open nature of wireless channels and the dominant LoS links in the LAENet, communication becomes particularly vulnerable to a diverse range of anomalous behaviors such as abnormal jamming, abnormal transmission power, and covert eavesdropping \cite{8883128,8883127}. Specifically, malicious jammers sense spectrum activity and dynamically adapt their interference patterns to mislead the UAV into taking suboptimal or harmful actions \cite{8999433,9200570}. In parallel, abnormal power emissions, either due to device faults, selfish behavior, or malicious intent, can violate spectrum policies, introduce harmful interference, and disrupt cooperative spectrum sharing \cite{8113526}. Additionally, the pervasive risk of eavesdropping is that adversaries exploit the UAV’s uplink or downlink transmissions to intercept sensitive data \cite{10325641,10114676}. Thus, it is essential to detect and mitigate these abnormal activities in the LAENet. Different from previously reviewed approaches such as anti-eavesdropping (Section III-A) and anti-jamming (Section IV-A), anomaly detection is a method used to identify and mitigate unexpected deviations from or irregularities in normal operational patterns by monitoring communication channels in the LAENet \cite{10.1145/1541880.1541882,10623395}.

Jamming anomalies generally aim to disrupt the normal operation of UAV communication links, such as by injecting disruptive signals to interfere with the legitimate communication process. The study in \cite{9741304} proposes a novel Self-Awareness (SA) module to leverage the radio to detect abnormal behaviors caused by jamming attacks for Cognitive UAV communications. The SA module unsupervisedly learns a generative model using a Hierarchical Dynamic Bayesian Network (HDBN) \cite{balaji2011bayesian} to represent the joint distribution of random variables characterizing the radio environment at different levels of abstraction and across time, where the Modified Bayesian Filtering \cite{8455592} is used to integrate multilevel abnormality measurements for online predictions of radio environmental states at different levels. Since jamming can disrupt and shift the distributions of the radio environment, the abnormalities can be detected by calculating the Kullback-Leibler Divergence (KLD) and Dhattacharyya distance (DB) \cite{pardo2018statistical}  between predictive messages and diagnostic messages. The predictive messages are generated by the HDBN to capture the expected patterns of normal signals, and diagnostic messages reflect the actual state of the signal. The jammer's impact is characterized by calculating generalized errors based on shifts in amplitude, phase, and frequency of signals, allowing the radio to predict future activities of the jammer. The SA module achieves a near 100\% abnormality detection accuracy, approximately 12\% higher than the traditional energy detector-based scheme.

\begin{figure*}[!t]
\centering
\includegraphics[width=7.1in]{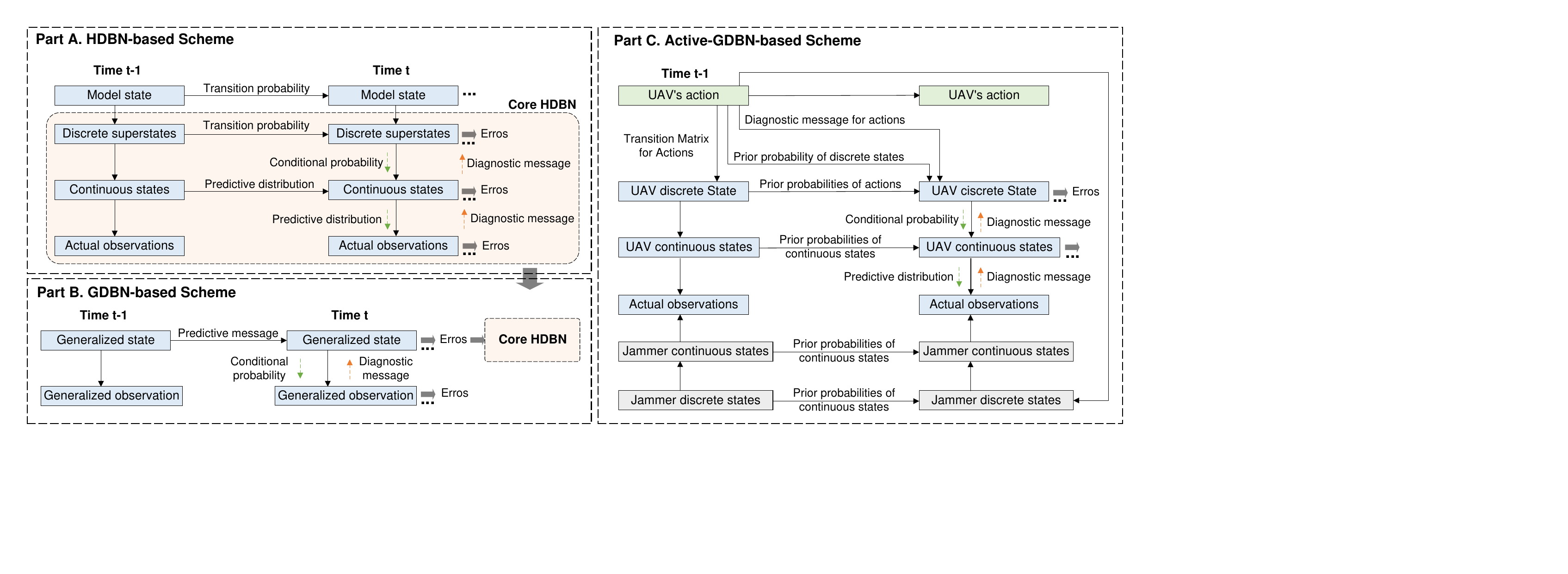}
\caption{The overall architecture illustrates jamming anomaly detection to enhance communication integrity in the LAEnet. Part A presents an HDBN-based scheme focusing on hierarchical dynamic models to predict and detect abnormal signals caused by jammers. It details the transition probabilities between model states and the prediction of continuous states based on discrete superstates. Part B introduces a GDBN-based scheme, extending the HDBN approach by incorporating generalized states and observations, allowing for more nuanced predictions and error calculations. Part C depicts an Active-GDBN-based scheme, integrating UAV actions into the model to actively infer and adapt to the environment, thereby optimizing resource allocation and anti-jamming measures.}
\label{Anti-jammingfig}
\end{figure*} 

Different from the previous work \cite{9741304}, which introduced the SA module using HDBN for anomaly detection, the authors in \cite{9858012} propose a Generalized Dynamic Bayesian Network (GDBN)-based framework to enhance the SA module by further classifying the detected anomalies caused by multiple jammers. A generalized state-space model \cite{balaji2011bayesian} is used to represent the evolving radio environment as a GDBN model learned in an unsupervised manner. Different from the KLD/DB metric in [1], Kullback-Leibler divergence and Bhattacharyya distance are used as abnormality measurements between predicted and observed signals to detect jamming. Once an abnormality indicative of jamming is detected, the UAV extracts the interfering signal and compares it with pre-learned GDBN models (each representing a different jamming modulation scheme). By evaluating which pre-learned model best explains the extracted jamming signal, the UAV can not only detect the presence of a jammer but also classify its modulation type. Simulation results show that the GDBN-based method achieves an overall classification accuracy of 98\% at SNR = 10 dB, outperforming LSTM (88\%), CNN (67\%), and SAE (90\%).

Based on the study \cite{9858012}, the authors in \cite{9829873} propose an Active-GDBN to model the dynamic interaction between the UAV and jammer for anomaly detection. Similar to \cite{9858012}, the generalized state-space model \cite{balaji2011bayesian} is used to capture the features and dynamic evolution of UAV signals to represent the radio environment. Differently from passive detection and classification of jamming signals in \cite{9858012}, the Active-GDBN achieves active anomaly detection by incorporating the UAV’s actions into the inference process. Specifically, the UAV employs a Modified Markov Jump Particle Filter (M-MJPF) \cite{9322583} to link the UAV’s actions to environmental states and observations. Meanwhile, the UAV dynamically adjusts physical resource block selections to evade jamming by encoding jammer behavior and updating beliefs. The Active-GDBN achieves about 25\% to 37.5\% faster convergence on anomaly detection probability than the Q-learning-based baseline under various jamming types.

Different from previous works \cite{9741304,9858012,9829873} that detect jamming anomalies based on the statistical distribution divergence of the signal, study \cite{9634169} focuses on detecting anomalies by exploiting the time delays, shifts, and modulation of the signal characteristics. Firstly, achieving blind channel estimation involves constructing cyclic correlation matrices to identify distinct Doppler shifts and time delays associated with transmissions by exploiting the inherent almost-cyclostationary (ACS) properties of UAV and jammer signals (e.g., periodic statistics from OFDM modulation). Then, this blind estimation process is combined with a widely linear minimum mean square error (WL-MMSE) filter to provide an initial estimate of the symbol vector by leveraging the non-circular statistics of the received signal, where the initial estimate includes contributions from both the UAV and the jammer. Finally, a post-sorting algorithm (PSA) is employed to iteratively decode and cancel the jammer’s contribution by ranking and removing symbols with the highest signal-to-disturbance-plus-noise ratio (SDNR). Simulation results demonstrate that the proposed scheme can effectively detect and separate the jamming signals from UAV signals without requiring prior knowledge of the jammer’s characteristics, even when the jammer’s power is twice as strong as the UAV’s power.

\begin{table*}[]
\centering

\caption{Summary of Injection Defense for Communication Integrity \\ CIRCLES DESCRIBE THE METHODS; CORRECT MARKERS AND CROSS MARKERS REPRESENT PROS AND CONS RESPECTIVELY.}
\label{tab:my-table}
\centering
\begin{tabular}{|c|c|c|l|}
\hline
\textbf{Injection type} & \textbf{Reference} & \textbf{Algorithm} & \multicolumn{1}{c|}{\textbf{Pros \& Cons}} \\ \hline
\multirow{8}{*}{\begin{tabular}[c]{@{}c@{}}Jamming \\ signal \\ \end{tabular}} 
& \cite{ 9741304} & HDBN & \begin{tabular}[c]{@{}l@{}}\textcolor{blue!20}{\ding{108}} HDBN-based jamming signal extraction and suppression \\ \textcolor{green}{\ding{51}} Autonomous learning from raw I/Q data enables adaptability to dynamic jamming patterns \\ \textcolor{red}{\ding{55}} Assume the jammer's output power remains constant during attacks \end{tabular} \\ \cline{2-4} 
& \cite{9634169} & {\begin{tabular}[c]{@{}c@{}}SIC\end{tabular}} & \begin{tabular}[c]{@{}l@{}}\textcolor{blue!20}{\ding{108}} SIC with blind channel estimation for detecting and eliminating jamming signals \\ \textcolor{green}{\ding{51}} Eliminate jamming signals regardless of the mobility patterns of jammers \\ \textcolor{red}{\ding{55}} Rely on sufficient cyclostationary features in the received signal \end{tabular} \\ \cline{2-4} 
& \cite{10530539} & {\begin{tabular}[c]{@{}c@{}} DBF \end{tabular}}  & \begin{tabular}[c]{@{}l@{}}\textcolor{blue!20}{\ding{108}} DBF algorithm for nullifying jamming signals \\ \textcolor{green}{\ding{51}} Effective suppression of jamming signals while maintaining carrier phase integrity \\ \textcolor{red}{\ding{55}} May be limited to specific GNSS frequency bands \end{tabular} \\ \hline
\multirow{4}{*}{\begin{tabular}[c]{@{}c@{}} Spoofing \\ signal \end{tabular}} 
& \cite{10.1145/3395351.3401703,sathaye2022semperfi} & {\begin{tabular}[c]{@{}c@{}}API \& LSR\end{tabular}} & \begin{tabular}[c]{@{}l@{}}\textcolor{blue!20}{\ding{108}} SIC combined with API and LSR to recover legitimate signals from spoofing attacks \\ \textcolor{green}{\ding{51}} SemperFi with a single antenna does not require additional hardware \\ \textcolor{red}{\ding{55}} Limited to attackers with a power advantage lower than 15 dB \end{tabular} \\ \cline{2-4} 
& \cite{8045998} & {\begin{tabular}[c]{@{}c@{}}Subspace \\ projection  \end{tabular}}  & \begin{tabular}[c]{@{}l@{}}\textcolor{blue!20}{\ding{108}} Subspace projection for nullifying spoofing signals \\ \textcolor{green}{\ding{51}} Low parameter dependency, requiring only code delays and carrier frequencies \\  \textcolor{red}{\ding{55}} Suppression performance declines if spoofing and legitimate signals have similar code delays \end{tabular} \\ \hline
\end{tabular}

\end{table*}

In addition to jamming anomalies, which cause interference and security threats in the LAENet, abnormal power emissions in UAV communication networks also represent a critical type of anomaly, potentially leading to severe disruption of communication quality and violation of spectrum policies. The work in \cite{8688501} proposes a cloud-based surveillance framework to address the detection of abnormal power emissions, where the cloud server assigns spectrum resources to the UAVs and shares UAVs’ spectrum usage information with the surveillance center. The surveillance center assigns the detection task to $K$ surveillance nodes (SNs) for local detection of abnormal power emission, where the detection rule is based on the Lagrange multiplier method and the generalized likelihood ratio test. After local decisions, $K$ SNs report results to the surveillance center, where cooperative detection of abnormal power emission is performed using the decision rule that declares an abnormal event when at least $L$ out of $K$ nodes detect an abnormality, where the optimal global threshold of $L$ is determined by solving the constraints on the global false-alarm probabilities. Simulation results show that the global detection probability exceeds 90\% when transmit power deviation exceeds 0.02W (allocated power is 0.01W).

Besides the threats of jamming and abnormal power emission, another critical anomaly that requires detection is eavesdropping in the LAENet, where malicious devices covertly intercept sensitive information during UAV-to-ground and UAV-to-UAV transmissions \cite{9713997,10114676}. Note that most previous works on anti-eavesdropping focused on measuring secure performance through secrecy rate and/or secrecy outage probability (such as \cite{10759093,10233023}) rather than emphasizing the detection of eavesdropping attacks. The work in \cite{8854240} explores anomaly detection for eavesdropping attacks in UAV-aided wireless systems using unsupervised learning. Two datasets are prepared: artificial training data (ATD), simulated without eavesdropping based on CSI (all labeled normal), and a practical dataset extracted from received signal features (mean and variance of amplitude). Two types of unsupervised learning methods are designed for anomaly detection. One-class SVM maps data to a high-dimensional space, defining a normal region where outliers are detected. K-means clustering classifies test data into two clusters, labeling the one nearest to the ATD center as normal. %The work \cite{8854240} investigates eavesdropping attack anomaly detection in UAV-aided wireless systems using unsupervised learning. Two datasets are prepared for unsupervised learning in eavesdropping. The artificial training data (ATD) is created by simulating the signal transmission process without eavesdropping based on CSI, where each ATD point is labeled as normal. The practical structured dataset is generated by extracting features from the received signals, including the mean and variance of the received signal's amplitude. One-class SVM and K-means clustering are proposed for anomaly detection. One-class SVM is used to identify abnormalities by mapping the input space to a higher-dimensional feature space and creating a region that contains most of the ATD points. Test data points are labeled as outliers if they fall outside this region. K-means clustering divides the test data points into two clusters labeled by the ATD, where the cluster closest to the ATD center is labeled as normal.

{\bf Lesson Learned} For jamming anomalies, the statistical distribution divergence detection and signal structural feature-based detection, such as HDBN, GDBN, and ACS, are used to model the dynamic environment and detect deviations from learned normal patterns. For abnormal transmission power detection, a cloud-based surveillance framework supports a statistical distribution detection approach to monitor and identify power emission outliers. Leveraging its high computing power, the cloud enables cooperative analysis through multi-source data aggregation, dynamically optimizes detection thresholds using global information, and maintains a feedback loop for adaptive anomaly detection. For eavesdropping detection, unsupervised learning techniques, including One-Class SVM and K-means clustering, achieve the identification of anomalies in received signals. These approaches effectively achieve anomaly detection and demonstrate excellent performance. However, challenges remain, including the reliance on high-quality training data and the complexity of maintaining real-time adaptability in dynamic spectrum environments. Currently, Generative AI such as GANs and generative diffusion models presents a promising research direction for anomaly detection, as demonstrated in the use of generalized models in HDBN and the artificial data generation for training ML and clustering models in \cite{xie2025multi,wang2024generative}. Generative AI could further enrich training datasets and provide a high-level generative model to enhance anomaly detection in the dynamic and uncertain LAENet.

\subsection{Injection Defense}

\begin{figure*}[!t]
\centering
\includegraphics[width=7.1in]{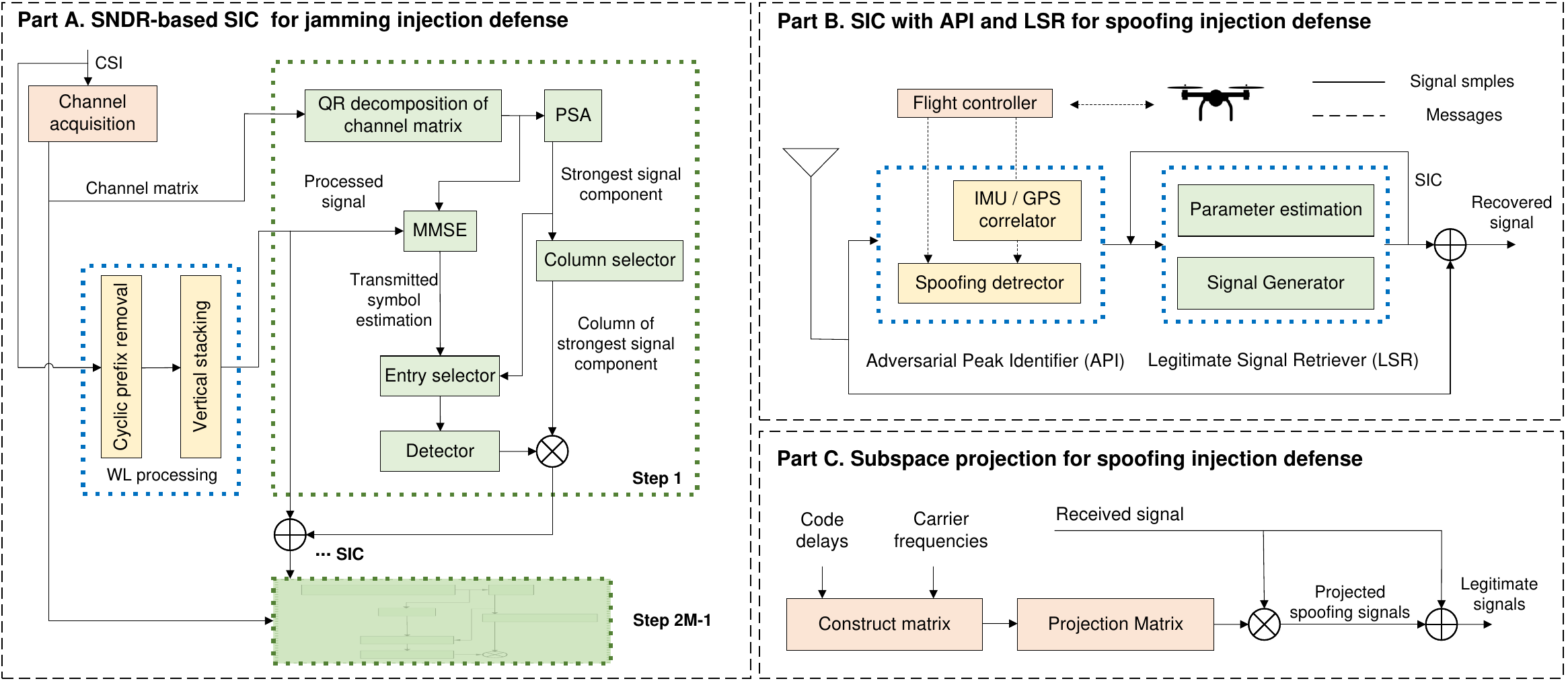}
\caption{The overall architecture of injection defense mechanisms for UAVs in smart city applications. Part A presents the SIC architecture that processes channel state information to defend against jamming injection attacks \cite{9634169}. Part B shows an SIC architecture integrated with API and LSR modules, which subtracts injection signals from the received signal to recover normal signals \cite{10.1145/3395351.3401703,sathaye2022semperfi}. Part C depicts a subspace projection-based architecture for spoofing injection defense, where the received signal is projected onto the orthogonal null space of the spoofing signals to eliminate them \cite{8045998}.}
\label{injectiondefensefig}
\end{figure*}

The low-altitude economy is highly dependent on open communication and network architecture with dense communication links, which brings injection attacks as a significant threat to UAV communication integrity \cite{8883128,9900257}. These attacks involve the deliberate injection of malicious signals, such as jamming and spoofing signals, to disrupt or manipulate legitimate communications \cite{10368002,10750142}. Jamming signal injection can make legitimate signals unrecognizable by emitting high-power electromagnetic interference to degrade signal reception \cite{9741304}. Additionally, spoofing signal injection can transmit high-power signals to overshadow legitimate GNSS signals. Therefore, eliminating injection signals or separating them from legitimate signals is crucial for ensuring communication integrity in the LAENet.

The UAV's communication can be severely disrupted by jammers that exploit LoS propagation to inject jamming signals into the transmission channel, which may effectively mask legitimate signals and render them unrecognizable \cite{9634169}. The work in \cite{9741304} proposes an HDBN-based injection defense scheme to extract and remove the jammer’s signal. This work first utilizes the HDBN to detect abnormal behaviors caused by jamming attacks, as mentioned earlier in Section V-A. Once the jammer's presence is confirmed, its signal characteristics are analyzed across multiple levels of continuous in-phase (I) and quadrature (Q) components and observation-level state vectors \cite{950789}. The extracted jammer signal is then separated from the received observation using frequency-domain subtraction \cite{5967984}, component-wise I/Q processing, and adaptive filtering \cite{950789}. The corrected signal is subsequently demodulated and decoded using techniques and error correction coding to restore the original signal. To maintain resilience against evolving jamming tactics, the system continuously updates the HDBN model to improve suppression commands. Simulation results show that the mean square error (MSE) of suppression commands decreases as the JSR increases, meaning that jammers attacking with higher power can be better estimated than jammers attacking with lower power.

Different from the work in \cite{9741304}, which separates the jamming signal by analyzing its I/Q characteristics, the study in \cite{9634169} proposes a Serial Interference Cancellation (SIC) scheme based on SDNR to eliminate injected anomalous signals in UAV communications, as shown in part A of Fig. \ref{injectiondefensefig}. First, blind channel estimation and a WL-MMSE filter are used to identify UAV and jammer signals (as detailed in Section V-A). Then, the PSA ranks detected symbols based on SDNR, where the jamming signals rank higher in SDNR due to their higher emitted power. The SIC \cite{1207260,298053} is subsequently designed for progressively eliminating jamming signals. Specifically, the high-rank jamming symbol is decoded, reconstructed using estimated channel parameters, and subtracted from the received signal. The process continues iteratively to eliminate previously detected jamming signals until all UAV symbols are successfully recovered, with the receiver dynamically updating channel estimation to adapt to jammer mobility and environmental changes. Simulation results demonstrate that the UAV signal can be recovered with low bit error rates ($ <\mathop {10}\nolimits^{ - 4} $) even when the power of the jammer is double that of the UAV.

Jamming attacks not only affect U2G and UAV-to-UAV communications but also cause RF interference, leading to UAVs failing to track GNSS signals in the LAENet. In light of this, the work in \cite{10530539} proposes a self-calibrating digital beamforming (DBF) algorithm to effectively nullify jamming signals while preserving high-precision carrier phase measurements. It calibrates the antenna array’s steering vectors and RF channel characteristics. Once calibration is complete, the system performs jamming detection and direction estimation by analyzing interference patterns across the antenna array. Then, the minimum power distortionless response (MPDR) optimization rule is used to calculate optimal beamforming weights, which aim to create nulls in the beam pattern corresponding to the directions of jamming signals, effectively suppressing them. The calculated beamforming weights are applied to the received signals to produce the beamformer output, which effectively suppresses jamming signals while preserving the carrier phase integrity of the desired signals. The proposed scheme achieves up to 80 dB Jammer-to-Signal Ratio (JSR) suppression, significantly outperforming the conventional Power Inversion (PI) scheme.

In addition to jamming signals, spoofing attacks can easily transmit fake signals to manipulate GNSS signals due to their open signal structure and weak signal strength \cite{7445815}. One type of method is based on signal encryption or data encryption to prevent malicious spoofers from injecting illegitimate signals \cite{6494400,8490218,wesson2012practical}. However, they may not be suitable for resource-constrained UAVs in the LAENet. Therefore, defending against spoofing signal injection based on different signal characteristics is a promising solution. The authors in \cite{10.1145/3395351.3401703,sathaye2022semperfi} propose an anti-spoofing system, called SemperFi, to autonomously recover legitimate signals during active spoofing for UAVs. The system employs two core modules: the Adversarial Peak Identifier (API) and the Legitimate Signal Retriever (LSR), as shown in part B of Fig. \ref{injectiondefensefig}. The API detects spoofed signals by correlating inertial measurement unit (IMU) data with calculated position-velocity-time (PVT) solutions \cite{10.1145/2973750.2973753}. The LSR module replicates the spoofing signal once it is identified. Then, similar to the study in \cite{9634169}, the SIC is applied to subtract the replica from the composite received signal that contains legitimate and spoofing signals. SemperFi enters an iterative refinement process if spoofing signals persist after initial cancellation, where replication, subtraction, and reassessment are performed until the spoofing detector no longer triggers an alarm, indicating sufficient attenuation or elimination of spoofing. %Experimental evaluations on real-world datasets show that SemperFi achieves 100-meter positioning accuracy during spoofing attacks and withstands adversarial signals with a 15 dB power advantage.

Besides recovering legitimate signals by subtracting spoofing signals from the received signal \cite{9634169,10.1145/3395351.3401703,sathaye2022semperfi}, projecting the signal is also a viable injection defense strategy. In the study \cite{8045998}, the GNSS receiver's spoofing mitigation algorithm employs a subspace projection-based interference cancellation method to effectively eliminate spoofing signals, as shown in part C of Fig. \ref{injectiondefensefig}. Specifically, the receiver on UAVs acquires and tracks incoming signals, identifying potential spoofing signals and reconstructing them based on their power levels, pseudo-random noise (PRN) code delays, and carrier frequencies. Then, the receiver uses these reconstructed spoofing signals to construct a spoofing subspace, which represents all possible linear combinations of spoofing signal characteristics. To effectively remove spoofing signals from the received signal, the receiver performs orthogonal projection to obtain a cleaned signal by mapping the received signal onto a complementary null space that is mathematically orthogonal to the spoofing subspace. Simulation results show that shorter projection lengths suppress spoofing signals more effectively than longer projections, achieving a 20 dB gain in Signal-to-Interference Ratio (SIR).

{\bf Lesson Learned} The above-mentioned studies have demonstrated the effectiveness for mitigating injection signals, such as jamming and spoofing attacks, thereby enhancing UAV communication reliability and security. These advancements leverage techniques that not only detect malicious signal interference but also enable autonomous recovery. One key advantage is that non-cooperative detection techniques, such as blind estimation \cite{9634169} and self-awareness models \cite{9741304}, allow for efficient attack identification without requiring prior knowledge of the attacker's signal characteristics to adapt to dynamic and adversarial environments. However, several challenges remain in that beamforming-based or spatial filtering techniques rely on multi-antenna configurations \cite{10530539,9634169}, limiting their applicability in cost-sensitive or small UAV systems. Future work should explore lightweight and energy-efficient implementations of injection defense to support stable UAV signal integrity protection. Additionally, more intelligent injection defense strategies combining optimization methods, RL, and ML could enhance resilience against more sophisticated adversaries.

\section{Future Research Directions }
 
\subsection{Energy-efficient Physical Layer Security}

Future work can focus on exploring more unique physical characteristics of wireless communication ,such as exploiting channel characteristics and implementing simple coding schemes, to develop secure and low-energy protocols. Meanwhile, drones in the LAENet need to develop adaptive power control strategies that dynamically adjust transmission power based on channel conditions and security requirements to minimize unnecessary energy consumption \cite{10833728}. Moreover, dynamic trajectory optimization is equally important for energy efficiency \cite{10589561}. Future research can explore enabling UAVs to learn attack patterns in real time, share secure trajectory models across swarms, and dynamically adjust flight paths based on real-time security and power consumption feedback.

\subsection{Multi-drone Collaboration for Secure Communication}

Future research on secure physical layer communication in the LAENet should move beyond existing dual-UAV collaboration models and explore distributed multi-UAV coordination (or UAV swarms) for enhanced resilience against jamming, spoofing, and unauthorized access \cite{10516683}. For example, UAV swarms can collaboratively emit interference signals to obscure unauthorized receivers, thereby enhancing the confidentiality of communications \cite{10.1145/3703625}. Additionally, the integration of adaptive trust-based mutual authentication protocols among UAVs is essential \cite{9946859}. Multiple UAVs with mutually verified identities can enable dynamic and secure spectrum-sharing mechanisms to optimize resource utilization in the LAENet.

\subsection{AI-driven Security Defense Strategy}
Existing AI-based security strategies mainly focus on training AI models to identify anomalous signals while having some limitations. The resource-constrained drones are unable to train high-quality AI models, making the integration of edge computing a promising approach for model training \cite{10833728}. Note that AI models may be difficult to generalize in recognizing various anomalous signals because they are pre-trained on previously collected datasets of fixed size. Future work can explore leveraging GAN or diffusion models to generate datasets based on real-time captured anomalous signals \cite{10454003}. Furthermore, emerging generative AI technologies, such as the diffusion model for secure network topology generation in low-altitude domains \cite{wang2024empowering,wang2024generative}, AI agents for human-aerial vehicle secure interaction \cite{10648594}, and mixture of experts for robust wireless communications \cite{zhang2024optimizing,zhao2025generative}, can be explored to achieve a more autonomous and intelligent LAENet.

\subsection{Space-Air-Ground Integrated Security Architecture} 
Future research can explore establishing a multi-domain physical layer security framework for LAENet to connect space, air, and ground layers, providing seamless communication coverage and cost-effective network access \cite{10572013,10612249}. A potential key research direction is the development of a coordinated multi-tier security mechanism, where satellites, UAVs, and terrestrial base stations collaboratively enhance physical layer security through dynamic resource allocation and interference management based on real-time CSI and environmental conditions, such as UAV mobility, channel fading, and spectrum constraints.

\subsection{6G-Enabled Secure UAV Communication}

The advent of 6G networks presents new opportunities for LAENet. Terahertz (THz) communication can offer ultra-high-speed data transmission capabilities for LAENet \cite{10819441}. Future research can explore the integration of THz with advanced beamforming techniques to focus signals on legitimate users, thereby enhancing security and reducing the risk of interception.  Furthermore, Reconfigurable Intelligent Surfaces (RIS) play a crucial role in strengthening physical layer security by intelligently controlling wireless signal propagation \cite{10299716,9768113}. Future work can investigate RIS-based secure beamforming strategies to mitigate adversary interception, and leverage optimization techniques and DRL to adaptively adjust beamforming against eavesdropping or jamming attacks.

%DRL can be employed to dynamically optimize both beamforming and UAV trajectory in response to real-time security threats.

\section{Conclusion}

This paper has presented a comprehensive survey on secure physical layer communications in the LAENet, emphasizing the importance of safeguarding confidentiality, availability, and integrity in communications. It introduced the concept and architecture of the LAENet and outlined the associated security issues in physical layer communication. Then, the survey provided in-depth reviews of countermeasures for anti-eavesdropping strategies, authentication schemes, anti-jamming strategies, spoofing defenses, anomaly detection, and injection defense. Finally, the paper proposed a set of forward-looking future research directions. These discussions highlighted the critical role of secure physical layer communication in supporting the development of the LAENet and offered valuable insights for ongoing advancements in this emerging domain.

\bibliographystyle{IEEEtran}
\bibliography{ref}

\begin{thebibliography}{100}
\providecommand{\url}[1]{#1}
\csname url@samestyle\endcsname
\providecommand{\newblock}{\relax}
\providecommand{\bibinfo}[2]{#2}
\providecommand{\BIBentrySTDinterwordspacing}{\spaceskip=0pt\relax}
\providecommand{\BIBentryALTinterwordstretchfactor}{4}
\providecommand{\BIBentryALTinterwordspacing}{\spaceskip=\fontdimen2\font plus
\BIBentryALTinterwordstretchfactor\fontdimen3\font minus \fontdimen4\font\relax}
\providecommand{\BIBforeignlanguage}[2]{{%
\expandafter\ifx\csname l@#1\endcsname\relax
\typeout{** WARNING: IEEEtran.bst: No hyphenation pattern has been}%
\typeout{** loaded for the language `#1'. Using the pattern for}%
\typeout{** the default language instead.}%
\else
\language=\csname l@#1\endcsname
\fi
#2}}
\providecommand{\BIBdecl}{\relax}
\BIBdecl

\bibitem{10759668}
Z.~Li, Z.~Gao, K.~Wang, Y.~Mei, C.~Zhu, L.~Chen, X.~Wu, and D.~Niyato, ``Unauthorized uav countermeasure for low-altitude economy: Joint communications and jamming based on mimo cellular systems,'' \emph{IEEE Internet of Things Journal}, vol.~12, no.~6, pp. 6659--6672, 2025.

\bibitem{zhao2025generative}
C.~Zhao, J.~Wang, R.~Zhang, D.~Niyato, G.~Sun, H.~Du, D.~I. Kim, and A.~Jamalipour, ``Generative ai-enabled wireless communications for robust low-altitude economy networking,'' \emph{arXiv preprint arXiv:2502.18118}, 2025.

\bibitem{9765519}
H.~A.~H. Alobaidy, R.~Nordin, M.~J. Singh, N.~F. Abdullah, A.~Haniz, K.~Ishizu, T.~Matsumura, F.~Kojima, and N.~Ramli, ``Low-altitude-platform-based airborne iot network (lap-ain) for water quality monitoring in harsh tropical environment,'' \emph{IEEE Internet of Things Journal}, vol.~9, no.~20, pp. 20\,034--20\,054, 2022.

\bibitem{china2023central2}
\BIBentryALTinterwordspacing
China holds central economic work conference to plan for 2024. Accessed: Dec. 12, 2023. [Online]. Available: \url{https://english.www.gov.cn/news/202312/12/content_WS657860aec6d0868f4e8e21c2.html}
\BIBentrySTDinterwordspacing

\bibitem{8869712}
J.~Qiu, D.~Grace, G.~Ding, M.~D. Zakaria, and Q.~Wu, ``Air-ground heterogeneous networks for 5g and beyond via integrating high and low altitude platforms,'' \emph{IEEE Wireless Communications}, vol.~26, no.~6, pp. 140--148, 2019.

\bibitem{9645207}
H.~Ahmadinejad and A.~Falahati, ``Forming a two-tier heterogeneous air-network via combination of high and low altitude platforms,'' \emph{IEEE Transactions on Vehicular Technology}, vol.~71, no.~2, pp. 1989--2001, 2022.

\bibitem{7572034}
N.~Hossein~Motlagh, T.~Taleb, and O.~Arouk, ``Low-altitude unmanned aerial vehicles-based internet of things services: Comprehensive survey and future perspectives,'' \emph{IEEE Internet of Things Journal}, vol.~3, no.~6, pp. 899--922, 2016.

\bibitem{10833672}
H.~Yang, M.~Zheng, Z.~Shao, Y.~Jiang, and Z.~Xiong, ``Intelligent computation offloading and trajectory planning for 3d target search in low-altitude economy scenarios,'' \emph{IEEE Wireless Communications Letters}, pp. 1--1, 2025.

\bibitem{8742658}
R.~Shakeri, M.~A. Al-Garadi, A.~Badawy, A.~Mohamed, T.~Khattab, A.~K. Al-Ali, K.~A. Harras, and M.~Guizani, ``Design challenges of multi-uav systems in cyber-physical applications: A comprehensive survey and future directions,'' \emph{IEEE Communications Surveys \& Tutorials}, vol.~21, no.~4, pp. 3340--3385, 2019.

\bibitem{10916952}
Y.~Zhang, X.~Gao, N.~Ye, D.~Niyato, Z.~Han, and K.~Yang, ``Joint uav deployment, power allocation, and coalition formation for physical layer security in heterogeneous networks,'' \emph{IEEE Transactions on Vehicular Technology}, pp. 1--15, 2025.

\bibitem{9745441}
Z.~Liu, Y.~Cao, P.~Gao, X.~Hua, D.~Zhang, and T.~Jiang, ``Multi-uav network assisted intelligent edge computing: Challenges and opportunities,'' \emph{China Communications}, vol.~19, no.~3, pp. 258--278, 2022.

\bibitem{9200679}
Y.~Liu, X.~Gong, J.~Chen, S.~Chen, and Y.~Yang, ``Rotation-invariant siamese network for low-altitude remote-sensing image registration,'' \emph{IEEE Journal of Selected Topics in Applied Earth Observations and Remote Sensing}, vol.~13, pp. 5746--5758, 2020.

\bibitem{10879807}
G.~Cheng, X.~Song, Z.~Lyu, and J.~Xu, ``Networked isac for low-altitude economy: Coordinated transmit beamforming and uav trajectory design,'' \emph{IEEE Transactions on Communications}, pp. 1--1, 2025.

\bibitem{10681882}
G.~Cheng, X.~Song, Z.~{Lyu}, and J.~Xu, ``Networked isac for low-altitude economy: Transmit beamforming and uav trajectory design,'' in \emph{2024 IEEE/CIC International Conference on Communications in China (ICCC)}, 2024, pp. 78--83.

\bibitem{zheng2025uav}
X.~Zheng, G.~Sun, J.~Li, J.~Wang, Q.~Wu, D.~Niyato, and A.~Jamalipour, ``Uav swarm-enabled collaborative post-disaster communications in low altitude economy via a two-stage optimization approach,'' \emph{arXiv preprint arXiv:2501.05742}, 2025.

\bibitem{CNLAEHS}
\BIBentryALTinterwordspacing
China's low-altitude economy soars at high speed. Accessed: Dec. 19, 2024. [Online]. Available: \url{https://www.chinadaily.com.cn/a/202412/19/WS6763b8b7a310f1265a1d3d24.html}
\BIBentrySTDinterwordspacing

\bibitem{CNLAETFNE}
\BIBentryALTinterwordspacing
China's low-altitude economy takes flight: A new engine for innovation-driven growth. Accessed: Mar. 17, 2025. [Online]. Available: \url{https://www.chinadaily.com.cn/a/202412/19/WS6763b8b7a310f1265a1d3d24.html}
\BIBentrySTDinterwordspacing

\bibitem{USFATMC}
\BIBentryALTinterwordspacing
Flying air taxis move closer to us takeoff with issuing of faa rule. Accessed: Oct. 22, 2024. [Online]. Available: \url{https://www.usnews.com/news/business/articles/2024-10-22/flying-air-taxis-move-closer-to-us-takeoff-with-issuing-of-faa-rule}
\BIBentrySTDinterwordspacing

\bibitem{10.1145/3485272}
\BIBentryALTinterwordspacing
A.~Rugo, C.~A. Ardagna, and N.~E. Ioini, ``A security review in the uavnet era: Threats, countermeasures, and gap analysis,'' \emph{ACM Comput. Surv.}, vol.~55, no.~1, Jan. 2022. [Online]. Available: \url{https://doi.org/10.1145/3485272}
\BIBentrySTDinterwordspacing

\bibitem{10.1145/3703625}
\BIBentryALTinterwordspacing
X.~Wang, Z.~Zhao, L.~Yi, Z.~Ning, L.~Guo, F.~R. Yu, and S.~Guo, ``A survey on security of uav swarm networks: Attacks and countermeasures,'' \emph{ACM Comput. Surv.}, vol.~57, no.~3, Nov. 2024. [Online]. Available: \url{https://doi.org/10.1145/3703625}
\BIBentrySTDinterwordspacing

\bibitem{10793113}
O.~Ceviz, S.~Sen, and P.~Sadioglu, ``A survey of security in uavs and fanets:issues, threats, analysis of attacks, and solutions,'' \emph{IEEE Communications Surveys \& Tutorials}, pp. 1--1, 2024.

\bibitem{HADI2023103607}
\BIBentryALTinterwordspacing
H.~J. Hadi, Y.~Cao, K.~U. Nisa, A.~M. Jamil, and Q.~Ni, ``A comprehensive survey on security, privacy issues and emerging defence technologies for uavs,'' \emph{Journal of Network and Computer Applications}, vol. 213, p. 103607, 2023. [Online]. Available: \url{https://www.sciencedirect.com/science/article/pii/S1084804523000267}
\BIBentrySTDinterwordspacing

\bibitem{9488323}
V.~Hassija, V.~Chamola, A.~Agrawal, A.~Goyal, N.~C. Luong, D.~Niyato, F.~R. Yu, and M.~Guizani, ``Fast, reliable, and secure drone communication: A comprehensive survey,'' \emph{IEEE Communications Surveys \& Tutorials}, vol.~23, no.~4, pp. 2802--2832, 2021.

\bibitem{10.1145/3673225}
\BIBentryALTinterwordspacing
B.~Zolfaghari, M.~Abbasmollaei, F.~Hajizadeh, N.~Yanai, and K.~Bibak, ``Secure uav (drone) and the great promise of ai,'' \emph{ACM Comput. Surv.}, vol.~56, no.~11, Jul. 2024. [Online]. Available: \url{https://doi.org/10.1145/3673225}
\BIBentrySTDinterwordspacing

\bibitem{10599134}
X.~Wei, J.~Ma, and C.~Sun, ``A survey on security of unmanned aerial vehicle systems: Attacks and countermeasures,'' \emph{IEEE Internet of Things Journal}, vol.~11, no.~21, pp. 34\,826--34\,847, 2024.

\bibitem{9946859}
M.~Adil, M.~A. Jan, Y.~Liu, H.~Abulkasim, A.~Farouk, and H.~Song, ``A systematic survey: Security threats to uav-aided iot applications, taxonomy, current challenges and requirements with future research directions,'' \emph{IEEE Transactions on Intelligent Transportation Systems}, vol.~24, no.~2, pp. 1437--1455, 2023.

\bibitem{KUMAR2024110695}
\BIBentryALTinterwordspacing
N.~Kumar and A.~Chaudhary, ``Surveying cybersecurity vulnerabilities and countermeasures for enhancing uav security,'' \emph{Computer Networks}, vol. 252, p. 110695, 2024. [Online]. Available: \url{https://www.sciencedirect.com/science/article/pii/S1389128624005279}
\BIBentrySTDinterwordspacing

\bibitem{9900257}
J.~Wang, X.~Wang, R.~Gao, C.~Lei, W.~Feng, N.~Ge, S.~Jin, and T.~Q.~S. Quek, ``Physical layer security for uav communications: A comprehensive survey,'' \emph{China Communications}, vol.~19, no.~9, pp. 77--115, 2022.

\bibitem{8675384}
A.~Fotouhi, H.~Qiang, M.~Ding, M.~Hassan, L.~G. Giordano, A.~Garcia-Rodriguez, and J.~Yuan, ``Survey on uav cellular communications: Practical aspects, standardization advancements, regulation, and security challenges,'' \emph{IEEE Communications Surveys \& Tutorials}, vol.~21, no.~4, pp. 3417--3442, 2019.

\bibitem{10236463}
M.~Adil, H.~Song, S.~Mastorakis, H.~Abulkasim, A.~Farouk, and Z.~Jin, ``Uav-assisted iot applications, cybersecurity threats, ai-enabled solutions, open challenges with future research directions,'' \emph{IEEE Transactions on Intelligent Vehicles}, vol.~9, no.~4, pp. 4583--4605, 2024.

\bibitem{10003076}
W.~U. Khan, E.~Lagunas, Z.~Ali, M.~A. Javed, M.~Ahmed, S.~Chatzinotas, B.~Ottersten, and P.~Popovski, ``Opportunities for physical layer security in uav communication enhanced with intelligent reflective surfaces,'' \emph{IEEE Wireless Communications}, vol.~29, no.~6, pp. 22--28, 2022.

\bibitem{10496520}
J.~Wang, H.~Du, D.~Niyato, M.~Zhou, J.~Kang, and H.~Vincent~Poor, ``Acceleration estimation of signal propagation path length changes for wireless sensing,'' \emph{IEEE Transactions on Wireless Communications}, vol.~23, no.~9, pp. 11\,476--11\,492, 2024.

\bibitem{8233654}
T.~Wang, C.-K. Wen, H.~Wang, F.~Gao, T.~Jiang, and S.~Jin, ``Deep learning for wireless physical layer: Opportunities and challenges,'' \emph{China Communications}, vol.~14, no.~11, pp. 92--111, 2017.

\bibitem{10599123}
J.~Wang, H.~Du, D.~Niyato, J.~Kang, S.~Cui, X.~Shen, and P.~Zhang, ``Generative ai for integrated sensing and communication: Insights from the physical layer perspective,'' \emph{IEEE Wireless Communications}, vol.~31, no.~5, pp. 246--255, 2024.

\bibitem{10110330}
S.~Li, L.~Xiao, Y.~Liu, G.~Liu, P.~Xiao, and T.~Jiang, ``Performance analysis for orthogonal time frequency space modulation systems with generalized waveform,'' \emph{China Communications}, vol.~20, no.~4, pp. 57--72, 2023.

\bibitem{9927483}
N.~Xie, W.~Xiong, M.~Sha, T.~Hu, P.~Zhang, L.~Huang, and D.~Niyato, ``Physical layer authentication with high compatibility using an encoding approach,'' \emph{IEEE Transactions on Communications}, vol.~70, no.~12, pp. 8270--8285, 2022.

\bibitem{LIU2021589}
\BIBentryALTinterwordspacing
S.~Liu, T.~Wang, and S.~Wang, ``Toward intelligent wireless communications: Deep learning - based physical layer technologies,'' \emph{Digital Communications and Networks}, vol.~7, no.~4, pp. 589--597, 2021. [Online]. Available: \url{https://www.sciencedirect.com/science/article/pii/S2352864821000742}
\BIBentrySTDinterwordspacing

\bibitem{10461999}
Y.~Zhang, Y.~Peng, X.~Tang, L.~Xiao, and T.~Jiang, ``Large-scale fading decoding aided user-centric cell-free massive mimo: Uplink error probability analysis and detector design,'' \emph{IEEE Transactions on Wireless Communications}, vol.~23, no.~8, pp. 10\,336--10\,349, 2024.

\bibitem{10163877}
H.~Du, J.~Wang, D.~Niyato, J.~Kang, Z.~Xiong, J.~Zhang, and X.~Shen, ``Semantic communications for wireless sensing: Ris-aided encoding and self-supervised decoding,'' \emph{IEEE Journal on Selected Areas in Communications}, vol.~41, no.~8, pp. 2547--2562, 2023.

\bibitem{9453787}
P.~Yang, X.~Xi, K.~Guo, T.~Q.~S. Quek, J.~Chen, and X.~Cao, ``Proactive uav network slicing for urllc and mobile broadband service multiplexing,'' \emph{IEEE Journal on Selected Areas in Communications}, vol.~39, no.~10, pp. 3225--3244, 2021.

\bibitem{10812989}
J.~Huang, A.~Wang, G.~Sun, J.~Li, J.~Wang, H.~Du, and D.~Niyato, ``Dual uav cluster-assisted maritime physical layer secure communications via collaborative beamforming,'' \emph{IEEE Internet of Things Journal}, pp. 1--1, 2024.

\bibitem{10.1145/3715319}
\BIBentryALTinterwordspacing
Z.~Duan, Z.~Chang, N.~Xie, W.~Sun, and D.~T. Niyato, ``Adaptive strategies in enhancing physical layer security: A comprehensive survey,'' \emph{ACM Comput. Surv.}, vol.~57, no.~7, Feb. 2025. [Online]. Available: \url{https://doi.org/10.1145/3715319}
\BIBentrySTDinterwordspacing

\bibitem{7875081}
Q.~Wang, Z.~Chen, W.~Mei, and J.~Fang, ``Improving physical layer security using uav-enabled mobile relaying,'' \emph{IEEE Wireless Communications Letters}, vol.~6, no.~3, pp. 310--313, 2017.

\bibitem{10678860}
S.~Liu, H.~Yang, M.~Zheng, L.~Xiao, Z.~Xiong, and D.~Niyato, ``Uav-enabled semantic communication in mobile edge computing under jamming attacks: An intelligent resource management approach,'' \emph{IEEE Transactions on Wireless Communications}, vol.~23, no.~11, pp. 17\,493--17\,507, 2024.

\bibitem{10352334}
S.~Bi, K.~Li, S.~Hu, W.~Ni, C.~Wang, and X.~Wang, ``Detection and mitigation of position spoofing attacks on cooperative uav swarm formations,'' \emph{IEEE Transactions on Information Forensics and Security}, vol.~19, pp. 1883--1895, 2024.

\bibitem{8883128}
X.~Sun, D.~W.~K. Ng, Z.~Ding, Y.~Xu, and Z.~Zhong, ``Physical layer security in uav systems: Challenges and opportunities,'' \emph{IEEE Wireless Communications}, vol.~26, no.~5, pp. 40--47, 2019.

\bibitem{10474115}
G.~Zhang, Q.~Hu, Y.~Zhang, Y.~Dai, and T.~Jiang, ``Lightweight cross-domain authentication scheme for securing wireless iot devices using backscatter communication,'' \emph{IEEE Internet of Things Journal}, vol.~11, no.~12, pp. 22\,021--22\,035, 2024.

\bibitem{8883124}
Q.~Wu, W.~Mei, and R.~Zhang, ``Safeguarding wireless network with uavs: A physical layer security perspective,'' \emph{IEEE Wireless Communications}, vol.~26, no.~5, pp. 12--18, 2019.

\bibitem{8883127}
H.-M. Wang, X.~Zhang, and J.-C. Jiang, ``Uav-involved wireless physical-layer secure communications: Overview and research directions,'' \emph{IEEE Wireless Communications}, vol.~26, no.~5, pp. 32--39, 2019.

\bibitem{8758975}
B.~Li, Z.~Fei, Y.~Zhang, and M.~Guizani, ``Secure uav communication networks over 5g,'' \emph{IEEE Wireless Communications}, vol.~26, no.~5, pp. 114--120, 2019.

\bibitem{9200889}
L.~Bai, L.~Zhu, J.~Liu, J.~Choi, and W.~Zhang, ``Physical layer authentication in wireless communication networks: A survey,'' \emph{Journal of Communications and Information Networks}, vol.~5, no.~3, pp. 237--264, 2020.

\bibitem{9279294}
N.~Xie, Z.~Li, and H.~Tan, ``A survey of physical-layer authentication in wireless communications,'' \emph{IEEE Communications Surveys \& Tutorials}, vol.~23, no.~1, pp. 282--310, 2021.

\bibitem{9203867}
Y.~Xu, T.~Zhang, D.~Yang, Y.~Liu, and M.~Tao, ``Joint resource and trajectory optimization for security in uav-assisted mec systems,'' \emph{IEEE Transactions on Communications}, vol.~69, no.~1, pp. 573--588, 2021.

\bibitem{10636964}
Y.~Zhang, Z.~Kuang, Y.~Feng, and F.~Hou, ``Task offloading and trajectory optimization for secure communications in dynamic user multi-uav mec systems,'' \emph{IEEE Transactions on Mobile Computing}, vol.~23, no.~12, pp. 14\,427--14\,440, 2024.

\bibitem{10572013}
Y.~Zhang, X.~Gao, H.~Yuan, K.~Yang, J.~Kang, P.~Wang, and D.~Niyato, ``Joint uav trajectory and power allocation with hybrid fso/rf for secure space–air–ground communications,'' \emph{IEEE Internet of Things Journal}, vol.~11, no.~19, pp. 31\,407--31\,421, 2024.

\bibitem{9453748}
W.~Wang, X.~Li, R.~Wang, K.~Cumanan, W.~Feng, Z.~Ding, and O.~A. Dobre, ``Robust 3d-trajectory and time switching optimization for dual-uav-enabled secure communications,'' \emph{IEEE Journal on Selected Areas in Communications}, vol.~39, no.~11, pp. 3334--3347, 2021.

\bibitem{9417318}
C.~Wen, L.~Qiu, and X.~Liang, ``Securing uav communication with mobile uav eavesdroppers: Joint trajectory and communication design,'' in \emph{2021 IEEE Wireless Communications and Networking Conference (WCNC)}, 2021, pp. 1--6.

\bibitem{9450021}
W.~Lu, Y.~Ding, Y.~Gao, S.~Hu, Y.~Wu, N.~Zhao, and Y.~Gong, ``Resource and trajectory optimization for secure communications in dual unmanned aerial vehicle mobile edge computing systems,'' \emph{IEEE Transactions on Industrial Informatics}, vol.~18, no.~4, pp. 2704--2713, 2022.

\bibitem{10310294}
F.~Lu, G.~Liu, W.~Lu, Y.~Gao, J.~Cao, N.~Zhao, and A.~Nallanathan, ``Resource and trajectory optimization for uav-relay-assisted secure maritime mec,'' \emph{IEEE Transactions on Communications}, vol.~72, no.~3, pp. 1641--1652, 2024.

\bibitem{10153699}
A.~S. Abdalla, A.~Behfarnia, and V.~Marojevic, ``Uav trajectory and multi-user beamforming optimization for clustered users against passive eavesdropping attacks with unknown csi,'' \emph{IEEE Transactions on Vehicular Technology}, vol.~72, no.~11, pp. 14\,426--14\,442, 2023.

\bibitem{10325641}
Y.~Ding, H.~Han, W.~Lu, Y.~Wang, N.~Zhao, X.~Wang, and X.~Yang, ``Ddqn-based trajectory and resource optimization for uav-aided mec secure communications,'' \emph{IEEE Transactions on Vehicular Technology}, vol.~73, no.~4, pp. 6006--6011, 2024.

\bibitem{9801656}
H.~Kang, X.~Chang, J.~Mišić, V.~B. Mišić, J.~Fan, and J.~Bai, ``Improving dual-uav aided ground-uav bi-directional communication security: Joint uav trajectory and transmit power optimization,'' \emph{IEEE Transactions on Vehicular Technology}, vol.~71, no.~10, pp. 10\,570--10\,583, 2022.

\bibitem{9161257}
Y.~Zhang, Z.~Mou, F.~Gao, J.~Jiang, R.~Ding, and Z.~Han, ``Uav-enabled secure communications by multi-agent deep reinforcement learning,'' \emph{IEEE Transactions on Vehicular Technology}, vol.~69, no.~10, pp. 11\,599--11\,611, 2020.

\bibitem{10287142}
Y.~Liu, C.~Huang, G.~Chen, R.~Song, S.~Song, and P.~Xiao, ``Deep learning empowered trajectory and passive beamforming design in uav-ris enabled secure cognitive non-terrestrial networks,'' \emph{IEEE Wireless Communications Letters}, vol.~13, no.~1, pp. 188--192, 2024.

\bibitem{10734220}
J.~Wang, R.~Wang, Z.~Zheng, R.~Lin, L.~Wu, and F.~Shu, ``Physical layer security enhancement in uav-assisted cooperative jamming for cognitive radio networks: A mappo-lstm deep reinforcement learning approach,'' \emph{IEEE Transactions on Vehicular Technology}, pp. 1--14, 2024.

\bibitem{9713997}
X.~Tang, N.~Liu, R.~Zhang, and Z.~Han, ``Deep learning-assisted secure uav-relaying networks with channel uncertainties,'' \emph{IEEE Transactions on Vehicular Technology}, vol.~71, no.~5, pp. 5048--5059, 2022.

\bibitem{10114676}
X.~Li, R.~Yao, Y.~Fan, P.~Wang, and J.~Xu, ``Secure efficiency map-enabled uav trajectory planning,'' \emph{IEEE Wireless Communications Letters}, vol.~12, no.~8, pp. 1324--1328, 2023.

\bibitem{10194980}
R.~Karmakar, G.~Kaddoum, and O.~Akhrif, ``A novel federated learning-based smart power and 3d trajectory control for fairness optimization in secure uav-assisted mec services,'' \emph{IEEE Transactions on Mobile Computing}, vol.~23, no.~5, pp. 4832--4848, 2024.

\bibitem{9584882}
Z.~Li, X.~Liao, J.~Shi, L.~Li, and P.~Xiao, ``Md-gan-based uav trajectory and power optimization for cognitive covert communications,'' \emph{IEEE Internet of Things Journal}, vol.~9, no.~12, pp. 10\,187--10\,199, 2022.

\bibitem{10700928}
S.~Jia, L.~Xiaomeng, L.~Xiaomin, T.~Zhuangzhuang, and H.~Junfan, ``Covert leo satellite communication aided by generative adversarial network based cooperative uav jamming,'' \emph{China Communications}, vol.~21, no.~9, pp. 27--39, 2024.

\bibitem{10759093}
C.~Zhang, G.~Sun, J.~Li, Q.~Wu, J.~Wang, D.~Niyato, and Y.~Liu, ``Multi-objective aerial collaborative secure communication optimization via generative diffusion model-enabled deep reinforcement learning,'' \emph{IEEE Transactions on Mobile Computing}, pp. 1--18, 2024.

\bibitem{9237145}
T.~Alladi, Naren, G.~Bansal, V.~Chamola, and M.~Guizani, ``Secauthuav: A novel authentication scheme for uav-ground station and uav-uav communication,'' \emph{IEEE Transactions on Vehicular Technology}, vol.~69, no.~12, pp. 15\,068--15\,077, 2020.

\bibitem{10146461}
R.~Karmakar, G.~Kaddoum, and O.~Akhrif, ``A puf and fuzzy extractor-based uav-ground station and uav-uav authentication mechanism with intelligent adaptation of secure sessions,'' \emph{IEEE Transactions on Mobile Computing}, vol.~23, no.~5, pp. 3858--3875, 2024.

\bibitem{10436373}
M.~Tanveer, A.~Aldosary, S.-u.-d. Khokhar, A.~K. Das, S.~A. Aldossari, and S.~A. Chaudhry, ``Paf-iod: Puf-enabled authentication framework for the internet of drones,'' \emph{IEEE Transactions on Vehicular Technology}, vol.~73, no.~7, pp. 9560--9574, 2024.

\bibitem{9672766}
S.~J. Maeng, Y.~Yapici, i.~Guvenc, A.~Bhuyan, and H.~Dai, ``Precoder design for physical-layer security and authentication in massive mimo uav communications,'' \emph{IEEE Transactions on Vehicular Technology}, vol.~71, no.~3, pp. 2949--2964, 2022.

\bibitem{10233023}
Y.~Zhou, Z.~Ma, H.~Liu, P.~L. Yeoh, Y.~Li, B.~Vucetic, and P.~Fan, ``A uav-aided physical layer authentication based on channel characteristics and geographical locations,'' \emph{IEEE Transactions on Vehicular Technology}, vol.~73, no.~1, pp. 1053--1064, 2024.

\bibitem{10834505}
Y.~Zhou, Y.~Wang, Z.~Ma, P.~Fan, and M.~Xiao, ``Physical layer authentication for uav communications under rayleigh and rician channels,'' \emph{IEEE Transactions on Wireless Communications}, pp. 1--1, 2025.

\bibitem{5751298}
Y.-S. Shiu, S.~Y. Chang, H.-C. Wu, S.~C.-H. Huang, and H.-H. Chen, ``Physical layer security in wireless networks: a tutorial,'' \emph{IEEE Wireless Communications}, vol.~18, no.~2, pp. 66--74, 2011.

\bibitem{10464352}
J.~Xu, D.~Li, Z.~Zhu, Z.~Yang, N.~Zhao, and D.~Niyato, ``Anti-jamming design for integrated sensing and communication via aerial irs,'' \emph{IEEE Transactions on Communications}, vol.~72, no.~8, pp. 4607--4619, 2024.

\bibitem{9271902}
B.~Duo, Q.~Wu, X.~Yuan, and R.~Zhang, ``Anti-jamming 3d trajectory design for uav-enabled wireless sensor networks under probabilistic los channel,'' \emph{IEEE Transactions on Vehicular Technology}, vol.~69, no.~12, pp. 16\,288--16\,293, 2020.

\bibitem{9200570}
Y.~Wu, W.~Yang, X.~Guan, and Q.~Wu, ``Energy-efficient trajectory design for uav-enabled communication under malicious jamming,'' \emph{IEEE Wireless Communications Letters}, vol.~10, no.~2, pp. 206--210, 2021.

\bibitem{9454372}
Y.~Wu, W.~Yang, X.~{Guan}, and Q.~Wu, ``Uav-enabled relay communication under malicious jamming: Joint trajectory and transmit power optimization,'' \emph{IEEE Transactions on Vehicular Technology}, vol.~70, no.~8, pp. 8275--8279, 2021.

\bibitem{7925694}
M.~A. Aref, S.~K. Jayaweera, and S.~Machuzak, ``Multi-agent reinforcement learning based cognitive anti-jamming,'' in \emph{2017 IEEE Wireless Communications and Networking Conference (WCNC)}, 2017, pp. 1--6.

\bibitem{8023829}
L.~Jia, F.~Yao, Y.~Sun, Y.~Xu, S.~Feng, and A.~Anpalagan, ``A hierarchical learning solution for anti-jamming stackelberg game with discrete power strategies,'' \emph{IEEE Wireless Communications Letters}, vol.~6, no.~6, pp. 818--821, 2017.

\bibitem{8314744}
X.~Liu, Y.~Xu, L.~Jia, Q.~Wu, and A.~Anpalagan, ``Anti-jamming communications using spectrum waterfall: A deep reinforcement learning approach,'' \emph{IEEE Communications Letters}, vol.~22, no.~5, pp. 998--1001, 2018.

\bibitem{9264659}
H.~Yang, Z.~Xiong, J.~Zhao, D.~Niyato, Q.~Wu, H.~V. Poor, and M.~Tornatore, ``Intelligent reflecting surface assisted anti-jamming communications: A fast reinforcement learning approach,'' \emph{IEEE Transactions on Wireless Communications}, vol.~20, no.~3, pp. 1963--1974, 2021.

\bibitem{9816050}
Z.~Yin, Y.~Lin, Y.~Zhang, Y.~Qian, F.~Shu, and J.~Li, ``Collaborative multiagent reinforcement learning aided resource allocation for uav anti-jamming communication,'' \emph{IEEE Internet of Things Journal}, vol.~9, no.~23, pp. 23\,995--24\,008, 2022.

\bibitem{10750022}
Y.~Ma, K.~Liu, Y.~Liu, X.~Wang, and Z.~Zhao, ``An intelligent game-based anti-jamming solution using adversarial populations for aerial communication networks,'' \emph{IEEE Transactions on Cognitive Communications and Networking}, pp. 1--1, 2024.

\bibitem{10614297}
Z.~Shao, H.~Yang, L.~Xiao, W.~Su, Y.~Chen, and Z.~Xiong, ``Deep reinforcement learning-based resource management for uav-assisted mobile edge computing against jamming,'' \emph{IEEE Transactions on Mobile Computing}, vol.~23, no.~12, pp. 13\,358--13\,374, 2024.

\bibitem{9739860}
Y.~Zhou, P.~L. Yeoh, K.~J. Kim, Z.~Ma, Y.~Li, and B.~Vucetic, ``Game theoretic physical layer authentication for spoofing detection in uav communications,'' \emph{IEEE Transactions on Vehicular Technology}, vol.~71, no.~6, pp. 6750--6755, 2022.

\bibitem{10851372}
Q.~Cheng, Y.~Zhou, H.~Liu, L.~Yang, Z.~Ma, and P.~Fan, ``Physical layer authentication in uav communications with channel randomness and jamming uncertainty,'' \emph{IEEE Transactions on Vehicular Technology}, pp. 1--6, 2025.

\bibitem{8946587}
A.~Eldosouky, A.~Ferdowsi, and W.~Saad, ``Drones in distress: A game-theoretic countermeasure for protecting uavs against gps spoofing,'' \emph{IEEE Internet of Things Journal}, vol.~7, no.~4, pp. 2840--2854, 2020.

\bibitem{10685447}
D.~She, W.~Wang, Z.~Yin, J.~Wang, and H.~Shan, ``Gps spoofing attack recognition for uavs with limited samples,'' \emph{IEEE Internet of Things Journal}, vol.~12, no.~1, pp. 250--261, 2025.

\bibitem{9845684}
Y.~Dang, C.~Benzaïd, B.~Yang, T.~Taleb, and Y.~Shen, ``Deep-ensemble-learning-based gps spoofing detection for cellular-connected uavs,'' \emph{IEEE Internet of Things Journal}, vol.~9, no.~24, pp. 25\,068--25\,085, 2022.

\bibitem{8999433}
X.~Wang, J.~Wang, Y.~Xu, J.~Chen, L.~Jia, X.~Liu, and Y.~Yang, ``Dynamic spectrum anti-jamming communications: Challenges and opportunities,'' \emph{IEEE Communications Magazine}, vol.~58, no.~2, pp. 79--85, 2020.

\bibitem{8113526}
L.~Zhang, G.~Ding, Q.~Wu, and Z.~Han, ``Spectrum sensing under spectrum misuse behaviors: A multi-hypothesis test perspective,'' \emph{IEEE Transactions on Information Forensics and Security}, vol.~13, no.~4, pp. 993--1007, 2018.

\bibitem{10368002}
S.~C. Hassler, U.~A. Mughal, and M.~Ismail, ``Cyber-physical intrusion detection system for unmanned aerial vehicles,'' \emph{IEEE Transactions on Intelligent Transportation Systems}, vol.~25, no.~6, pp. 6106--6117, 2024.

\bibitem{9741304}
A.~Krayani, A.~S. Alam, L.~Marcenaro, A.~Nallanathan, and C.~Regazzoni, ``An emergent self-awareness module for physical layer security in cognitive uav radios,'' \emph{IEEE Transactions on Cognitive Communications and Networking}, vol.~8, no.~2, pp. 888--906, 2022.

\bibitem{9858012}
A.~Krayani, A.~S. {Alam}, L.~{Marcenaro}, A.~{Nallanathan}, and C.~{Regazzoni}, ``Automatic jamming signal classification in cognitive uav radios,'' \emph{IEEE Transactions on Vehicular Technology}, vol.~71, no.~12, pp. 12\,972--12\,988, 2022.

\bibitem{9829873}
A.~Krayani, A.~S. Alam, L.~Marcenaro, A.~Nallanathan, and C.~Regazzoni, ``A novel resource allocation for anti-jamming in cognitive-uavs: An active inference approach,'' \emph{IEEE Communications Letters}, vol.~26, no.~10, pp. 2272--2276, 2022.

\bibitem{9634169}
D.~Darsena, G.~Gelli, I.~Iudice, and F.~Verde, ``Detection and blind channel estimation for uav-aided wireless sensor networks in smart cities under mobile jamming attack,'' \emph{IEEE Internet of Things Journal}, vol.~9, no.~14, pp. 11\,932--11\,950, 2022.

\bibitem{8688501}
L.~Zhang, G.~Ding, Q.~Wu, and P.~Liu, ``Detection of abnormal power emission in uav communication networks,'' \emph{IEEE Wireless Communications Letters}, vol.~8, no.~4, pp. 1179--1182, 2019.

\bibitem{8854240}
T.~M. Hoang, N.~M. Nguyen, and T.~Q. Duong, ``Detection of eavesdropping attack in uav-aided wireless systems: Unsupervised learning with one-class svm and k-means clustering,'' \emph{IEEE Wireless Communications Letters}, vol.~9, no.~2, pp. 139--142, 2020.

\bibitem{10530539}
Y.~An, R.~Kang, Y.~Ban, and S.~Yang, ``Beidou receiver based on anti-jamming antenna arrays with self-calibration for precise relative positioning,'' \emph{Journal of Systems Engineering and Electronics}, vol.~35, no.~5, pp. 1132--1147, 2024.

\bibitem{10.1145/3395351.3401703}
\BIBentryALTinterwordspacing
H.~Sathaye and A.~Ranganathan, ``Semperfi: a spoofer eliminating standalone gps receiver,'' in \emph{Proceedings of the 13th ACM Conference on Security and Privacy in Wireless and Mobile Networks}, ser. WiSec '20.\hskip 1em plus 0.5em minus 0.4em\relax New York, NY, USA: Association for Computing Machinery, 2020, p. 353–355. [Online]. Available: \url{https://doi.org/10.1145/3395351.3401703}
\BIBentrySTDinterwordspacing

\bibitem{sathaye2022semperfi}
H.~Sathaye, G.~LaMountain, P.~Closas, and A.~Ranganathan, ``Semperfi: Anti-spoofing gps receiver for uavs,'' in \emph{Network and Distributed Systems Security (NDSS) Symposium 2022}, 2022.

\bibitem{8045998}
S.~Han, L.~Chen, W.~Meng, and C.~Li, ``Improve the security of gnss receivers through spoofing mitigation,'' \emph{IEEE Access}, vol.~5, pp. 21\,057--21\,069, 2017.

\bibitem{ye2024integrated}
X.~Ye, Y.~Mao, X.~Yu, S.~Sun, L.~Fu, and J.~Xu, ``Integrated sensing and communications for low-altitude economy: A deep reinforcement learning approach,'' \emph{arXiv preprint arXiv:2412.04074}, 2024.

\bibitem{HUANG2024100694}
\BIBentryALTinterwordspacing
C.~Huang, S.~Fang, H.~Wu, Y.~Wang, and Y.~Yang, ``Low-altitude intelligent transportation: System architecture, infrastructure, and key technologies,'' \emph{Journal of Industrial Information Integration}, vol.~42, p. 100694, 2024. [Online]. Available: \url{https://www.sciencedirect.com/science/article/pii/S2452414X24001377}
\BIBentrySTDinterwordspacing

\bibitem{yang2024embodied}
Y.~Yang, Y.~Chen, J.~Wang, G.~Sun, and D.~Niyato, ``Embodied ai-empowered low altitude economy: Integrated sensing, communications, computation, and control (isc3),'' \emph{arXiv preprint arXiv:2412.19996}, 2024.

\bibitem{li2025aerial}
J.~Li, G.~Sun, Q.~Wu, S.~Liang, J.~Wang, D.~Niyato, and D.~I. Kim, ``Aerial secure collaborative communications under eavesdropper collusion in low-altitude economy: A generative swarm intelligent approach,'' \emph{arXiv preprint arXiv:2503.00721}, 2025.

\bibitem{10749978}
G.~Sun, W.~Xie, D.~Niyato, H.~Du, J.~Kang, J.~Wu, S.~Sun, and P.~Zhang, ``Generative ai for advanced uav networking,'' \emph{IEEE Network}, pp. 1--1, 2024.

\bibitem{10089851}
X.~Tang, X.~Li, R.~Yu, Y.~Wu, J.~Ye, F.~Tang, and Q.~Chen, ``Digital-twin-assisted task assignment in multi-uav systems: A deep reinforcement learning approach,'' \emph{IEEE Internet of Things Journal}, vol.~10, no.~17, pp. 15\,362--15\,375, 2023.

\bibitem{10720878}
X.~Tang, Q.~Chen, R.~Yu, and X.~Li, ``Digital twin-empowered task assignment in aerial mec network: A resource coalition cooperation approach with generative model,'' \emph{IEEE Transactions on Network Science and Engineering}, vol.~12, no.~1, pp. 13--27, 2025.

\bibitem{Jiang20236GNN}
\BIBentryALTinterwordspacing
Y.~Jiang, X.~Li, G.~Zhu, H.~Li, J.~Deng, and Q.~Shi, ``6g non-terrestrial networks enabled low-altitude economy: Opportunities and challenges,'' \emph{ArXiv}, vol. abs/2311.09047, 2023. [Online]. Available: \url{https://api.semanticscholar.org/CorpusID:265213350}
\BIBentrySTDinterwordspacing

\bibitem{8869876}
X.~Luo, Y.~Zhang, Z.~He, G.~Yang, and Z.~Ji, ``A two-step environment-learning-based method for optimal uav deployment,'' \emph{IEEE Access}, vol.~7, pp. 149\,328--149\,340, 2019.

\bibitem{10884737}
X.~Tang, Q.~Chen, W.~Weng, B.~Liao, J.~Wang, X.~Cao, and X.~Li, ``Dnn task assignment in uav networks: A generative ai enhanced multi-agent reinforcement learning approach,'' \emph{IEEE Internet of Things Journal}, pp. 1--1, 2025.

\bibitem{9453811}
H.~Yang, J.~Zhao, Z.~Xiong, K.-Y. Lam, S.~Sun, and L.~Xiao, ``Privacy-preserving federated learning for uav-enabled networks: Learning-based joint scheduling and resource management,'' \emph{IEEE Journal on Selected Areas in Communications}, vol.~39, no.~10, pp. 3144--3159, 2021.

\bibitem{9454158}
X.~Cai, T.~Izydorczyk, J.~Rodríguez-Piñeiro, I.~Z. Kovács, J.~Wigard, F.~M.~L. Tavares, and P.~E. Mogensen, ``Empirical low-altitude air-to-ground spatial channel characterization for cellular networks connectivity,'' \emph{IEEE Journal on Selected Areas in Communications}, vol.~39, no.~10, pp. 2975--2991, 2021.

\bibitem{10097717}
Y.~Zhao, F.~Zhou, L.~Feng, W.~Li, Y.~Sun, and M.~A. Imran, ``Backhaul-constrained coverage analysis of integrated high and low altitude platforms aerial communication system in post-disaster areas,'' \emph{IEEE Communications Letters}, vol.~27, no.~6, pp. 1629--1633, 2023.

\bibitem{9635590}
S.~H. Alsamhi, F.~A. Almalki, F.~Afghah, A.~Hawbani, A.~V. Shvetsov, B.~Lee, and H.~Song, ``Drones’ edge intelligence over smart environments in b5g: Blockchain and federated learning synergy,'' \emph{IEEE Transactions on Green Communications and Networking}, vol.~6, no.~1, pp. 295--312, 2022.

\bibitem{AHMAD2020107122}
\BIBentryALTinterwordspacing
A.~Ahmad, A.~A. Cheema, and D.~Finlay, ``A survey of radio propagation channel modelling for low altitude flying base stations,'' \emph{Computer Networks}, vol. 171, p. 107122, 2020. [Online]. Available: \url{https://www.sciencedirect.com/science/article/pii/S1389128619310692}
\BIBentrySTDinterwordspacing

\bibitem{9562043}
I.~Bozcan and E.~Kayacan, ``Context-dependent anomaly detection for low altitude traffic surveillance,'' in \emph{2021 IEEE International Conference on Robotics and Automation (ICRA)}, 2021, pp. 224--230.

\bibitem{9094586}
Y.~Liu, X.~Gong, and Y.~Yang, ``A multilayer fusion network with rotation- invariant and dynamic feature representation for multiview low-altitude image registration,'' \emph{IEEE Geoscience and Remote Sensing Letters}, vol.~18, no.~6, pp. 1019--1023, 2021.

\bibitem{8690950}
A.~Omri and M.~O. Hasna, ``Physical layer security analysis of uav based communication networks,'' in \emph{2018 IEEE 88th Vehicular Technology Conference (VTC-Fall)}, 2018, pp. 1--6.

\bibitem{samonas2014cia}
S.~Samonas and D.~Coss, ``The cia strikes back: Redefining confidentiality, integrity and availability in security.'' \emph{Journal of Information System Security}, vol.~10, no.~3, 2014.

\bibitem{10623395}
C.~Zhao, H.~Du, D.~Niyato, J.~Kang, Z.~Xiong, D.~I. Kim, X.~Shen, and K.~B. Letaief, ``Generative ai for secure physical layer communications: A survey,'' \emph{IEEE Transactions on Cognitive Communications and Networking}, vol.~11, no.~1, pp. 3--26, 2025.

\bibitem{8509094}
J.~M. Hamamreh, H.~M. Furqan, and H.~Arslan, ``Classifications and applications of physical layer security techniques for confidentiality: A comprehensive survey,'' \emph{IEEE Communications Surveys \& Tutorials}, vol.~21, no.~2, pp. 1773--1828, 2019.

\bibitem{shakiba2021physical}
M.~Shakiba-Herfeh, A.~Chorti, and H.~Vincent~Poor, ``Physical layer security: Authentication, integrity, and confidentiality,'' \emph{Physical layer security}, pp. 129--150, 2021.

\bibitem{9201322}
S.~Hu, Q.~Wu, and X.~Wang, ``Energy management and trajectory optimization for uav-enabled legitimate monitoring systems,'' \emph{IEEE Transactions on Wireless Communications}, vol.~20, no.~1, pp. 142--155, 2021.

\bibitem{8543573}
D.~Wang, B.~Bai, W.~Zhao, and Z.~Han, ``A survey of optimization approaches for wireless physical layer security,'' \emph{IEEE Communications Surveys \& Tutorials}, vol.~21, no.~2, pp. 1878--1911, 2019.

\bibitem{9070153}
M.~A. Arfaoui, M.~D. Soltani, I.~Tavakkolnia, A.~Ghrayeb, M.~Safari, C.~M. Assi, and H.~Haas, ``Physical layer security for visible light communication systems: A survey,'' \emph{IEEE Communications Surveys \& Tutorials}, vol.~22, no.~3, pp. 1887--1908, 2020.

\bibitem{9463880}
Z.~Yin, M.~Jia, N.~Cheng, W.~Wang, F.~Lyu, Q.~Guo, and X.~Shen, ``Uav-assisted physical layer security in multi-beam satellite-enabled vehicle communications,'' \emph{IEEE Transactions on Intelligent Transportation Systems}, vol.~23, no.~3, pp. 2739--2751, 2022.

\bibitem{7944621}
X.~Fang, N.~Zhang, S.~Zhang, D.~Chen, X.~Sha, and X.~Shen, ``On physical layer security: Weighted fractional fourier transform based user cooperation,'' \emph{IEEE Transactions on Wireless Communications}, vol.~16, no.~8, pp. 5498--5510, 2017.

\bibitem{10013691}
W.~Tian, X.~Ding, G.~Liu, Y.~Dai, and Z.~Han, ``A uav-assisted secure communication system by jointly optimizing transmit power and trajectory in the internet of things,'' \emph{IEEE Transactions on Green Communications and Networking}, vol.~7, no.~4, pp. 2025--2037, 2023.

\bibitem{IRRAM2022103431}
\BIBentryALTinterwordspacing
F.~Irram, M.~Ali, M.~Naeem, and S.~Mumtaz, ``Physical layer security for beyond 5g/6g networks: Emerging technologies and future directions,'' \emph{Journal of Network and Computer Applications}, vol. 206, p. 103431, 2022. [Online]. Available: \url{https://www.sciencedirect.com/science/article/pii/S108480452200087X}
\BIBentrySTDinterwordspacing

\bibitem{Lu2021}
\BIBentryALTinterwordspacing
W.~Lu, P.~Si, F.~Lu, B.~Li, Z.~Liu, S.~Hu, and Y.~Gong, ``Resource and trajectory optimization in uav-powered wireless communication system,'' \emph{Science China Information Sciences}, vol.~64, no.~4, p. 140304, Mar 2021, accessed: 2025-01-03. [Online]. Available: \url{https://doi.org/10.1007/s11432-020-3060-4}
\BIBentrySTDinterwordspacing

\bibitem{10.1145/3560261}
\BIBentryALTinterwordspacing
J.~Luo, Z.~Wang, M.~Xia, L.~Wu, Y.~Tian, and Y.~Chen, ``Path planning for uav communication networks: Related technologies, solutions, and opportunities,'' \emph{ACM Comput. Surv.}, vol.~55, no.~9, Jan. 2023. [Online]. Available: \url{https://doi-org.remotexs.ntu.edu.sg/10.1145/3560261}
\BIBentrySTDinterwordspacing

\bibitem{9061039}
A.~V. Savkin, H.~Huang, and W.~Ni, ``Securing uav communication in the presence of stationary or mobile eavesdroppers via online 3d trajectory planning,'' \emph{IEEE Wireless Communications Letters}, vol.~9, no.~8, pp. 1211--1215, 2020.

\bibitem{8643815}
X.~Zhou, Q.~Wu, S.~Yan, F.~Shu, and J.~Li, ``Uav-enabled secure communications: Joint trajectory and transmit power optimization,'' \emph{IEEE Transactions on Vehicular Technology}, vol.~68, no.~4, pp. 4069--4073, 2019.

\bibitem{9171468}
R.~Ding, F.~Gao, and X.~S. Shen, ``3d uav trajectory design and frequency band allocation for energy-efficient and fair communication: A deep reinforcement learning approach,'' \emph{IEEE Transactions on Wireless Communications}, vol.~19, no.~12, pp. 7796--7809, 2020.

\bibitem{8589002}
C.~Zhong, J.~Yao, and J.~Xu, ``Secure uav communication with cooperative jamming and trajectory control,'' \emph{IEEE Communications Letters}, vol.~23, no.~2, pp. 286--289, 2019.

\bibitem{10283826}
Y.~Bai, H.~Zhao, X.~Zhang, Z.~Chang, R.~Jäntti, and K.~Yang, ``Toward autonomous multi-uav wireless network: A survey of reinforcement learning-based approaches,'' \emph{IEEE Communications Surveys \& Tutorials}, vol.~25, no.~4, pp. 3038--3067, 2023.

\bibitem{10413638}
R.~Dong, B.~Wang, K.~Cao, J.~Tian, and T.~Cheng, ``Secure transmission design of ris enabled uav communication networks exploiting deep reinforcement learning,'' \emph{IEEE Transactions on Vehicular Technology}, vol.~73, no.~6, pp. 8404--8419, 2024.

\bibitem{10283889}
X.~Tang, T.~Jiang, J.~Liu, B.~Li, D.~Zhai, F.~R. Yu, and Z.~Han, ``Secure communication with uav-enabled aerial ris: Learning trajectory with reflection optimization,'' \emph{IEEE Transactions on Intelligent Vehicles}, pp. 1--10, 2023.

\bibitem{9448360}
J.~Duan, Y.~Guan, S.~E. Li, Y.~Ren, Q.~Sun, and B.~Cheng, ``Distributional soft actor-critic: Off-policy reinforcement learning for addressing value estimation errors,'' \emph{IEEE Transactions on Neural Networks and Learning Systems}, vol.~33, no.~11, pp. 6584--6598, 2022.

\bibitem{9403369}
W.~Chen, X.~Qiu, T.~Cai, H.-N. Dai, Z.~Zheng, and Y.~Zhang, ``Deep reinforcement learning for internet of things: A comprehensive survey,'' \emph{IEEE Communications Surveys \& Tutorials}, vol.~23, no.~3, pp. 1659--1692, 2021.

\bibitem{9606690}
F.~Tang, H.~Hofner, N.~Kato, K.~Kaneko, Y.~Yamashita, and M.~Hangai, ``A deep reinforcement learning-based dynamic traffic offloading in space-air-ground integrated networks (sagin),'' \emph{IEEE Journal on Selected Areas in Communications}, vol.~40, no.~1, pp. 276--289, 2022.

\bibitem{10545344}
N.~Yang, S.~Chen, H.~Zhang, and R.~Berry, ``Beyond the edge: An advanced exploration of reinforcement learning for mobile edge computing, its applications, and future research trajectories,'' \emph{IEEE Communications Surveys \& Tutorials}, pp. 1--1, 2024.

\bibitem{8382166}
Q.~Mao, F.~Hu, and Q.~Hao, ``Deep learning for intelligent wireless networks: A comprehensive survey,'' \emph{IEEE Communications Surveys \& Tutorials}, vol.~20, no.~4, pp. 2595--2621, 2018.

\bibitem{10348530}
P.~Consul, I.~Budhiraja, and D.~Garg, ``A hybrid secure resource allocation and trajectory optimization approach for mobile edge computing using federated learning based on web 3.0,'' \emph{IEEE Transactions on Consumer Electronics}, vol.~70, no.~1, pp. 1167--1179, 2024.

\bibitem{hou2025split}
X.~Hou, J.~Wang, Z.~Zhang, J.~Wang, L.~Liu, and Y.~Ren, ``Split federated learning for uav-enabled integrated sensing, computation, and communication,'' \emph{arXiv preprint arXiv:2504.01443}, 2025.

\bibitem{10477870}
K.~Heo, W.~Lee, and K.~Lee, ``Uav-assisted wireless-powered secure communications: Integration of optimization and deep learning,'' \emph{IEEE Transactions on Wireless Communications}, vol.~23, no.~9, pp. 10\,530--10\,545, 2024.

\bibitem{10463689}
U.~A. Mughal, Y.~Alkhrijah, A.~Almadhor, and C.~Yuen, ``Deep learning for secure uav-assisted ris communication networks,'' \emph{IEEE Internet of Things Magazine}, vol.~7, no.~2, pp. 38--44, 2024.

\bibitem{9416564}
R.~Dong, B.~Wang, and K.~Cao, ``Deep learning driven 3d robust beamforming for secure communication of uav systems,'' \emph{IEEE Wireless Communications Letters}, vol.~10, no.~8, pp. 1643--1647, 2021.

\bibitem{8755300}
M.~Chen, U.~Challita, W.~Saad, C.~Yin, and M.~Debbah, ``Artificial neural networks-based machine learning for wireless networks: A tutorial,'' \emph{IEEE Communications Surveys \& Tutorials}, vol.~21, no.~4, pp. 3039--3071, 2019.

\bibitem{9816078}
M.~T. Nguyen and L.~B. Le, ``Multi-uav trajectory control, resource allocation, and noma user pairing for uplink energy minimization,'' \emph{IEEE Internet of Things Journal}, vol.~9, no.~23, pp. 23\,728--23\,740, 2022.

\bibitem{8907392}
X.~Liao, J.~Shi, Z.~Li, L.~Zhang, and B.~Xia, ``A model-driven deep reinforcement learning heuristic algorithm for resource allocation in ultra-dense cellular networks,'' \emph{IEEE Transactions on Vehicular Technology}, vol.~69, no.~1, pp. 983--997, 2020.

\bibitem{9069247}
X.~Liao, J.~Si, J.~Shi, Z.~Li, and H.~Ding, ``Generative adversarial network assisted power allocation for cooperative cognitive covert communication system,'' \emph{IEEE Communications Letters}, vol.~24, no.~7, pp. 1463--1467, 2020.

\bibitem{8456560}
Y.~Zhou, P.~L. Yeoh, H.~Chen, Y.~Li, R.~Schober, L.~Zhuo, and B.~Vucetic, ``Improving physical layer security via a uav friendly jammer for unknown eavesdropper location,'' \emph{IEEE Transactions on Vehicular Technology}, vol.~67, no.~11, pp. 11\,280--11\,284, 2018.

\bibitem{10419041}
H.~Cao, C.~Tan, Z.~Gao, Y.~Xu, G.~Chen, P.-A. Heng, and S.~Z. Li, ``A survey on generative diffusion models,'' \emph{IEEE Transactions on Knowledge and Data Engineering}, vol.~36, no.~7, pp. 2814--2830, 2024.

\bibitem{8382275}
D.~Chen, N.~Zhang, N.~Cheng, K.~Zhang, Z.~Qin, and X.~Shen, ``Physical layer based message authentication with secure channel codes,'' \emph{IEEE Transactions on Dependable and Secure Computing}, vol.~17, no.~5, pp. 1079--1093, 2020.

\bibitem{9551779}
G.~Bansal and B.~Sikdar, ``S-maps: Scalable mutual authentication protocol for dynamic uav swarms,'' \emph{IEEE Transactions on Vehicular Technology}, vol.~70, no.~11, pp. 12\,088--12\,100, 2021.

\bibitem{8390918}
B.~Chatterjee, D.~Das, S.~Maity, and S.~Sen, ``Rf-puf: Enhancing iot security through authentication of wireless nodes using in-situ machine learning,'' \emph{IEEE Internet of Things Journal}, vol.~6, no.~1, pp. 388--398, 2019.

\bibitem{9018072}
G.~Bansal, N.~Naren, V.~Chamola, B.~Sikdar, N.~Kumar, and M.~Guizani, ``Lightweight mutual authentication protocol for v2g using physical unclonable function,'' \emph{IEEE Transactions on Vehicular Technology}, vol.~69, no.~7, pp. 7234--7246, 2020.

\bibitem{9745033}
C.~Pu, A.~Wall, K.-K.~R. Choo, I.~Ahmed, and S.~Lim, ``A lightweight and privacy-preserving mutual authentication and key agreement protocol for internet of drones environment,'' \emph{IEEE Internet of Things Journal}, vol.~9, no.~12, pp. 9918--9933, 2022.

\bibitem{ZHANG2024110118}
\BIBentryALTinterwordspacing
Z.~Zhang, C.~Hsu, M.~H. Au, L.~Harn, J.~Cui, Z.~Xia, and Z.~Zhao, ``Prlap-iod: A puf-based robust and lightweight authentication protocol for internet of drones,'' \emph{Computer Networks}, vol. 238, p. 110118, 2024. [Online]. Available: \url{https://www.sciencedirect.com/science/article/pii/S1389128623005637}
\BIBentrySTDinterwordspacing

\bibitem{7420748}
J.~Liu and X.~Wang, ``Physical layer authentication enhancement using two-dimensional channel quantization,'' \emph{IEEE Transactions on Wireless Communications}, vol.~15, no.~6, pp. 4171--4182, 2016.

\bibitem{9592624}
X.~Lu, J.~Lei, Y.~Shi, and W.~Li, ``Improved physical layer authentication scheme based on wireless channel phase,'' \emph{IEEE Wireless Communications Letters}, vol.~11, no.~1, pp. 198--202, 2022.

\bibitem{9335644}
N.~Xie, J.~Chen, and L.~Huang, ``Physical-layer authentication using multiple channel-based features,'' \emph{IEEE Transactions on Information Forensics and Security}, vol.~16, pp. 2356--2366, 2021.

\bibitem{10293903}
Y.~Zhou, Z.~Ma, H.~Liu, P.~L. Yeoh, Y.~Li, and B.~Vucetic, ``Signal-to-noise ratio based physical layer authentication in uav communications,'' in \emph{2023 IEEE 34th Annual International Symposium on Personal, Indoor and Mobile Radio Communications (PIMRC)}, 2023, pp. 1--6.

\bibitem{10591496}
Y.~Shang, Y.~Peng, R.~Ye, and J.~Lee, ``Ris-assisted secure uav communication scheme against active jamming and passive eavesdropping,'' \emph{IEEE Transactions on Intelligent Transportation Systems}, vol.~25, no.~11, pp. 16\,953--16\,963, 2024.

\bibitem{9493713}
Y.~Wu, X.~Guan, W.~Yang, and Q.~Wu, ``Uav swarm communication under malicious jamming: Joint trajectory and clustering design,'' \emph{IEEE Wireless Communications Letters}, vol.~10, no.~10, pp. 2264--2268, 2021.

\bibitem{9367220}
Z.~Shen, K.~Xu, and X.~Xia, ``Beam-domain anti-jamming transmission for downlink massive mimo systems: A stackelberg game perspective,'' \emph{IEEE Transactions on Information Forensics and Security}, vol.~16, pp. 2727--2742, 2021.

\bibitem{9989422}
X.~Li, J.~Chen, X.~Ling, and T.~Wu, ``Deep reinforcement learning-based anti-jamming algorithm using dual action network,'' \emph{IEEE Transactions on Wireless Communications}, vol.~22, no.~7, pp. 4625--4637, 2023.

\bibitem{9777258}
L.~Jia, N.~Qi, F.~Chu, S.~Fang, X.~Wang, S.~Ma, and S.~Feng, ``Game-theoretic learning anti-jamming approaches in wireless networks,'' \emph{IEEE Communications Magazine}, vol.~60, no.~5, pp. 60--66, 2022.

\bibitem{8664589}
F.~Yao and L.~Jia, ``A collaborative multi-agent reinforcement learning anti-jamming algorithm in wireless networks,'' \emph{IEEE Wireless Communications Letters}, vol.~8, no.~4, pp. 1024--1027, 2019.

\bibitem{9079458}
E.~Schmidt, N.~Gatsis, and D.~Akopian, ``A gps spoofing detection and classification correlator-based technique using the lasso,'' \emph{IEEE Transactions on Aerospace and Electronic Systems}, vol.~56, no.~6, pp. 4224--4237, 2020.

\bibitem{9760395}
B.~Pardhasaradhi and L.~R. Cenkeramaddi, ``Gps spoofing detection and mitigation for drones using distributed radar tracking and fusion,'' \emph{IEEE Sensors Journal}, vol.~22, no.~11, pp. 11\,122--11\,134, 2022.

\bibitem{9844986}
Z.~Chen, J.~Li, J.~Li, X.~Zhu, and C.~Li, ``Gnss multiparameter spoofing detection method based on support vector machine,'' \emph{IEEE Sensors Journal}, vol.~22, no.~18, pp. 17\,864--17\,874, 2022.

\bibitem{9760100}
X.~Chen, D.~He, X.~Yan, W.~Yu, and T.-K. Truong, ``Gnss interference type recognition with fingerprint spectrum dnn method,'' \emph{IEEE Transactions on Aerospace and Electronic Systems}, vol.~58, no.~5, pp. 4745--4760, 2022.

\bibitem{9348030}
Y.~Dang, C.~Benzaïd, Y.~Shen, and T.~Taleb, ``Gps spoofing detector with adaptive trustable residence area for cellular based-uavs,'' in \emph{GLOBECOM 2020 - 2020 IEEE Global Communications Conference}, 2020, pp. 1--6.

\bibitem{10.1145/1541880.1541882}
\BIBentryALTinterwordspacing
V.~Chandola, A.~Banerjee, and V.~Kumar, ``Anomaly detection: A survey,'' \emph{ACM Comput. Surv.}, vol.~41, no.~3, Jul. 2009. [Online]. Available: \url{https://doi.org/10.1145/1541880.1541882}
\BIBentrySTDinterwordspacing

\bibitem{balaji2011bayesian}
B.~Balaji and K.~Friston, ``Bayesian state estimation using generalized coordinates,'' \emph{Signal processing, sensor fusion, and target recognition XX}, vol. 8050, pp. 716--727, 2011.

\bibitem{8455592}
M.~Baydoun, D.~Campo, V.~Sanguineti, L.~Marcenaro, A.~Cavallaro, and C.~Regazzoni, ``Learning switching models for abnormality detection for autonomous driving,'' in \emph{2018 21st International Conference on Information Fusion (FUSION)}, 2018, pp. 2606--2613.

\bibitem{pardo2018statistical}
L.~Pardo, \emph{Statistical inference based on divergence measures}.\hskip 1em plus 0.5em minus 0.4em\relax Chapman and Hall/CRC, 2018.

\bibitem{9322583}
A.~Krayani, M.~Baydoun, L.~Marcenaro, A.~S. Alam, and C.~Regazzoni, ``Self-learning bayesian generative models for jammer detection in cognitive-uav-radios,'' in \emph{GLOBECOM 2020 - 2020 IEEE Global Communications Conference}, 2020, pp. 1--7.

\bibitem{xie2025multi}
W.~Xie, G.~Sun, J.~Wang, H.~Du, J.~Kang, K.~Huang, and V.~Leung, ``Multi-objective aerial irs-assisted isac optimization via generative ai-enhanced deep reinforcement learning,'' \emph{arXiv preprint arXiv:2502.10687}, 2025.

\bibitem{wang2024generative}
J.~Wang, H.~Du, Y.~Liu, G.~Sun, D.~Niyato, S.~Mao, D.~I. Kim, and X.~Shen, ``Generative ai based secure wireless sensing for isac networks,'' \emph{arXiv preprint arXiv:2408.11398}, 2024.

\bibitem{10750142}
X.~Wang, C.~P. Tan, Y.~Wang, and X.~Wang, ``Defending uav networks against covert attacks using auxiliary signal injections,'' \emph{IEEE Transactions on Automation Science and Engineering}, pp. 1--13, 2024.

\bibitem{950789}
M.~Valkama, M.~Renfors, and V.~Koivunen, ``Advanced methods for i/q imbalance compensation in communication receivers,'' \emph{IEEE Transactions on Signal Processing}, vol.~49, no.~10, pp. 2335--2344, 2001.

\bibitem{5967984}
J.~Zhang and Y.~R. Zheng, ``Frequency-domain turbo equalization with soft successive interference cancellation for single carrier mimo underwater acoustic communications,'' \emph{IEEE Transactions on Wireless Communications}, vol.~10, no.~9, pp. 2872--2882, 2011.

\bibitem{1207260}
P.~Madhani, P.~Axelrad, K.~Krumvieda, and J.~Thomas, ``Application of successive interference cancellation to the gps pseudolite near-far problem,'' \emph{IEEE Transactions on Aerospace and Electronic Systems}, vol.~39, no.~2, pp. 481--488, 2003.

\bibitem{298053}
P.~Patel and J.~Holtzman, ``Analysis of a simple successive interference cancellation scheme in a ds/cdma system,'' \emph{IEEE Journal on Selected Areas in Communications}, vol.~12, no.~5, pp. 796--807, 1994.

\bibitem{7445815}
M.~L. Psiaki and T.~E. Humphreys, ``Gnss spoofing and detection,'' \emph{Proceedings of the IEEE}, vol. 104, no.~6, pp. 1258--1270, 2016.

\bibitem{6494400}
T.~E. Humphreys, ``Detection strategy for cryptographic gnss anti-spoofing,'' \emph{IEEE Transactions on Aerospace and Electronic Systems}, vol.~49, no.~2, pp. 1073--1090, 2013.

\bibitem{8490218}
Z.~Wu, R.~Liu, and H.~Cao, ``Ecdsa-based message authentication scheme for beidou-ii navigation satellite system,'' \emph{IEEE Transactions on Aerospace and Electronic Systems}, vol.~55, no.~4, pp. 1666--1682, 2019.

\bibitem{wesson2012practical}
K.~Wesson, M.~Rothlisberger, and T.~Humphreys, ``Practical cryptographic civil gps signal authentication,'' \emph{NAVIGATION: Journal of the Institute of Navigation}, vol.~59, no.~3, pp. 177--193, 2012.

\bibitem{10.1145/2973750.2973753}
\BIBentryALTinterwordspacing
A.~Ranganathan, H.~\'{O}lafsd\'{o}ttir, and S.~Capkun, ``Spree: a spoofing resistant gps receiver,'' in \emph{Proceedings of the 22nd Annual International Conference on Mobile Computing and Networking}, ser. MobiCom '16.\hskip 1em plus 0.5em minus 0.4em\relax New York, NY, USA: Association for Computing Machinery, 2016, p. 348–360. [Online]. Available: \url{https://doi.org/10.1145/2973750.2973753}
\BIBentrySTDinterwordspacing

\bibitem{10833728}
M.~Ahmed, A.~A. Soofi, S.~Raza, F.~Khan, S.~Ahmad, W.~U. Khan, M.~Asif, F.~Xu, and Z.~Han, ``Advancements in ris-assisted uav for empowering multiaccess edge computing: A survey,'' \emph{IEEE Internet of Things Journal}, vol.~12, no.~6, pp. 6325--6346, 2025.

\bibitem{10589561}
G.~K. Pandey, D.~S. Gurjar, S.~Yadav, Y.~Jiang, and C.~Yuen, ``Uav-assisted communications with rf energy harvesting: A comprehensive survey,'' \emph{IEEE Communications Surveys \& Tutorials}, pp. 1--1, 2024.

\bibitem{10516683}
P.~Cao, L.~Lei, S.~Cai, G.~Shen, X.~Liu, X.~Wang, L.~Zhang, L.~Zhou, and M.~Guizani, ``Computational intelligence algorithms for uav swarm networking and collaboration: A comprehensive survey and future directions,'' \emph{IEEE Communications Surveys \& Tutorials}, vol.~26, no.~4, pp. 2684--2728, 2024.

\bibitem{10454003}
P.~Li, H.~Zhang, Y.~Wu, L.~Qian, R.~Yu, D.~Niyato, and X.~Shen, ``Filling the missing: Exploring generative ai for enhanced federated learning over heterogeneous mobile edge devices,'' \emph{IEEE Transactions on Mobile Computing}, vol.~23, no.~10, pp. 10\,001--10\,015, 2024.

\bibitem{wang2024empowering}
J.~Wang, Y.~Liu, H.~Du, D.~Niyato, J.~Kang, H.~Zhou, and D.~I. Kim, ``Empowering wireless networks with artificial intelligence generated graph,'' \emph{arXiv preprint arXiv:2405.04907}, 2024.

\bibitem{10648594}
M.~Xu, D.~Niyato, J.~Kang, Z.~Xiong, S.~Mao, Z.~Han, D.~I. Kim, and K.~B. Letaief, ``When large language model agents meet 6g networks: Perception, grounding, and alignment,'' \emph{IEEE Wireless Communications}, vol.~31, no.~6, pp. 63--71, 2024.

\bibitem{zhang2024optimizing}
R.~Zhang, H.~Du, D.~Niyato, J.~Kang, Z.~Xiong, P.~Zhang, and D.~I. Kim, ``Optimizing generative ai networking: A dual perspective with multi-agent systems and mixture of experts,'' \emph{arXiv preprint arXiv:2405.12472}, 2024.

\bibitem{10612249}
A.~H. Arani, P.~Hu, and Y.~Zhu, ``Uav-assisted space-air-ground integrated networks: A technical review of recent learning algorithms,'' \emph{IEEE Open Journal of Vehicular Technology}, vol.~5, pp. 1004--1023, 2024.

\bibitem{10819441}
N.~T.~T. Van, N.~L. Tuan, N.~C. Luong, T.~H. Nguyen, S.~Feng, S.~Gong, D.~Niyato, and D.~I. Kim, ``Network access selection for urllc and embb applications in sub-6ghz-mmwave-thz networks: Game theory versus multi-agent reinforcement learning,'' \emph{IEEE Transactions on Communications}, pp. 1--1, 2024.

\bibitem{10299716}
Q.~Yuan, L.~Xiao, C.~He, P.~Xiao, and T.~Jiang, ``Deep learning-based hybrid precoding for ris-aided broadband terahertz communication systems in the face of beam squint,'' \emph{IEEE Wireless Communications Letters}, vol.~13, no.~2, pp. 303--307, 2024.

\bibitem{9768113}
G.~Geraci, A.~Garcia-Rodriguez, M.~M. Azari, A.~Lozano, M.~Mezzavilla, S.~Chatzinotas, Y.~Chen, S.~Rangan, and M.~D. Renzo, ``What will the future of uav cellular communications be? a flight from 5g to 6g,'' \emph{IEEE Communications Surveys \& Tutorials}, vol.~24, no.~3, pp. 1304--1335, 2022.

\end{thebibliography}

\vfill

\end{document}